\documentclass[12pt]{article}
\PassOptionsToPackage{hyphens}{url}
\usepackage[pagebackref=true,colorlinks]{hyperref}
\usepackage{amsmath,amssymb,amsthm,mathtools,amsrefs}
\usepackage{csquotes}
\usepackage{tikz}
\usetikzlibrary{arrows.meta}
\usetikzlibrary{decorations.pathmorphing,shapes}
\usetikzlibrary{calc}
\usepackage{pgfplots}
\pgfplotsset{compat=newest}
\usepackage{comment}
\usepackage{adjustbox}
\usepackage{stmaryrd}
\usepackage{fancyhdr}
\usepackage{soul}
\usepackage{dsfont}%this is for mathbb{1} but you write mathds{1}
\usepackage[normalem]{ulem}
\usepackage{longtable}
\usepackage[bottom]{footmisc}%this forces footnotes to appear on the bottom of page
\usepackage{caption}
\usepackage{subfig} %this is to plot figures vertically
\usepackage{wrapfig}%this is to allow figures in line with text
\usepackage{pictexwd, dcpic}
\usepackage{float}
\usepackage{enumerate}
\usepackage[all]{xy} %This is to make higher categorical stuff
\hypersetup{
	colorlinks=true,%set true if you want colored links
    linktoc=all,%set to all if you want both sections and subsections linked
    linkcolor=blue,  %choose some color if you want links to stand out
    citecolor=blue,
    }
\makeatletter
\renewcommand*\env@matrix[1][\arraystretch]{%
  \edef\arraystretch{#1}%
  \hskip -\arraycolsep
  \let\@ifnextchar\new@ifnextchar
  \array{*\c@MaxMatrixCols c}}
\makeatother

\theoremstyle{plain}
\newtheorem{theorem}[equation]{Theorem}
\newtheorem{lemma}[equation]{Lemma}
\newtheorem{proposition}[equation]{Proposition}
\newtheorem{corollary}[equation]{Corollary}
\theoremstyle{definition}
\newtheorem{definition}[equation]{Definition}
\newtheorem{construction}[equation]{Construction}
\newtheorem{question}[equation]{Question}
\newtheorem{problem}[equation]{Problem}
\newtheorem{example}[equation]{Example}
\newtheorem{exercise}[equation]{Exercise}
\newtheorem*{answer}{Answer}
\newtheorem*{solution}{Solution}
\newtheorem{remark}[equation]{Remark}

\newtheorem{notation}[equation]{Notation}

\numberwithin{equation}{section}
\newcommand\define[1]{\emph{\textbf{#1}}}%italicize and bold-face %this seems like a good alternative

\let\a=\alpha \let\b=\beta \let\g=\gamma \let\de=\delta 
 \let\h=\eta \let\q=\theta \let\i=\iota 
\let\l=\lambda 
\let\s=\sigma \let\t=\tau    
\let\w=\omega       \let\D=\Delta  
  \let\S=\Sigma   
\let\C=\Chi \let\W=\Omega
\def\vf{\varphi}

\newcommand{\be}{\begin{equation}}
\newcommand{\ee}{\end{equation}}
\def\ba{\begin{align}} %previously this was ``array''
\def\ea{\end{align}}
\newcommand{\bea}{\begin{eqnarray}}
\newcommand{\eea}{\end{eqnarray}}
\newcommand{\bx}{\begin{example}}
\newcommand{\ex}{\end{example}}
\newcommand{\bex}{\begin{exercise}}
\newcommand{\eex}{\end{exercise}}
\newcommand{\ban}{\begin{answer}}
\newcommand{\ean}{\end{answer}}
\newcommand{\bt}{\begin{theorem}}
\newcommand{\et}{\end{theorem}}
\newcommand{\bc}{\begin{corollary}}
\newcommand{\ec}{\end{corollary}}
\newcommand{\blem}{\begin{lemma}}
\newcommand{\elem}{\end{lemma}}
\newcommand{\bp}{\begin{problem}}
\newcommand{\ep}{\end{problem}}
\newcommand{\bn}{\begin{proposition}}
\newcommand{\en}{\end{proposition}}
\newcommand{\bd}{\begin{definition}}
\newcommand{\ed}{\end{definition}}
\newcommand{\bcon}{\begin{construction}}
\newcommand{\econ}{\end{construction}}
\newcommand{\bq}{\begin{question}}
\newcommand{\eq}{\end{question}}
\newcommand{\bprf}{\begin{proof}}
\newcommand{\eprf}{\end{proof}}
\newcommand{\br}{\begin{remark}}
\newcommand{\er}{\end{remark}}
\newcommand{\bs}{\begin{solution}}
\newcommand{\es}{\end{solution}}
\newcommand{\beqs}{\begin{eqnarray}}
\newcommand{\eeqs}{\end{eqnarray}}

\newcommand{\<}{\langle}
\renewcommand{\>}{\rangle}

\newcommand{\id}{\mathrm{id}}

\newcommand{\mC}{\mathcal{C}}
\newcommand{\mM}{\mathcal{M}}

\newcommand{\tr}{{\rm tr} }

\def\R{{{\mathbb R}}}
\def\C{{{\mathbb C}}}

\def\N{{{\mathbb N}}}

\def\B{{{\mathbb B}}}

\def\Re{{{\mathfrak{Re}}}}

\def\mA{{{\mathcal{A}}}}

\def\mB{{{\mathcal{B}}}}

\def\mN{{{\mathcal{N}}}}

\newcommand{\FinProb}{\mathbf{FinProb}}

\newcommand{\stoch}{\;\xy0;/r.25pc/:(-3,0)*{}="1";(3,0)*{}="2";{\ar@{~>}"1";"2"|(1.06){\hole}};\endxy\!}
\newcounter{sarrow}
\newcommand\xstoch[1]{%
\stepcounter{sarrow}%
\mathrel{\begin{tikzpicture}[baseline= {( $ (current bounding box.south) + (0,-0.1ex) $ )}]
\node[inner sep=.5ex] (\thesarrow) {\;$\scriptstyle #1$\;};
\path[draw,{<[scale=1.5,width=3,length=2]}-,decorate,
  decoration={snake,amplitude=0.3mm,segment length=2.1mm,pre=lineto,pre length=1pt}] 
    (\thesarrow.south east) -- (\thesarrow.south west);
\end{tikzpicture}}%
}

\newcommand{\ds}{\displaystyle}
\newcommand{\ben}{\renewcommand{\theenumi}{\alph{enumi}} 
\renewcommand{\labelenumi}{(\theenumi)}\begin{enumerate}}
\newcommand{\een}{\end{enumerate}}

\newcommand\blfootnote[1]{%
  \begingroup
  \renewcommand\thefootnote{}\footnote{#1}%
  \addtocounter{footnote}{-1}%
  \endgroup
}

\usepackage{tocbasic}
\DeclareTOCStyleEntry[
  beforeskip=.2em plus 1pt,% default is 1em plus 1pt
  pagenumberformat=\textbf
]{tocline}{section}
\title{
Non-commutative disintegrations: \\
existence and uniqueness in finite dimensions}
\author{Arthur J.\ Parzygnat
and Benjamin P.\ Russo}
\date{\today}

\newcommand{\Addresses}{{% additional braces for segregating \footnotesize
  \bigskip
  \footnotesize

  A.~Parzygnat, \textsc{Institut des Hautes \'Etudes Scientifiques,
    35 Route de Chartres, 91440, Bures-sur-Yvette, France}\par\nopagebreak
  \textit{E-mail address}, A.~Parzygnat: \texttt{parzygnat@ihes.fr}

  \medskip

  B.~Russo, \textsc{Department of Mathematics, Farmingdale State College SUNY,
    Farmingdale, New York 11735}\par\nopagebreak
  \textit{E-mail address}, B.~Russo: \texttt{russobp@farmingdale.edu}

}}

\begin{document}
\emergencystretch 2em
%\sloppy%this forces text to not go into the margins, though some argue it is not ideal https://tex.stackexchange.com/questions/241343/what-is-the-meaning-of-fussy-sloppy-emergencystretch-tolerance-hbadness/241355#241355    &     https://tex.stackexchange.com/questions/9107/how-can-i-make-my-text-never-go-over-the-right-margin-by-always-hyphenating-or-b
\maketitle 

\begin{abstract}
Motivated by advances in categorical probability, 
we introduce non-commutative almost everywhere (a.e.) equivalence 
and disintegrations in the setting of $C^*$-algebras. 
We show that $C^*$-algebras (resp.\ $W^*$-algebras)
and a.e.\ equivalence classes of 2-positive (resp.\ positive) unital maps form a category. 
We prove non-commutative disintegrations are a.e.\ unique whenever they exist. 
We provide an explicit characterization for when disintegrations exist in the setting of finite-dimensional $C^*$-algebras, and we give formulas for the associated disintegrations. 
\blfootnote{\emph{2020 Mathematics Subject Classification.} 
46L53 (Primary); %Non-commutative probability and statistics
%60B05 (Primary) %Probability theory: Probability measures on topological spaces
%46J10, %Banach algebras of continuous functions, function algebras
47B65, %(Secondary)%Operator Theory: Positive operators and order-bounded operators
%18A40 (Secondary). %Category theory: Adjoint functors
81R15, %(Secondary). %operator algebraic methods in quantum mechanics
%46M99 (Secondary)%Functional analysis: methods of category theory in functional analysis<--- that is for 2010, 2020 has expanded this section 
46M15 (Secondary).%Categories, functors in functional analysis
}
\blfootnote{
\emph{Key words and phrases.} 
Bayesian inverse;
categorical quantum mechanics;
conditional expectation;
optimal hypothesis;
pre-Hilbert module;
quantum measurement;
quantum probability;
regular conditional probability.
%density matrix;
%von Neumann algebra;
%completely positive;
%$C^*$-algebra;
%non-commutative measure theory; 
%non-commutative probability theory; 
%disintegration;
%conditional probability;
%Markov kernel;
%transition kernel;
%stochastic map;
%probability monad;
%Giry monad; 
%Kleisli category;
%quantum operation;
%reversible quantum channel;
%probabilistic programming;
%quantum programming;
%L\"uders projection;
%applied category theory 
}
\end{abstract}

\tableofcontents

%\pagebreak
%%%%%%%%%%%%%%%%%%%%%%%%%%%%%%%%%%%%%

%%%%%%%%%%%%%%%%%%%%%%%%%%%%%%%%%%%%%
\section{Introduction and outline} 
%%%%%%%%%%%%%%%%%%%%%%%%%%%%%%%%%%%%%
%
%%%%%%%%%%%%%%%%%%%%%%%%%%%%%%%%%%%%%%
%\subsection{Main purpose} 
%%%%%%%%%%%%%%%%%%%%%%%%%%%%%%%%%%%%%%

Regular conditional probabilities, 
optimal hypotheses, 
disintegrations of one measure over another consistent with a measure-preserving map, 
conditional expectations, 
perfect error-correcting codes, and 
sufficient statistics
are all examples of a single mathematical notion. We call this notion a \emph{disintegration}. Although we only make the connection between our definition of disintegration and the first three examples listed, relationships to the other notions are described in~\cite{PaBayes}, and further connections will be made in subsequent work. 
In this paper, our primary focus is to provide necessary and sufficient conditions for the existence and uniqueness of disintegrations in the setting of finite-dimensional $C^*$-algebras. 

Developing this and related ideas is part of a larger program in extending Bayesian statistics to the \emph{non-commutative} setting~\cites{PaRuBayes,PaBayes,PaConditionals} in such a way so that it is compatible with a recently developed \emph{categorical} framework for classical statistics~\cites{ChJa18,Fr19}.  
These recent advances in classical
categorical functional analysis and measure theory provide a suitable notion of disintegration~\cites{La62,Se73,Sw74,Gi82,We94,BFL,CDDG17,ChJa18,Ja18,Fr19}, whose diagrammatic formulation can be transferred from 
a category of probability spaces to a category of states on $C^*$-algebras. 
This is achieved by utilizing a fully faithful (contravariant) functor from the former to the latter~\cites{Pa17,FuJa15}. 
This categorical perspective offers a candidate for 
generalizing disintegrations to non-commutative probability theory
without relying on the specific measure-theoretic details of classical probability theory. 
Since a disintegration is a special kind of \emph{Bayesian inverse}~\cites{ChJa18,PaBayes,PaRuBayes}, this article serves as a step towards a theory of non-commutative Bayesian inversion. 

Briefly, the definition of a disintegration 
of a state $\w$ over another state $\xi$ consistent with a 
unital $^{*}$-homomorphism $F$ preserving these states
is a completely positive unital map $R$ in the \emph{reverse} direction that is both state preserving and a left inverse of $F$ modulo the null space of $\xi$.
If $^{*}$-homomorphisms are written as straight arrows $\to$ 
and completely positive unital maps are written as squiggly arrows $\stoch,$
this definition of a disintegration can be summarized diagrammatically as
\be
\xy0;/r.25pc/:
(0,-7.5)*+{\C}="C";
(-12.5,7.5)*+{\mA}="H";
(12.5,7.5)*+{\mB}="K";
{\ar@{~>}"H";"C"_{\w}};
{\ar@{~>}"K";"C"^{\xi}};
{\ar@/_0.75pc/"K";"H"_{F}};
{\ar@{~>}@/_0.75pc/"H";"K"_{R}};
\endxy
\quad\text{ such that }\quad
\xy0;/r.25pc/:
(0,-7.5)*+{\C}="C";
(-10,7.5)*+{\mA}="H";
(10,7.5)*+{\mB}="K";
{\ar@{~>}"H";"C"_{\w}};
{\ar@{~>}"K";"C"^{\xi}};
{\ar@{~>}"H";"K"^{R}};
{\ar@{=}(-3,0);(5,4.5)};
\endxy
\quad\text{and}\quad
\xy0;/r.25pc/:
(0,-7.5)*+{\mA}="C";
(-10,7.5)*+{\mB}="H";
(10,7.5)*+{\mB}="K";
{\ar"H";"C"_{F}};
{\ar@{~>}"C";"K"_{R}};
{\ar"H";"K"^{\id_{\mB}}};
{\ar@{=}(-3,0);(5,4.5)_{\xi}};
\endxy
\ee
in the category of finite-dimensional $C^*$-algebras and completely positive unital maps. The right-most diagram commutes almost everywhere (a.e.), in a sense that we make precise in this article. We introduce and develop non-commutative a.e.\ equivalence in order to properly address the uniqueness properties of disintegrations. 

The interpretation of completely positive unital maps as quantum conditional probabilities is not new \cite{Le06}, but 
we take this perspective further and include the relationships between states and partially reversible dynamics analogous to what \emph{regular} conditional probabilities accomplish in classical statistics. 
Our core result is Theorem~\ref{thm:diagonalimpliesseparable}, which specializes to the case where $\mA$ and $\mB$ are matrix algebras and $F$ sends $B\in\mB$ to $\mathrm{diag}(B,\dots,B)$.
If we express our states $\omega$ and $\xi$ in terms of density matrices $\rho$ and $\sigma$, respectively, Theorem~\ref{thm:diagonalimpliesseparable} says that a unique disintegration exists
if and only if there exists a density matrix $\t$ such that $\rho=\t\otimes\s$. This is closely related to a well-known result on the existence of state-preserving conditional expectations~\cite{Pe08}, but our notion generalizes it due to our weakened assumption of a.e.\ equivalence. 

Our subsequent results are generalizations of this theorem and culminate in Theorem~\ref{thm:theoremaegregiumarbitraryF}, which assumes $\mA$ and $\mB$ are arbitrary finite-dimensional $C^*$-algebras and $F$ is an arbitrary unital $^*$-homomorphism. 
We provide explicit formulas for disintegrations 
and we analyze several examples, including one involving entanglement, which has its origins in the work of Einstein, Podolsky, and Rosen~\cite{EPR}.
In Example~\ref{ex:classicaldisintegrations},
we show how the standard classical theorem on the existence
and uniqueness of disintegrations (Theorem~\ref{thm:classicalmeasprestocondition}) is a direct corollary of our theorem. We conclude by exploring consequences of our characterization theorem in the context of measurement in quantum information theory. Finally, Appendix~\ref{app:optimalhyparedis} reviews stochastic maps (Markov kernels) and justifies our usage of the terminology `disintegration' by showing that the diagrammatic notion agrees with a general measure-theoretic one.

%%%%%%%%%%%%%%%%%%%%%%%%%%%%%%%%%%%%%
\section{Non-commutative a.e.\ equivalence} 
\label{sec:noncae}
%%%%%%%%%%%%%%%%%%%%%%%%%%%%%%%%%%%%%

For classical probability spaces, 
a.e.\ equivalence specifies the degree of uniqueness of disintegrations, Bayesian inverses, and conditional distributions. 
The same is true in the quantum/non-commutative setting. 
In this section, we first recall some relevant definitions involving states on $C^*$-algebras and completely positive maps from Paulsen~\cite{Pa02} and Sakai~\cite{Sa71} to establish notation and terminology. Afterwards, we define a.e.\ equivalence for linear maps between $C^*$-algebras in Definition~\ref{defn:aestates}. We provide a more computationally useful definition for finite-dimensional $C^*$-algebras in Lemma~\ref{lem:aeidentitysupport}. 
In the rest of this section, we analyze several properties of a.e.\ equivalence. 

\bd
A \define{$C^*$-algebra} is an algebra $\mA$ 
equipped with a unit $1_{\mA},$ an involution $^*:\mA\to\mA$, and a norm 
$\lVert\;\cdot\;\rVert:\mA\to\R$ 
such that it is a $^{*}$-algebra, 
it is closed with respect to the topology induced by its norm,
and it satisfies the $C^*$-identity, 
which says $\lVert a^*a\rVert=\lVert a\rVert^2$ for all $a\in\mA.$
Given a $C^*$-algebra $\mA,$ 
a \define{positive element} of $\mA$ 
is an element $a\in\mA$ for which there exists an $x\in\mA$
such that $a=x^*x.$ 
The set of positive elements in $\mA$ 
is denoted by $\mA^{+}.$
Given another $C^*$-algebra $\mB,$ a \define{positive map}
$\vf:\mB\stoch\mA$ is a linear map such that 
$\vf(\mB^{+})\subseteq\mA^{+}.$ 
A linear map $\vf:\mB\stoch\mA$ is \define{unital} iff $\vf(1_{\mB})=1_{\mA}.$ 
A \define{state} on a 
$C^*$-algebra $\mA$ is a positive
linear unital functional $\w:\mA\stoch\C.$
A \define{$^{*}$-homomorphism}
from $\mA$ to $\mB$ is a function $f:\mA\to\mB$ 
preserving the $C^*$-algebra structure, namely $f$ is linear, 
$f$ is multiplicative $f(aa')=f(a)f(a'),$
$f$ is unital $f(1_{\mA})=1_{\mB},$ 
and $f(a^*)=f(a)^*$ for all $a,a'\in\mA.$  
If $\w:\mA\stoch\C$ and $\xi:\mB\stoch\C$ are states, then a linear map
$\vf:\mB\stoch\mA$ is said to be \define{state-preserving} whenever $\vf\circ\w=\xi$, and the notation $(\mB,\xi)\xstoch{\varphi}(\mA,\omega)$ will be used to indicate this. 
\ed

All $C^*$-algebras and $^*$-homomorphisms will be unital unless specified otherwise. 
Note that positive (and linear) maps on $C^*$-algebras are denoted with squiggly arrows
$\stoch$, while $^{*}$-homomorphisms are denoted with straight arrows $\to.$

\bx
For each $n\in\N,$ let $\mathcal{M}_{n}(\C)$ denote the set of 
$n\times n$ complex matrices. The involution applied to $A\in\mathcal{M}_{n}(\C)$ 
is given by 
the conjugate transpose and is written as $A^{\dag}$ instead of 
$A^{*}$ to be consistent with the standard notation used in quantum theory.
A \define{matrix algebra} is a $C^*$-algebra
of the form $\mathcal{M}_{n}(\C)$ for some $n\in\N.$
If $\mA$ is another $C^*$-algebra, then 
$\mathcal{M}_{n}(\C)\otimes\mA\cong\mathcal{M}_{n}(\mA)$, 
the algebra of $n\times n$
matrices with entries in $\mA,$ admits a $C^*$-algebra structure
by matrix operations and a norm that can be obtained in many ways (cf.\ Chapter~1 in Paulsen~\cite{Pa02}). 
\ex

Our convention for the tensor product (also called the Kronecker product) of matrices will be 
\be
\label{eq:kroneckerproduct}
\begin{bmatrix}
a_{11}&\cdots&a_{1m}\\
\vdots&&\vdots\\
a_{m1}&\cdots&a_{mm}
\end{bmatrix}
\otimes
\begin{bmatrix}
b_{11}&\cdots&b_{1n}\\
\vdots&&\vdots\\
b_{n1}&\cdots&b_{nn}
\end{bmatrix}
=
\begin{bmatrix}
a_{11}\!\begin{bmatrix}
b_{11}&\cdots&b_{1n}\\
\vdots&&\vdots\\
b_{n1}&\cdots&b_{nn}
\end{bmatrix}&\cdots&a_{1m}\!\begin{bmatrix}
b_{11}&\cdots&b_{1n}\\
\vdots&&\vdots\\
b_{n1}&\cdots&b_{nn}
\end{bmatrix}\\
\vdots&&\vdots\\
a_{m1}\!\begin{bmatrix}
b_{11}&\cdots&b_{1n}\\
\vdots&&\vdots\\
b_{n1}&\cdots&b_{nn}
\end{bmatrix}&\cdots&a_{mm}\!\begin{bmatrix}
b_{11}&\cdots&b_{1n}\\
\vdots&&\vdots\\
b_{n1}&\cdots&b_{nn}
\end{bmatrix}
\end{bmatrix}
,
\ee
which is induced by the isomorphism $\C^{m}\otimes\C^{n}\to\C^{mn}$ determined by 
\be
\label{eq:tensorproductiso}
\vec{e}_{1}\otimes\vec{e}_{1}\mapsto\vec{e}_{1},
\quad\dots\quad
\vec{e}_{1}\otimes\vec{e}_{n}\mapsto\vec{e}_{n},
\quad
\vec{e}_{2}\otimes\vec{e}_{1}\mapsto\vec{e}_{n+1},
\quad\dots\quad
\vec{e}_{m}\otimes\vec{e}_{n}\mapsto\vec{e}_{mn}.
\ee
Here, $\vec{e}_{i}$ denotes the standard $i$-th unit vector in $\C^{n}$ regardless of $n$.

\bd
Let $\mA$ and $\mB$ be $C^*$-algebras. 
Given $n\in\N,$ 
a linear map $\vf:\mB\stoch\mA$ is \define{$n$-positive} iff 
$\id_{\mathcal{M}_{n}(\C)}\otimes\vf:\mathcal{M}_{n}(\C)\otimes\mB\stoch\mathcal{M}_{n}(\C)\otimes\mA$ 
is positive. The map 
$\vf$ is \define{completely positive} iff $\vf$ is $n$-positive for all $n\in\N.$
A completely positive (unital) map will be abbreviated as
a CP (CPU) map. 
\ed

The Choi--Kraus theorem gives a characterization of completely
positive maps between \emph{matrix} algebras. This will be used often, 
so we state it here to set notation~\cites{Ch75,Kr83}. 

\bt%[Kraus' Theorem~\cite{Kr83}]
\label{thm:Kraustheorem}
Fix $n,m\in\N.$ A linear map $R:\mathcal{M}_{n}(\C)\stoch\mathcal{M}_{m}(\C)$ is completely positive
if and only if there exists a finite collection 
$\{R_{i}:\C^{n}\to\C^{m}\}$ of linear maps such that 
\be
\label{eq:krausdecomp}
R=\sum_{i}\mathrm{Ad}_{R_{i}}.
\ee
Here, $\mathrm{Ad}_{R_{i}}(A):=R_{i}AR_{i}^{\dag}$ 
for all $A\in\mathcal{M}_{n}(\C).$ The map $R$ is CPU if and only if, 
in addition, 
$\sum_{i}R_{i}R_{i}^{\dag}=\mathds{1}_{m}.$
\et

A collection $\{R_{i}\}$ satisfying (\ref{eq:krausdecomp}) 
is called a \define{Kraus decomposition} for $R.$ 

\br
\label{rmk:notationQIT}
The standard assumption in quantum information theory is to work with
completely positive \emph{trace-preserving} maps instead of
\emph{unital} maps. 
The former class, typically called \emph{quantum operations/channels} (cf.\ Section 8.2 in Nielsen and Chuang~\cite{NiCh11}) is used in the Schr\"odinger representation when transforming physical states, while the latter is used in the Heisenberg representation when transforming physical observables. 
We briefly explain the relationship between the two. A completely positive map 
$R:\mathcal{M}_{n}(\C)\stoch\mathcal{M}_{m}(\C)$ is
unital if and only if the dual map 
$R^*:\mathcal{M}_{m}(\C)\stoch\mathcal{M}_{n}(\C)$
is trace-preserving. The dual map $R^*$ is defined with respect
to the \define{Hilbert--Schmidt} (a.k.a.\ \define{Frobenius}) 
inner product on a matrix algebra, which is given by 
$\<A,B\>:=\tr(A^{\dag}B)$
for all square matrices (of the same dimension) $A$ and $B.$ 
Therefore, $R^*$ is the unique map satisfying 
$\<R^*(A),B\>=\<A,R(B)\>$
for all $A\in\mathcal{M}_{m}(\C)$
and all $B\in\mathcal{M}_{n}(\C).$ 
If $R=\sum_{i}\mathrm{Ad}_{R_{i}}$ is a Kraus decomposition of
$R,$ then $R^*=\sum_{i}\mathrm{Ad}_{R_{i}^{\dag}}$ is a Kraus
decomposition of $R^*$ because
$\<R^*(A),B\>=\<A,R(B)\>
=\sum_{i}\tr(A^{\dag}R_{i}BR_{i}^{\dag})
=\sum_{i}\tr(R_{i}^{\dag}A^{\dag}R_{i}B)
=\sum_{i}\<R_{i}^{\dag}AR_{i},B\>$ 
by the cyclicity of the
trace. By a similar argument, 
$\<R^*(\mathds{1}_{m}),B\>=\<\mathds{1}_{m},R(B)\>=\tr(R^*(B))$
for all $B\in\mathcal{M}_{n}(B).$
From this, it follows that
$R$ is unital if and only if $R^*$ is trace-preserving. 

In particular, if $\w:\mathcal{M}_{n}(\C)\stoch\C$ is a state, then
its dual $\w^*:\C\stoch\mathcal{M}_{n}(\C)$ is determined by 
the image $\w^{*}(1)$ of the unit $1$ in $\C$, which is positive. Furthermore, 
since $\w$ is unital, $\w^{*}$ is trace-preserving. Hence, 
$\tr(\w^{*}(1))=1.$ In other words, $\w^{*}(1)$ is a trace $1$
positive matrix. This is called the \define{density matrix}
associated to $\w.$ 
Finally, $\w=\tr(\w^*(1)\;\cdot\;)$ as states on $\mathcal{M}_{n}(\C).$
\er

We now proceed to defining a.e.\ equivalence of linear maps on $C^*$-algebras. 

\bd
\label{defn:aestates}
Let $\mA$ and $\mB$ be 
$C^*$-algebras,
let $F,F':\mB\stoch\mA$ be two 
linear maps, and let $\w:\mA\stoch\C$ be a 
state on $\mA$ (or more generally a positive linear functional). Let 
\be
\mathcal{N}_{\w}:=\big\{a\in\mA\;:\;\w(a^*a)=0\big\}
\ee
denote the \define{null space} of $\w.$ 
Since $\mathcal{N}_{\w}$ is a left ideal of $\mA$ 
(see Construction~3.1 in \cite{PaGNS} for details)
denote the quotient vector space by $\mA/\mathcal{N}_{\w}.$
The maps $F$ and $F'$ are said to be 
\define{equal almost everywhere (a.e.) with respect to $\w$}
or \define{equal $\w$-a.e.} iff the diagram (in the category of vector spaces and linear maps)
\be
\label{eq:omegaequalae}
\xy0;/r.25pc/:
(15,0)*+{\mA/\mathcal{N}_{\w}}="AN";
(-15,0)*+{\mB}="B";
(0,7.5)*+{\mA}="AF";
(0,-7.5)*+{\mA}="AG";
{\ar@{~>}"B";"AF"^{F}};
{\ar@{~>}"B";"AG"_{F'}};
{\ar@{->>}"AF";"AN"};
{\ar@{->>}"AG";"AN"};
\endxy
\ee
commutes, i.e.\ iff $F(b)-F'(b)\in\mathcal{N}_{\w}$ for all $b\in\mB.$ 
The map $\mA\twoheadrightarrow\mA/\mathcal{N}_{\w}$ in 
(\ref{eq:omegaequalae}) is the quotient map of $\mA$ onto $\mA/\mathcal{N}_{\w}.$ 
When $F$ and $F'$ are equal $\w$-a.e., the notation
$
F\underset{\raisebox{.6ex}[0pt][0pt]{\scriptsize$\w$}}{=}F'
$
will be used. 
\ed

The justification for the above terminology of a.e.\ equivalence is explained in the following illustrative example of finite probability spaces (cf.\ Appendix~\ref{app:optimalhyparedis} for terminology).

\bx
\label{ex:aecommutativealgebras}
Let $\mA:=\C^{X}$ and $\mB:=\C^{Y}$ be the commutative $C^*$-algebras of
complex-valued functions on the finite sets $X$ and $Y,$ respectively, 
let $P:\C^{X}\stoch\C$ be a state on $X$, and let $F,G:\C^{Y}\stoch \C^{X}$ be two 
positive unital maps. Then there exists a unique
probability measure $p$ on $X$ such that
%\be
%\label{eq:probabilitytostate}
%\sum_{x\in X}\vf(x)p_{x}=P(\vf)\qquad\forall\;\vf\in \C^{X},
%\ee
$\sum\limits_{x\in X}\vf(x)p_{x}=P(\vf)$ for all $\vf\in \C^{X}$ (see Section~2.6 of \cite{Pa17} for details). 
Namely, 
$
p_{x}:=P(e_{x}),
$
where $e_{x}$ is the function on $X$ defined by 
$X\ni x'\mapsto e_{x}(x'):=\de_{xx'}.$
Similarly, there exist unique stochastic maps 
$f,g:X\stoch Y$ such that 
\be
\big(F(\psi)\big)(x)=\sum_{y\in Y}\psi(y)f_{yx}
\qquad\forall\;\psi\in \C^{Y},\;\forall\;x\in X,
\ee
namely 
\be
X\ni x\mapsto \Big(Y\ni y\mapsto f_{yx}:=F(e_{y})(x)\Big)
\ee
and similarly for $G$ with $g.$
One can show that 
the null space of $P$ is given by
\be
\mathcal{N}_{P}:=\big\{\vf\in\C^{X}\;:\;P(\vf^*\vf)=0\big\}
=\big\{\vf\in\C^{X}\;:\;\vf\big|_{X\setminus N_{p}}=0\big\}
=\mathrm{span}\left(\bigcup_{x\in N_{p}}\big\{e_{x}\big\}\right),
\ee
where $N_{p}\subseteq X$ is the measure-theoretic null space of $p$
and $\vf|_{X\setminus N_{p}}$ denotes the restriction of $\vf$ to 
$X\setminus N_{p}.$
Hence, the quotient $\C^{X}/\mathcal{N}_{P}$ is isomorphic to functions on 
$X\setminus N_{p}$ by the isomorphism  
\be
\begin{split}
\C^{X}/\mathcal{N}_{P}&\to\C^{X\setminus N_{p}}\\
[\vf]&\mapsto\vf\big|_{X\setminus N_{p}}.
\end{split}
\ee
As a result, the two positive unital maps $F,G:\C^{Y}\stoch\C^{X}$ 
are equal $P$-a.e.\ if and only if 
the associated stochastic maps
$\tilde{f},\tilde{g}:X\setminus N_{p}\stoch Y$ defined by 
the restrictions of $f$ and $g$ to $X\setminus N_{p},$ respectively, 
are equal.
This precisely means $f\underset{\raisebox{.6ex}[0pt][0pt]{\scriptsize$p$}}{=}g.$
\ex

We now proceed to establishing several important facts regarding non-commutative a.e.\ equivalence. First, if two maps are a.e.\ equivalent in terms of some state, then they pullback that state to the same state. 

\blem
\label{lem:aestatespullback}
Let $\mA$ and $\mB$ be $C^*$-algebras, let $\xi:\mB\stoch\C$ be a state (or more generally a positive functional),
and let $\vf,\psi:\mA\stoch\mB$
be linear maps. 
If $\vf\underset{\raisebox{.6ex}[0pt][0pt]{\scriptsize$\xi$}}{=}\psi,$ 
then $\xi\circ\vf=\xi\circ\psi$.
\elem

\bprf
Let $a\in\mA$. Then 
\be
\big|\xi\big(\vf(a)-\psi(a)\big)\big|^{2}
\le\xi\Big(\big(\vf(a)-\psi(a)\big)^*\big(\vf(a)-\psi(a)\big)\Big)
=0
\ee
by the Cauchy--Schwarz inequality for positive functionals (cf.\ Proposition~5.2.1 in Fillmore~\cite{Fi96}) %set a=1 and b=\vf(a)-\psi(a)
and because $\vf(a)-\psi(a)\in\mathcal{N}_{\xi}.$ Hence, 
$\xi\big(\vf(a)\big)=\xi\big(\psi(a)\big).$ Since $a$ was arbitrary, 
$\xi\circ\vf=\xi\circ\psi.$ 
\eprf

The support of a state will also be useful in when formulating and proving our disintegration theorem.

\blem
\label{lem:supportofstate}
Let $\w:\mA\stoch\C$ be a state on a finite-dimensional $C^*$-algebra $\mA$ (or more generally a $W^*$-algebra). Then there exists a unique projection $P_{\w}\in\mA$ (this means $P_{\w}^{*}=P_{\w}$ and $P_{\w}^{2}=P_{\w}$) such that $\mathcal{N}_{\w}=\mA(1_{\mA}-P_{\w}).$ Equivalently, $P_{\w}$ is characterized by
\be
\label{eq:supportstateidentities}
\w(a)=\w(aP_{\w})=\w(P_{\w}a)=(\w\circ\mathrm{Ad}_{P_{\w}})(a)\qquad\forall\;a\in\mA.
\ee
\elem

\bprf
See Section~1.14 of Sakai~\cite{Sa71}. 
\eprf

\br
\label{eq:Wstaralgebrassupport}
If $\mA$ is not a finite-dimensional $C^*$-algebra in Lemma~\ref{lem:supportofstate}, then such a projection $P_{\w}$ for a state $\w$ need not exist. Indeed, if $\mA=C(X),$ continuous complex-valued functions on a connected compact Hausdorff space $X,$ then there are no non-trivial projections and yet there are many states generating non-trivial null spaces. Such a projection does exist, however, if $\mA$ is a $W^*$-algebra. Hence, several (but not all) of the results that follow involving such supports also hold for $W^*$-algebras. 
\er

\bd
\label{defn:support}
Using the same notation from Lemma~\ref{lem:supportofstate}, 
$P_{\w}$ is called the \define{support of $\w$}. Its complement
will be denoted by 
$
P_{\w}^{\perp}:=1_{\mA}-P_{\w}.
$
\ed

\bx
\label{ex:densitymatrixsupport}
When $\mA=\mathcal{M}_{n}(\C)$ is a matrix algebra with a state 
$\w:\mathcal{M}_{n}(\C)\stoch\C,$ then 
$\w=\tr(\rho\;\cdot\;)$ for some unique density matrix $\rho\in\mathcal{M}_{n}(\C)$
(cf.\ Remark~\ref{rmk:notationQIT}).
In this case, 
$P_{\w}^{\perp}$
is the projection onto the zero eigenspace of $\rho$ 
and $P_{\w}$ satisfies $P_{\w}\rho=\rho=\rho P_{\w}.$ 
\ex

\blem
\label{lem:projectiondecomp}
Let $P$ be a projection in a $C^*$-algebra (or $W^*$-algebra) $\mB$ and let $B\in\mB.$ Then $BP=0$ implies 
$B\in\mB P^{\perp}.$
\elem

\bprf
This follows from the fact that 
every $B\in\mB$ can be uniquely expressed as a sum of four terms
\be
\label{eq:decomposelements}
B=(P+P^{\perp})B(P+P^{\perp})=PBP+PBP^{\perp}+P^{\perp}BP+P^{\perp}BP^{\perp}, 
\ee
the non-zero ones of which are linearly independent. 
\eprf

The decomposition~(\ref{eq:decomposelements}) will be used frequently in this work,%
%footnote{
\footnote{The usage of such a decomposition is certainly not new. More recently, they have made an appearance in the study of Pierce and corner algebras, also in the context of conditional expectations~\cite{Pluta13}. We thank Chris Heunen for informing us of this reference.
}
%end footnote
 particularly in conjunction with the support of a state. For example, we have the following alternative and computationally useful characterization of a.e.\ equivalence.

\blem
\label{lem:aeidentitysupport}
Let $\mA$ and $\mB$ be finite-dimensional $C^*$-algebras, let $\xi:\mB\stoch\C$ be a state,
and let $\vf,\psi:\mA\stoch\mB$
be 
linear maps. 
Let $P_{\xi}\in\mB$ denote the support of $\xi.$ 
\begin{enumerate}[i.]
\itemsep0pt
\item
\label{partioflem:aeidentitysupport}
Then $\vf\underset{\raisebox{.6ex}[0pt][0pt]{\scriptsize$\xi$}}{=}\psi$ if and only if\,%
%footnote
\footnote{In the case where $\mA=\mathcal{M}_{m}(\C)$ and $\mB=\mathcal{M}_{n}(\C)$, this says that the two maps $\vf(A)$ and $\psi(A)$ agree when restricted to the subspace $P_{\xi}\C^{n}\subseteq\C^{n}$.
}
%end footnote
$\vf(A)P_{\xi}=\psi(A)P_{\xi}$ for all $A\in\mA.$
\item
\label{partiioflem:aeidentitysupport}
If $\vf\underset{\raisebox{.6ex}[0pt][0pt]{\scriptsize$\xi$}}{=}\psi,$ 
then 
$
\mathrm{Ad}_{P_{\xi}}\circ\vf=\mathrm{Ad}_{P_{\xi}}\circ\psi.
$
\end{enumerate}
\elem

\bprf
For the first claim, if $\vf\underset{\raisebox{.6ex}[0pt][0pt]{\scriptsize$\xi$}}{=}\psi$, then $\vf(A)-\psi(A)=B P_{\xi}^{\perp}$ for some $B\in\mB$ since $\mathcal{N}_{\xi}=\mB P_{\xi}^{\perp}$ by Lemma~\ref{lem:supportofstate}. Multiplying by $P_{\xi}$ on the right gives item~\ref{partioflem:aeidentitysupport}. Conversely, if $\vf(A)P_{\xi}=\psi(A)P_{\xi}$ holds, then $\big(\vf(A)-\psi(A)\big)P_{\xi}=0$. 
Hence, $\vf(A)-\psi(A)\in\mB P_{\xi}^{\perp}$ by Lemma~\ref{lem:projectiondecomp}. 
Item~\ref{partiioflem:aeidentitysupport} follows from item~\ref{partioflem:aeidentitysupport} by multiplying $\vf(A)P_{\xi}=\psi(A)P_{\xi}$ on the left by $P_{\xi}.$ 
\eprf

Since CP maps between matrix algebras have particularly simple forms (cf.\ Theorem~\ref{thm:Kraustheorem}),
it will also be useful to have a more quantitative version of Lemma~\ref{lem:aeidentitysupport}.
To state it, we first recall a general fact about the relationship
between two Kraus decompositions of a CP map of matrix algebras. 

\blem
\label{lem:ChoiKraus}
Let $\vf:\mathcal{M}_{m}(\C)\stoch\mathcal{M}_{n}(\C)$ be a CP map and suppose  
\be
\sum_{i=1}^{p}\mathrm{Ad}_{V_{i}}=\vf=\sum_{j=1}^{q}\mathrm{Ad}_{W_{j}}
\ee
are two Kraus decompositions of $\vf$ with $p\ge q.$  
Then there exists a $q\times p$ matrix $U$ that is a coisometry (meaning $UU^{\dag}=\mathds{1}_{q}$, i.e.\ the rows of $U$ are orthonormal) such that 
$
V_{i}=\sum\limits_{j=1}^{q}u_{ji}W_{j}
$
for all $i\in\{1,2,\dots,p\}.$ 
Here $u_{ji}$ denotes the $ji$-th entry of $U.$ 
\elem

\bprf
%See Theorem 8.2. in \cite{NiCh11} for the case when $m=n.$ 
The reader is referred to Sections~6 and 7 of \cite{Pa18} for any unexplained details and terminology. First note that every such Kraus decomposition $\vf=\sum\limits_{i=1}^{p}\mathrm{Ad}_{V_{i}}$ can be expressed as a Stinespring representation $\vf=\pi\circ\mathrm{Ad}_{V},$ where 
$\pi$ and $V$ are defined by
\be
\mathcal{M}_{m}(\C)\ni A\xmapsto{\pi}\mathds{1}_{p}\otimes A\in\mathcal{M}_{p}(\C)\otimes\mathcal{M}_{m}(\C)
\ee
and
\be
\C^{p}\otimes\C^{m}\cong\C^{pm}\xrightarrow{V:=[\;V_{1}\;\cdots\;V_{p}\;]}\C^{n}, 
\ee
respectively (and similarly for $\vf=\sum\limits_{j=1}^{q}\mathrm{Ad}_{W_{j}}$ and $W:=\begin{bmatrix}W_{1}&\cdots&W_{q}\end{bmatrix}$). By the universal property of Stinespring representations (see Theorem~6.29 and the end of Section~7 in \cite{Pa18}), there exists a coisometry%
\footnote{Technically, the theorem referenced claims there exists a partial isometry. However, this partial isometry can be extended to a coisometry by similar techniques to those employed in Example~7.27 and Theorem~7.30  in \cite{Pa18}.}
$\C^{p}\xrightarrow{U}\C^{q}$ such that the diagram 
\be
\label{eq:commutativityStinespring}
\xy0;/r.25pc/:
(-10,0)*+{\C^{n}}="n";
(10,7.5)*+{\C^{q}\otimes\C^{m}}="qm";
(10,-7.5)*+{\C^{p}\otimes\C^{m}}="pm";
{\ar"pm";"n"^{V}};
{\ar"qm";"n"_{W}};
{\ar"pm";"qm"_{U\otimes\mathds{1}_{m}}};
\endxy
\ee
commutes. Writing 
\be
U\otimes\mathds{1}_{m}=\begin{bmatrix}u_{11}\mathds{1}_{m}&\cdots&u_{1p}\mathds{1}_{m}\\
\vdots&&\vdots\\
u_{q1}\mathds{1}_{m}&\cdots&u_{qp}\mathds{1}_{m}\\
\end{bmatrix}
,
\ee
we see that commutativity of (\ref{eq:commutativityStinespring}) gives
\be
\begin{bmatrix}V_{1}&\cdots&V_{p}\end{bmatrix}
=\begin{bmatrix}W_{1}&\cdots&W_{q}\end{bmatrix}(U\otimes\mathds{1}_{n}),
\ee
which is the result claimed. 
\eprf

\blem
\label{lem:CPUtoidentity}
Fix a positive integer $n,$ let $\xi:\mathcal{M}_{n}(\C)\stoch\C$ be a state, and 
let $\vf:\mathcal{M}_{n}(\C)\stoch\mathcal{M}_{n}(\C)$ be a CPU map such that
$\vf\underset{\raisebox{.6ex}[0pt][0pt]{\scriptsize$\xi$}}{=}\id_{\mathcal{M}_{n}(\C)}.$
Let $P_{\xi}\in\mathcal{M}_{n}(\C)$ denote the support of $\xi$ and 
let $\vf$ have a Kraus decomposition $\vf=\sum\limits_{i=1}^{p}\mathrm{Ad}_{V_{i}}.$ 
Then there exist complex numbers $\{\a_{i}\}_{i\in\{1,\dots,p\}}$ such that 
\be
P_{\xi}V_{i}=\a_{i}P_{\xi}\qquad\forall\;i\in\{1,\dots,p\}
\quad\text{ and }\quad
\sum_{i=1}^{p}|\a_{i}|^2=1. 
\ee
\elem

\bprf
Since $\vf\underset{\raisebox{.6ex}[0pt][0pt]{\scriptsize$\xi$}}{=}\id_{\mathcal{M}_{n}(\C)},$ Lemma~\ref{lem:aeidentitysupport} implies
\be
\mathrm{Ad}_{P_{\xi}}\circ \vf
=\sum_{i=1}^{p}\mathrm{Ad}_{P_{\xi}V_{i}}
=\mathrm{Ad}_{P_{\xi}}.
\ee
Furthermore, Lemma~\ref{lem:ChoiKraus} implies there exists a $1\times p$ coisometry 
$U$ such that 
$P_{\xi}V_{i}=u_{1i}P_{\xi},$ where $u_{1i}$ is the $1i$-th entry
of $U.$ 
Set $\a_{i}:=u_{1i}.$
Since $U$ is a coisometry, 
the first (and only) column of $U$ is a unit vector, i.e.\ 
$\sum\limits_{i=1}^{p}|u_{1i}|^2=1,$ 
which proves the claim.
\eprf

\br
\label{rmk:PVbreakup}
Using the notation of Lemma~\ref{lem:CPUtoidentity}, 
$P_{\xi}V_{i}=\a_{i}P_{\xi}$ says 
\be
\label{eq:Vibreaksuptothree}
V_{i}=\a_{i}P_{\xi}+P_{\xi}^{\perp}V_{i}P_{\xi}+P_{\xi}^{\perp}V_{i}P_{\xi}^{\perp}.
\ee
If we choose a basis in which the density matrix $\xi^*(1)$ is diagonal with its non-zero eigenvalues all appearing on the top left, then~(\ref{eq:Vibreaksuptothree}) reads
\be
\label{eq:Vigeneraldiagonalbasis}
V_{i}=\begin{bmatrix}\a_{i}\mathds{1}_{r}&0\\V_{i}^{\mathrm{bl}}&V_{i}^{\mathrm{br}}\end{bmatrix},
\ee
where $r$ is the rank of $\xi^*(1),$ and where $V_{i}^{\mathrm{bl}}$ is an
$(n-r)\times r$ matrix while $V_{i}^{\mathrm{br}}$ is an $(n-r)\times(n-r)$ matrix. 
\er

It turns out that Lemma~\ref{lem:CPUtoidentity}
holds even when the support $P_{\xi}$ from the equations is removed (cf.\ Theorem~\ref{thm:CPUtoidentity} below). To prove this, it seems convenient to recall the notion of a pre-Hilbert $C^*$-algebra module due to Paschke~\cite{Pa73}.

\bd
\label{defn:Paschke}
Let $\mA$ be a (unital) $C^*$-algebra. A \define{pre-Hilbert $\mA$-module} is a 
left%
%\footnote
\footnote{Paschke defines a \emph{right} module structure instead of a left one. This does change some properties, but we have modified them appropriately. One such property is the Paschke--Cauchy--Schwarz inequality
in (\ref{eq:prehilbCSinequality}). 
}
%end footnote
 $\mA$-module $\mathcal{E}$ together with a linear-conjugate linear map%
%\footnote
\footnote{This means $\<\!\<s+t,u\>\!\>=\<\!\<s,u\>\!\>+\<\!\<t,u\>\!\>$ 
for all $s,t,u\in\mathcal{E}.$ The other properties usually associated with sesqui-linearity (with conjugate linearity in the second coordinate) follow from the other conditions in the definition since the algebra $\mA$ is unital.}
%end footnote
$\<\!\<\;\cdot\;,\;\cdot\;\>\!\>:\mathcal{E}\times\mathcal{E}\to\mA$ satisfying the following properties
\begin{enumerate}[i.]
\itemsep0pt
\item
\label{item:Paschkesgez}
$\<\!\<s,s\>\!\>\ge0$ for all $s\in\mathcal{E},$
\item
$\<\!\<s,t\>\!\>=\<\!\<t,s\>\!\>^*$ for all $s,t\in\mathcal{E},$
\item
$\<\!\<as,t\>\!\>=a\<\!\<s,t\>\!\>$ for all $s,t\in\mathcal{E}$ and $a\in\mA,$ and
\item
$\<\!\<s,s\>\!\>=0$ if and only if $s=0$ (this is called \define{non-degeneracy} of $\<\!\<\;\cdot\;,\;\cdot\;\>\!\>$). 
\end{enumerate}
$\<\!\<\;\cdot\;,\;\cdot\;\>\!\>$ is called the 
\define{$\mA$-valued inner product} on $\mathcal{E}.$ 
\ed

\br
\label{rmk:paschkeswap}
It follows from this definition that $\<\!\<s,at\>\!\>=\<\!\<s,t\>\!\>a^*$ for all
$s,t\in\mathcal{E}$ and $a\in\mA.$ 
\er

\blem
\label{lem:Paschke}
Let $\mA$ be a $C^*$-algebra and $\mathcal{E}$ a pre-Hilbert module over $\mA.$ 
Then
\be
\mathcal{E}\ni s\mapsto
\lVert s\rVert_{\mathcal{E}}:=\sqrt{\lVert\<\!\<s,s\>\!\>\rVert}
\ee
defines a norm on $\mathcal{E}.$ Furthermore, 
\be
\label{eq:prehilbCSinequality}
\<\!\<t,s\>\!\>\<\!\<s,t\>\!\>\le\lVert t\rVert_{\mathcal{E}}^2\<\!\<s,s\>\!\>
\ee
for all $s,t\in\mathcal{E}.$ 
\elem

\bprf
See Proposition~2.3 in Paschke~\cite{Pa73} or Section~3.14 in Fillmore~\cite{Fi96}.
\eprf

\bx
\label{ex:vectorsofmatrices}
Fix $n,m,p\in\N.$ Let ${}_{m}\mathcal{M}_{n}(\C)$ denote the vector space 
of $m\times n$ complex matrices. Set $\mA:=\mathcal{M}_{m}(\C)$ and 
$\mathcal{E}:={}_{m}\mathcal{M}_{n}(\C)^{p},$ the vector space direct sum of $p$ copies of ${}_{m}\mathcal{M}_{n}(\C)$. Denote elements of 
$\mathcal{E}$ by $\vec{A}:=(A_{1},\dots,A_{p})$ so that
$A_{i}\in{}_{m}\mathcal{M}_{n}(\C)$ for all $i\in\{1,\dots,p\}.$ 
Define the 
left $\mA$-module structure on $\mathcal{E}$ to be 
$B\vec{A}:=(BA_{1},\dots,BA_{p})$ for all $\vec{A}\in\mathcal{E}$ and $B\in\mA.$ 
Define the $\mA$-valued inner product by
\be
\label{eq:Paschkematrixinnerproduct}
\mathcal{E}\times\mathcal{E}\ni(\vec{A},\vec{B})\mapsto
\left\<\!\!\left\<\vec{A},\vec{B}\right\>\!\!\right\>:=\sum_{i=1}^{p}A_{i}B_{i}^{\dag}.
\ee
Straightforward matrix algebra shows $\mathcal{E}$
is indeed a pre-Hilbert $\mA$-module
with these structures.
In fact, $\mathcal{E}$ is also a \emph{right} $\mathcal{M}_{n}(\C)$-module
satisfying 
\be
\left\<\!\!\left\<\vec{A}C,\vec{B}\right\>\!\!\right\>=\left\<\!\!\left\<\vec{A},\vec{B}C^{\dag}\right\>\!\!\right\>\qquad\forall\;\vec{A},\vec{B}\in\mathcal{E}\text{ and }C\in\mathcal{M}_{n}(\C). 
\ee
However, $\mathcal{E}$ is \emph{not} a (right) pre-Hilbert module with 
respect to this action. 
\ex

\bt
\label{thm:CPUtoidentity}
Fix $n\in\N,$
let $\xi:\mathcal{M}_{n}(\C)\stoch\C$ be a state, and 
let $\vf:\mathcal{M}_{n}(\C)\stoch\mathcal{M}_{n}(\C)$ be a CPU map such that
$\vf\underset{\raisebox{.6ex}[0pt][0pt]{\scriptsize$\xi$}}{=}\id_{\mathcal{M}_{n}(\C)}.$
Let $\vf$ have a Kraus decomposition $\vf=\sum\limits_{i=1}^{p}\mathrm{Ad}_{V_{i}}.$ 
Then there exist complex numbers $\{\a_{i}\}_{i\in\{1,\dots,p\}}$ such that 
\be
V_{i}=\a_{i}\mathds{1}_{n}\qquad\forall\;i\in\{1,\dots,p\}
\quad\text{ and }\quad
\sum_{i=1}^{p}|\a_{i}|^2=1. 
\ee
In particular, $\vf=\id_{\mathcal{M}_{n}(\C)}.$
\et

\bprf
Let $P_{\xi}\in\mathcal{M}_{n}(\C)$ denote the support of $\xi$. 
In order to proceed avoiding as many indices and sums as possible, 
we will first introduce
a certain pre-Hilbert $C^*$-algebra module based on the number $p$ of Kraus operators assumed for $\vf.$
Set $\mathcal{E}:=\mathcal{M}_{n}(\C)^{p}$ and equip this
with the pre-Hilbert $\mathcal{M}_{n}(\C)$-module structure
from Example~\ref{ex:vectorsofmatrices}.
By Remark~\ref{rmk:PVbreakup}, 
\be
\label{eq:Vigeneralvector}
\vec{V}=P_{\xi}\vec{\a}+\vec{V}^{\mathrm{bl}}+\vec{V}^{\mathrm{br}},
\ee
where 
\be
\vec{\a}:=\big(\a_{1}\mathds{1}_{n},\dots,\a_{p}\mathds{1}_{n}\big),\quad
\vec{V}^{\mathrm{bl}}:=P_{\xi}^{\perp}\vec{V}P_{\xi},\quad\text{ and }\quad
\vec{V}^{\mathrm{br}}:=P_{\xi}^{\perp}\vec{V}P_{\xi}^{\perp}.
\ee
Note that $C\vec{\a}=\vec{\a}C$ for all $C\in\mathcal{M}_{n}(\C).$ 
The identities
\be
\label{eq:Vibasicidentities}
\<\!\<\vec{\a},\vec{\a}\>\!\>=\mathds{1}_{n},\qquad
\<\!\<P_{\xi}\vec{\a},\vec{V}^{\mathrm{br}}\>\!\>={0},\quad\text{and}\quad
\<\!\<\vec{V}^{\mathrm{bl}},\vec{V}^{\mathrm{br}}\>\!\>={0}
\ee
follow directly from the definitions. 
The fact that $\vf$ is unital means $\mathds{1}_{n}=\sum\limits_{i}V_{i}V_{i}^{\dag}.$ In terms of the $\mathcal{M}_{n}(\C)$-valued inner product, 
this becomes 
\be
\label{eq:iunitalityfactorcase}
\begin{split}
\mathds{1}_{n}&=\<\!\<\vec{V},\vec{V}\>\!\>
\overset{\text{(\ref{eq:Vigeneralvector})}}{=\joinrel=\joinrel=}\<\!\<P_{\xi}\vec{\a}+\vec{V}^{\mathrm{bl}}+\vec{V}^{\mathrm{br}},P_{\xi}\vec{\a}+\vec{V}^{\mathrm{bl}}+\vec{V}^{\mathrm{br}}\>\!\>\\
&\overset{\text{(\ref{eq:Vibasicidentities})}}{=\joinrel=\joinrel=}\underbrace{\<\!\<P_{\xi}\vec{\a},P_{\xi}\vec{\a}\>\!\>}_{P_{\xi}}+\<\!\<P_{\xi}\vec{\a},\vec{V}^{\mathrm{bl}}\>\!\>+\<\!\<\vec{V}^{\mathrm{bl}},P_{\xi}\vec{\a}\>\!\>+\<\!\<\vec{V}^{\mathrm{bl}},\vec{V}^{\mathrm{bl}}\>\!\>+\<\!\<\vec{V}^{\mathrm{br}},\vec{V}^{\mathrm{br}}\>\!\>.
\end{split}
\ee
Since
\be
P_{\xi}\vec{V}^{\mathrm{bl}}=\vec{0}
\quad\text{ and }\quad
P_{\xi}\vec{V}^{\mathrm{br}}=\vec{0},
\ee
it follows that 
\be
\label{eq:iPkillsVblandVbr}
\<\!\<\vec{X},\vec{V}^{\mathrm{bl}}\>\!\>P_{\xi}=0
\quad\text{ and }\quad
\<\!\<\vec{X},\vec{V}^{\mathrm{br}}\>\!\>P_{\xi}=0
\qquad\forall\;\vec{X}\in\mathcal{E}
\ee
by the properties of the pre-Hilbert module structure
(see Remark~\ref{rmk:paschkeswap}). Hence, multiplying
(\ref{eq:iunitalityfactorcase}) by $P_{\xi}$ on the right and simplifying 
gives
$
0=\<\!\<\vec{V}^{\mathrm{bl}},P_{\xi}\vec{\a}\>\!\>
$
and similarly $\<\!\<P_{\xi}\vec{\a},\vec{V}^{\mathrm{bl}}\>\!\>=0.$ 
Furthermore, $\<\!\<\vec{V}^{\mathrm{bl}},P_{\xi}^{\perp}\vec{\a}\>\!\>=0$ follows immediately from the definition of $\vec{V}^{\mathrm{bl}}$. Putting these two together gives
\be
\label{eq:Viblalphaiszero}
\<\!\<\vec{V}^{\mathrm{bl}},\vec{\a}\>\!\>={0}.
\ee
Hence, the unitality of $R$ condition (\ref{eq:iunitalityfactorcase}) simplifies to 
\be
\label{eq:iunitalityfactorcasesimpler}
P_{\xi}^{\perp}=\<\!\<\vec{V}^{\mathrm{bl}},\vec{V}^{\mathrm{bl}}\>\!\>+\<\!\<\vec{V}^{\mathrm{br}},\vec{V}^{\mathrm{br}}\>\!\>.
\ee
Now, write $A\in\mathcal{M}_{n}(\C)$ as (cf.\ Equation~(\ref{eq:decomposelements}))
\be
\label{eq:idecomposeA}
A=P_{\xi}AP_{\xi}+P_{\xi}AP_{\xi}^{\perp}+P_{\xi}^{\perp}AP_{\xi}+P_{\xi}^{\perp}AP_{\xi}^{\perp}
\ee
in terms of the support $P_{\xi}$ of $\xi$ and its orthogonal complement $P_{\xi}^{\perp}$. 
Using this decomposition, 
\be
\label{eq:idisintexplicit}
\begin{split}
\vf(A)P_{\xi}&=\sum_{i=1}^{p}V_{i}AV_{i}^{\dag}P_{\xi}
=\<\!\<\vec{V}A,\vec{V}\>\!\>P_{\xi}
\overset{\text{Rmk~\ref{rmk:paschkeswap}}}{=\joinrel=\joinrel=\joinrel=\joinrel=}\<\!\<\vec{V}A,P_{\xi}\vec{V}\>\!\>
\overset{\text{(\ref{eq:Vigeneralvector})}}{=\joinrel=\joinrel=}\<\!\<\vec{V}A,P_{\xi}\vec{\a}\>\!\>\\
&\overset{\text{(\ref{eq:Paschkematrixinnerproduct})}}{=\joinrel=\joinrel=}\<\!\<\vec{V},\vec{\a}\>\!\>AP_{\xi}
\overset{\text{(\ref{eq:Vibasicidentities}) \& (\ref{eq:Viblalphaiszero})}}{=\joinrel=\joinrel=\joinrel=\joinrel=\joinrel=\joinrel=\joinrel=}\left(P_{\xi}+\<\!\<\vec{V}^{\mathrm{br}},\vec{\a}\>\!\>\right)AP_{\xi}\\
&\overset{\text{(\ref{eq:idecomposeA})}}{=\joinrel=\joinrel=}\left(P_{\xi}+\<\!\<\vec{V}^{\mathrm{br}},\vec{\a}\>\!\>\right)\Big(P_{\xi}AP_{\xi}+P_{\xi}^{\perp}AP_{\xi}\Big)\overset{\text{(\ref{eq:iPkillsVblandVbr})}}{=\joinrel=\joinrel=}P_{\xi}AP_{\xi}+\<\!\<\vec{V}^{\mathrm{br}},\vec{\a}\>\!\>P_{\xi}^{\perp}AP_{\xi}
\end{split}
\ee
for all $A\in\mathcal{M}_{n}(\C).$ But since $\vf\underset{\raisebox{.6ex}[0pt][0pt]{\scriptsize$\xi$}}{=}\id_{\mathcal{M}_{n}(\C)},$ this equals 
$P_{\xi}AP_{\xi}+P_{\xi}^{\perp}AP_{\xi}$ by Lemma~\ref{lem:aeidentitysupport} item~\ref{partioflem:aeidentitysupport}. 
Identifying terms, $\<\!\<\vec{V}^{\mathrm{br}},\vec{\a}\>\!\>P_{\xi}^{\perp}AP_{\xi}=P_{\xi}^{\perp}AP_{\xi}$ for all $A\in\mathcal{M}_{n}(\C),$ i.e.\
$\<\!\<\vec{V}^{\mathrm{br}},\vec{\a}\>\!\>$ acts as the identity on $n\times n$ matrices  
of the form $P_{\xi}^{\perp}AP_{\xi}.$ This combined with the fact that
$\<\!\<\vec{V}^{\mathrm{br}},\vec{\a}\>\!\>=P_{\xi}^{\perp}\<\!\<\vec{V}^{\mathrm{br}},\vec{\a}\>\!\>P_{\xi}^{\perp}$ 
implies 
\be
\label{eq:Vibralphaissupportcomp}
\<\!\<\vec{V}^{\mathrm{br}},\vec{\a}\>\!\>=P_{\xi}^{\perp}.
\ee
This implies
$
\<\!\<\vec{\a},\vec{V}^{\mathrm{br}}\>\!\>^*\<\!\<\vec{\a},\vec{V}^{\mathrm{br}}\>\!\>=P_{\xi}^{\perp}.
$
Hence, by the Paschke--Cauchy--Schwarz inequality (Lemma~\ref{lem:Paschke}), 
\be
P_{\xi}^{\perp}\le\lVert\vec{\a}\rVert^2_{\mathcal{E}}\<\!\<\vec{V}^{\mathrm{br}},\vec{V}^{\mathrm{br}}\>\!\>
\overset{\text{(\ref{eq:Vibasicidentities})}}{=\joinrel=\joinrel=}\<\!\<\vec{V}^{\mathrm{br}},\vec{V}^{\mathrm{br}}\>\!\>.
\ee
On the other hand, (\ref{eq:iunitalityfactorcasesimpler}) entails 
$
\<\!\<\vec{V}^{\mathrm{br}},\vec{V}^{\mathrm{br}}\>\!\>\le P_{\xi}^{\perp}
$
by condition~\ref{item:Paschkesgez} in Definition~\ref{defn:Paschke}.
These two inequalities force
\be
\label{eq:iVbrVbrVblVbl}
\<\!\<\vec{V}^{\mathrm{br}},\vec{V}^{\mathrm{br}}\>\!\>=P_{\xi}^{\perp}
\qquad\text{ and }\qquad
\<\!\<\vec{V}^{\mathrm{bl}},\vec{V}^{\mathrm{bl}}\>\!\>=0.
\ee
By non-degeneracy of the $\mathcal{M}_{n}(\C)$-valued inner product
on $\mathcal{E},$ this forces $\vec{V}^{\mathrm{bl}}=\vec{0},$ i.e.\ 
\be
P_{\xi}V_{i}P_{\xi}^{\perp}=0\qquad\forall\;i\in\{1,\dots,p\}.
\ee
Finally, using these relations and the properties of the $\mathcal{M}_{n}(\C)$-valued inner product, 
\be
\label{eq:iVbrisalphaid}
%\begin{split}
\<\!\<P_{\xi}^{\perp}\vec{\a}-\vec{V}^{\mathrm{br}},P_{\xi}^{\perp}\vec{\a}-\vec{V}^{\mathrm{br}}\>\!\>
\overset{\text{(\ref{eq:Vibasicidentities})}}{=\joinrel=\joinrel=}
P_{\xi}^{\perp}-\<\!\<\vec{\a},\vec{V}^{\mathrm{br}}\>\!\>
-\<\!\<\vec{V}^{\mathrm{br}},\vec{\a}\>\!\>
+\<\!\<\vec{V}^{\mathrm{br}},\vec{V}^{\mathrm{br}}\>\!\>
\overset{\text{(\ref{eq:Vibralphaissupportcomp}) \& (\ref{eq:iVbrVbrVblVbl})}}{=\joinrel=\joinrel=\joinrel=\joinrel=\joinrel=\joinrel=\joinrel=}0,
%\end{split}
\ee
which, by non-degeneracy of the $\mathcal{M}_{n}(\C)$-valued inner product, 
proves 
\be
\label{eq:VbrisPperpalpha}
\vec{V}^{\mathrm{br}}=P_{\xi}^{\perp}\vec{\a},
\ee
i.e.\ 
\be
P_{\xi}^{\perp}V_{i}P_{\xi}^{\perp}
=\a_{i}P_{\xi}^{\perp}\qquad\forall\;i\in\{1,\dots,p\}.
\ee
Putting this all together gives
\be
\vec{V}=\vec{\a},
\quad\text{i.e.}\quad
V_{i}=\a_{i}\mathds{1}_{n}\qquad\forall\;i\in\{1,\dots,p\}.
\ee
The fact that $\vf=\id_{\mathcal{M}_{n}(\C)}$ follows 
immediately from this result. 
\eprf

\br
It should be stressed how surprising Theorem~\ref{thm:CPUtoidentity} is. 
Even if $\xi$ is a pure state, so that its support is a rank one projection, a.e.\ equivalence of a CPU map to the identity is strong enough to enforce \emph{equality} of that CPU map to the identity, regardless of how large the dimension of the Hilbert space is. We feel this gives us a precise sense of how ``probability zero'' objects (the projection onto the orthogonal complement of the support of $\xi$ in this case) in quantum theory cannot be disregarded in the way that they can be in classical probability theory~\cite{Gr14II}. 
\er

\br
\label{rmk:aeequaldnimplyequal}
Theorem~\ref{thm:CPUtoidentity} might seem to suggest that 
if $\xi:\mathcal{M}_{n}(\C)\stoch\C$ is a state and 
if $\vf,\psi:\mathcal{M}_{m}(\C)\stoch\mathcal{M}_{n}(\C)$ 
are two CPU maps, then 
$\vf\underset{\raisebox{.6ex}[0pt][0pt]{\scriptsize$\xi$}}{=}\psi$ 
implies $\vf=\psi.$ The following example shows this is false in general. 
Let $s\in\{1,\dots,m-1\}$ and let $\chi:\mathcal{M}_{m}(\C)\stoch\C$ be the 
map that takes the trace of the bottom right part of a matrix, namely
\be
\chi(A):=\sum_{i=s+1}^{m}a_{ii},
\ee
where $a_{ii}$ is the $ii$-th entry of $A.$ 
Note that this map is positive and therefore CP since the codomain of $\chi$ is $\C$ (cf.\ Theorem~3 in Stinespring~\cite{St55}). Note, however, that $\chi$ is not unital. Similarly, the trace map $\tr:\mathcal{M}_{m}(\C)\stoch\C$ is CP (but not unital).
Now, consider the following two maps
\be
\begin{split}
\mathcal{M}_{m}(\C)&\stoch\mathcal{M}_{n}(\C)\\
A&\xmapsto{\,\;\vf\;}\frac{1}{m}\tr(A)\mathds{1}_{n} \quad\text{ and}\\
A&\xmapsto{\,\;\psi\;}\frac{1}{m}\tr(A)P_{\xi}+\frac{1}{m-s}\chi(A)P_{\xi}^{\perp}.\\
\end{split}
\ee
Note that $\vf$ and $\psi$ are not equal. 
Nevertheless, $\vf$ and $\psi$ are $\xi$-a.e.\ equivalent because $\vf(A)P_{\xi}=\psi(A)P_{\xi}$ for all $A\in\mathcal{M}_{m}(\C).$ A simple calculation shows $\psi$ and $\vf$ are unital. Furthermore, they are both completely positive as their $p$-ampliations are
\be
\vf_{p}(\;\cdot\;)=\frac{1}{m}\tr_{p}(\;\cdot\;)\otimes\mathds{1}_{n}
%\ee
%and
%\be
\qquad\text{ and }\qquad
\psi_{p}(\;\cdot\;)=\frac{1}{m}\tr_{p}(\;\cdot\;)\otimes
P_{\xi}+\frac{1}{m-s}\chi_{p}(\;\cdot\;)\otimes P_{\xi}^{\perp},
\ee
respectively. 
Here, $\tr_{p}$ and $\chi_{p}$ are the $p$-ampliations of $\tr$ and $\chi,$ which are positive. 
\er

\br
\label{rmk:unitalnecessaryforaeidentity}
The conclusion of Theorem~\ref{thm:CPUtoidentity} is false if 
$\vf$ is assumed to only be CP but not unital. 
A simple counter-example is the CP map 
\be
\begin{split}
\mathcal{M}_{n}(\C)&\stoch\mathcal{M}_{n}(\C)\\
A&\xmapsto{\,\;\vf\;}A+\tr(A)P_{\xi}^{\perp}.
\end{split}
\ee
Here, $\vf$ is $\xi$-a.e.\ equivalent to $\id_{\mathcal{M}_{n}(C)}$ but is not equal to it.
\er

\br
Using the same notation and assumptions as in Theorem~\ref{thm:CPUtoidentity}, if $\vf$ is $\xi$-a.e.\ equivalent to a $^*$-isomorphism, then it equals that $^*$-isomorphism. However, if $\xi:\mathcal{M}_{mp}(\C)\stoch\C$ is a state and $\vf:\mathcal{M}_{m}(\C)\stoch\mathcal{M}_{mp}(\C)$ is a CPU map that is $\xi$-a.e.\ equivalent to a $^*$-homomorphism, then it is not necessarily equal to that $^*$-homomorphism (unless $p=1$). A simple counter-example is $\vf(B):=\mathrm{diag}(B,\tr(B)\mathds{1}_{m},\dots,\tr(B)\mathds{1}_{m})$ and $\xi$ the state represented by the density matrix $\frac{1}{m}\mathrm{diag}(\mathds{1}_{m},0,\dots,0)$. Then $\vf$ is $\xi$-a.e.\ equivalent to the $^*$-homomorphism $B\mapsto\mathrm{diag}(B,\dots,B)$, but it is not equal to it (unless $m=1$ or $p=1$). 
\er

The following corollary of Theorem~\ref{thm:CPUtoidentity} is similar to a fact used frequently
in the area of reversible quantum operations (cf.\ the proof of Theorem~2.1 in Nayak and Sen~\cite{NaSe07}).

\bc
\label{cor:ChoiKrausIdentity}
Let $F:\mathcal{M}_{m}(\C)\stoch\mathcal{M}_{n}(\C)$ and $R:\mathcal{M}_{n}(\C)\stoch\mathcal{M}_{m}(\C)$ be CPU maps with  
Kraus decompositions 
\be
R=\sum_{i=1}^{p}\mathrm{Ad}_{R_{i}}
\qquad
\text{and}
\qquad
F=\sum_{j=1}^{q}\mathrm{Ad}_{F_{j}}.
\ee
If $R\circ F\underset{\raisebox{.6ex}[0pt][0pt]{\scriptsize$\xi$}}{=}\id_{\mathcal{M}_{m}(\C)}$ for some state $\xi:\mathcal{M}_{n}(\C)\stoch\C$,
then there exist complex numbers $\{\a_{ij}\}_{\substack{i\in\{1,\dots,p\}\\j\in\{1,\dots,q\}}}$ such that 
\be
\label{eq:PRFP}
R_{i}F_{j}=\a_{ij}\mathds{1}_m
\qquad
\text{and}
\qquad
\sum_{i,j}|\a_{ij}|^2=1.
\ee
In particular, $R\circ F=\id_{\mathcal{M}_{m}(\C)}.$ 
\ec

\bprf
This follows immediately from Theorem~\ref{thm:CPUtoidentity}. 
\eprf

%%%%%%%%%%%%%%%%%%%%%%%%%%%%%%%%%%%%%
\section{Categories of C*-algebras, states, and morphisms} 
\label{sec:finprobtocalg}
%%%%%%%%%%%%%%%%%%%%%%%%%%%%%%%%%%%%%

We prove that \emph{non-commutative probability spaces}, 
$C^*$-algebras equipped with states, and a.e.\ equivalence classes of 
CPU maps 
(in fact, 2-positive unital maps) form a category. In fact, finite-dimensional $C^*$-algebras and a.e.\ equivalence classes of positive unital maps form a category.
The following Cauchy--Schwarz type inequality, due to Kadison~\cite{Ka52},
for positive unital and 2-positive unital maps
is useful in proving many of these claims.

\blem
\label{lem:KadisonSchwarz}
Let $\mA$ and $\mB$ be $C^*$-algebras and let $\vf:\mA\stoch\mB$ be
a positive unital map. 
\begin{enumerate}[i.]
\itemsep0pt
\item
\label{item:kadison}
If $a\in\mA$ is self-adjoint, then 
$
\vf(a)^2\le\vf(a^2).
$
\item
If $\vf$ is 2-positive, then 
$\vf(a)^*\vf(a)\le\vf(a^*a)$ for all $a\in\mA$.
\end{enumerate}
\elem

\bprf
See Theorem~1.3.1 and Corollary~1.3.2 in \cite{St13} and Proposition~3.3 in \cite{Pa02}. 
\eprf

\bn
\label{prop:aecompositionalgebraswelldefined}
Let $\mC,\mB$, and $\mA$ be $C^*$-algebras, let $\w:\mA\stoch\C$ be a state on $\mA,$ 
and let $G,G':\mC\stoch\mB$ and $F,F':\mB\stoch\mA$ be 2-positive (or Schwarz-positive) unital maps. If $F\underset{\raisebox{.6ex}[0pt][0pt]{\scriptsize$\w$}}{=}F'$ 
and 
$G\underset{\raisebox{.6ex}[0pt][0pt]{\scriptsize$\xi$}}{=}G'$, where $\xi:=\w\circ F,$  
then 
$F\circ G\underset{\raisebox{.6ex}[0pt][0pt]{\scriptsize$\w$}}{=}F'\circ G'.$
\en

\bprf
By assumption, the diagrams 
\be
\label{eq:aecpcompositionwelldefined}
\xy0;/r.25pc/:
(15,0)*+{\mB/\mathcal{N}_{\xi}}="AN";
(-15,0)*+{\mC}="B";
(0,7.5)*+{\mB}="AF";
(0,-7.5)*+{\mB}="AG";
{\ar@{~>}"B";"AF"^{G}};
{\ar@{~>}"B";"AG"_{G'}};
{\ar@{->>}"AF";"AN"};
{\ar@{->>}"AG";"AN"};
\endxy
\qquad
\text{and}
\qquad
\xy0;/r.25pc/:
(15,0)*+{\mA/\mathcal{N}_{\w}}="AN";
(-15,0)*+{\mB}="B";
(0,7.5)*+{\mA}="AF";
(0,-7.5)*+{\mA}="AG";
{\ar@{~>}"B";"AF"^{F}};
{\ar@{~>}"B";"AG"_{F'}};
{\ar@{->>}"AF";"AN"};
{\ar@{->>}"AG";"AN"};
\endxy
\ee
both commute. 
For the composite, we have
\be
\xy0;/r.25pc/:
(15,0)*+{\mB/\mathcal{N}_{\xi}}="BN";
(45,0)*+{\mA/\mathcal{N}_{\w}}="AN";
(-15,0)*+{\mC}="C";
(0,7.5)*+{\mB}="BF";
(0,-7.5)*+{\mB}="BG";
(15,15)*+{\mA}="AF";
(15,-15)*+{\mA}="AG";
{\ar@{~>}"C";"BF"^{G}};
{\ar@{~>}"C";"BG"_{G'}};
{\ar@{~>}"BF";"AF"^{F}};
{\ar@{~>}"BG";"AG"_{F'}};
{\ar@{->>}"BF";"BN"};
{\ar@{->>}"BG";"BN"};
{\ar@{->>}"AF";"AN"};
{\ar@{->>}"AG";"AN"};
\endxy
\ee
The left part of this diagram commutes
by commutativity of the left diagram in 
(\ref{eq:aecpcompositionwelldefined}). 
It would be convenient to have a function 
$\mB/\mathcal{N}_{\xi}\to\mA/\mathcal{N}_{\w}$
to fill in the diagram.
In this regard, let $\widetilde{F},\widetilde{F'}:\mB/\mathcal{N}_{\xi}\to\mA/\mathcal{N}_{\w}$ 
be the functions defined by 
\be
\mB/\mathcal{N}_{\xi}\ni[b]_{\xi}
\mapsto\widetilde{F}\big([b]_{\xi}\big):=\big[F(b)\big]_{\w}
\qquad
\text{and}
\qquad
\mB/\mathcal{N}_{\xi}\ni[b]_{\xi}
\mapsto\widetilde{F'}\big([b]_{\xi}\big):=\big[F'(b)\big]_{\w}.
\ee
To see that $\widetilde{F}$ is well-defined, let $b\in\mathcal{N}_{\xi},$
i.e.\ $\xi(b^*b)=0.$ 
Then
\be
\w\big(F(b)^*F(b)\big)\le\w\big(F(b^*b)\big)=\xi(b^*b)=0
\ee
by Lemma~\ref{lem:KadisonSchwarz} applied to $F$ and the fact that 
$F$ is state-preserving so that $\w\circ F=\xi.$
A similar conclusion can be made
for $\widetilde{F'}.$ Since $F(b)^*F(b)\ge0$
and $\w$ is a positive functional, this shows $F(b)\in\mN_{\w},$
which proves $\widetilde{F}$ and $\widetilde{F'}$ are well-defined.
In fact, by commutativity of the right diagram in 
(\ref{eq:aecpcompositionwelldefined}), $\widetilde{F}=\widetilde{F'}.$
Hence, all the subdiagrams in the diagram 
\be
\xy0;/r.25pc/:
(15,0)*+{\mB/\mathcal{N}_{\xi}}="BN";
(45,0)*+{\mA/\mathcal{N}_{\w}}="AN";
(-15,0)*+{\mC}="C";
(0,7.5)*+{\mB}="BF";
(0,-7.5)*+{\mB}="BG";
(15,15)*+{\mA}="AF";
(15,-15)*+{\mA}="AG";
{\ar@{~>}"C";"BF"^{G}};
{\ar@{~>}"C";"BG"_{G'}};
{\ar@{~>}"BF";"AF"^{F}};
{\ar@{~>}"BG";"AG"_{F'}};
{\ar@{->>}"BF";"BN"};
{\ar@{->>}"BG";"BN"};
{\ar@{->>}"AF";"AN"};
{\ar@{->>}"AG";"AN"};
{\ar@{->}@<0.5ex>"BN";"AN"^{\widetilde{F}}};
{\ar@{->}@<-0.5ex>"BN";"AN"_{\widetilde{F'}}};
\endxy
\ee
commute so that 
$F\circ G\underset{\raisebox{.6ex}[0pt][0pt]{\scriptsize$\w$}}{=}F'\circ G'.$
\eprf

Assuming finite-dimensionality, we can prove more. Although the previous proposition is enough for the sequel, the following theorem is an interesting result in its own right.

\bt
\label{thm:compositionpositiveunital}
Let $\mC,\mB$, and $\mA$ be finite-dimensional $C^*$-algebras, 
let $\w:\mA\stoch\C$ be a state on $\mA,$
and let $G,G':\mC\stoch\mB$ and $F,F':\mB\stoch\mA$ be positive unital maps
with $G\underset{\raisebox{.6ex}[0pt][0pt]{\scriptsize$\xi$}}{=}G'$
and
$F\underset{\raisebox{.6ex}[0pt][0pt]{\scriptsize$\w$}}{=}F',$ where
$\xi:=\w\circ F=\w\circ F'$. 
Then 
$F\circ G\underset{\raisebox{.6ex}[0pt][0pt]{\scriptsize$\w$}}{=}F'\circ G'.$
\et

We will break up this proof into several lemmas, some of which are of independent interest. For the first lemma (the proof of which is immediate), recall that if $\vf:\mA\stoch\mB$ is a linear map between $C^*$-algebras, then $\vf$ is \define{self-adjoint} iff $\vf(a)^*=\vf(a^*)$ for all $a\in\mA$. Also, a vector subspace $V\subseteq\mB$ is \define{self-adjoint} iff $v\in V$ implies $v^*\in V$. 

\blem
\label{lem:selfadjointimage}
Let $\mA$ and $\mB$ be $C^*$-algebras and let $\vf:\mA\stoch\mB$ be a linear map. If $\vf$ is self-adjoint, then the image is a self-adjoint subspace of $\mB$. 
\elem

\blem
\label{lem:bottomright}
Let $\mA$ and $\mB$ be finite-dimensional $C^*$-algebras, let $\xi:\mB\stoch\C$ be a state, and let $\vf:\mA\stoch\mB$ be a self-adjoint linear map. If $\vf$ is $\xi$-a.e.\ equivalent to $0$, then $\mathrm{Im}(\vf)\subseteq P_{\xi}^{\perp}\mB P_{\xi}^{\perp}.$ 
\elem

\bprf
[Proof of Lemma~\ref{lem:bottomright}]
By assumption, $\mathrm{Im}(\vf)\subseteq\mB P_{\xi}^{\perp}.$
If $\vf(a)=P_{\xi}bP_{\xi}^{\perp}+P_{\xi}^{\perp}bP_{\xi}^{\perp}$ for some $b\in\mB$, then $\vf(a^*)=\vf(a)^*\in\mN_{\xi}$ as well by Lemma~\ref{lem:selfadjointimage}. But $\vf(a)^*=P_{\xi}^{\perp}b^*P_{\xi}+P_{\xi}^{\perp}b^*P_{\xi}^{\perp}.$ Hence $P_{\xi}^{\perp}b^*P_{\xi}=0.$ By taking the adjoint of this, we get $P_{\xi}bP_{\xi}^{\perp}=0.$ Thus, $\vf(a)=P_{\xi}^{\perp}bP_{\xi}^{\perp}$ for some $b\in\mB$.
\eprf

\blem
\label{lem:decomposeF}
$F:\mB\stoch\mA$ be positive map between $C^*$-algebras and let $P$ be a projection in $\mA.$ If $b\in\mB,$ then $F(b)$ can be uniquely decomposed as
\be
F(b)=F^{\mathrm{tl}}(b)+F^{\mathrm{tr}}(b)+F^{\mathrm{bl}}(b)+F^{\mathrm{br}}(b),
\ee
where 
\be
F^{\mathrm{tl}}(b):=PF(b)P,\;\;
F^{\mathrm{tr}}(b):=PF(b)P^{\perp},\;\;
F^{\mathrm{bl}}(b):=P^{\perp}F(b)P,\;\;
F^{\mathrm{br}}(b):=P^{\perp}F(b)P^{\perp}.
\ee
Furthermore, 
$F^{\mathrm{tr}}(b)^*=F^{\mathrm{bl}}(b^*)$ for all $b\in\mB$
and the maps $F^{\mathrm{tl}},F^{\mathrm{br}}:\mB\stoch\mA$ are positive. 
\elem

\bprf
The decomposition itself is just (\ref{eq:decomposelements}). 
From this and self-adjointness of $F,$
\be
F^{\mathrm{tr}}(b)^*
=\big(PF(b)P^{\perp}\big)^*
=P^{\perp}F(b)^*P
=P^{\perp}F(b^*)P
=F^{\mathrm{bl}}(b^*).
\ee
Furthermore, $F^{\mathrm{tl}}$ and $F^{\mathrm{br}}$ are positive maps since $F^{\mathrm{tl}}=\mathrm{Ad}_{P}\circ F$ and $F^{\mathrm{br}}=\mathrm{Ad}_{P^{\perp}}\circ F$ are composites of positive maps.
\eprf

\blem
\label{lem:postivebottomrighttobottomright}
Let $(\mB,\xi)$ and $(\mA,\w)$ be finite-dimensional $C^*$-algebras equipped with states and let $F:\mB\stoch\mA$ be positive unital and state-preserving. Then 
$
F\big(P_{\xi}^{\perp}\mB P_{\xi}^{\perp}\big)\subseteq P_{\w}^{\perp}\mA P_{\w}^{\perp}.
$
In particular $F\big(P_{\xi}^{\perp}\mB P_{\xi}^{\perp}\big)\subseteq\mathcal{N}_{\w}.$
\elem

\bprf
Let $b\in P_{\xi}^{\perp}\mB P_{\xi}^{\perp}.$ 
First, assume $b$ is self-adjoint. Then 
\be
\w\big(F(b)^*F(b)\big)
=\w\big(F(b^*)F(b)\big)%since $F$ is self-adjoint
=\w\big(F(b)^2\big)%Since $b$ is self-adjoint
\le\w\big(F(b^2)\big)%by Kadison
=\xi(b^2)
=\xi(b^*b)
=0,
\ee
where the inequality follows from part~\ref{item:kadison} of Lemma~\ref{lem:KadisonSchwarz}, the equality after it follows from the fact that $F$ is state-preserving, and the final equality follows from $b\in\mathcal{N}_{\xi}.$ This proves that $F(b)\in\mathcal{N}_{\w}$ for self-adjoint $b\in P_{\xi}^{\perp}\mB P_{\xi}^{\perp}.$ Hence, 
$F^{\mathrm{tl}}(b)=0$
and $F^{\mathrm{bl}}(b)=0$
for $b$ self-adjoint by Lemma~\ref{lem:decomposeF}. 
Second, assume $b$ is skew-adjoint. Then $b=ib'$ for some self-adjoint $b'\in\mB$ (namely, $b':=-ib$). Then
\be
F^{\mathrm{tl}}(b)=F^{\mathrm{tl}}(ib')=iF^{\mathrm{tl}}(b')=0
\quad\text{ and similarly }\quad
F^{\mathrm{bl}}(b)=0
\ee
for skew-adjoint $b$ by the previous fact 
since $F^{\mathrm{tl}}$ and $F^{\mathrm{bl}}$ are linear. Since every $b$ can be decomposed as the linear combination of a self-adjoint and skew-adjoint element, this proves $F^{\mathrm{tl}}$ and $F^{\mathrm{bl}}$ are both equal to the zero map. Finally, for any $b\in P_{\xi}^{\perp}\mB P_{\xi}^{\perp},$ 
\be
F^{\mathrm{tr}}(b)
=F^{\mathrm{tr}}\big((b^*)^*\big)
=F^{\mathrm{bl}}(b^*)^*
=0
\ee
by Lemma~\ref{lem:decomposeF} and the facts just proved. 
\eprf

\bprf
[Proof of Theorem~\ref{thm:compositionpositiveunital}]
We are required to prove $F\big(G(c)\big)P_{\w}=F'\big(G'(c)\big)P_{\w}$ for all $c\in\mC.$ First, note that 
\be
\label{eq:splittingFGcPw}
\begin{split}
F'\big(G'(c)\big)P_{\w}&=F\big(G'(c)\big)P_{\w}\quad\text{ since $F\underset{\raisebox{.6ex}[0pt][0pt]{\scriptsize$\w$}}{=}F'$}\\
&=F\big(G'(c)P_{\xi}+G'(c)P_{\xi}^{\perp}\big)P_{\w}.
\end{split}
\ee
Therefore, 
\be
\begin{split}
F\big(G(c)\big)P_{\w}-F'\big(G'(c)\big)P_{\w}&=F\Big(\big(G(c)-G'(c)\big)P_{\xi}+\big(G(c)-G'(c)\big)P_{\xi}^{\perp}\Big)P_{\w}\quad\text{ by (\ref{eq:splittingFGcPw})}\\
&=F\Big(\big(G(c)-G'(c)\big)P_{\xi}^{\perp}\Big)P_{\w}\quad\text{ since $G\underset{\raisebox{.6ex}[0pt][0pt]{\scriptsize$\xi$}}{=}G'$}\\
&=F\Big(\mathrm{Ad}_{P_{\xi}^{\perp}}\big(G(c)-G'(c)\big)\Big)P_{\w}\quad\text{ by Lemma~\ref{lem:bottomright}}
\\
&=\mathrm{Ad}_{P_{\w}^{\perp}}\bigg(F\Big(\mathrm{Ad}_{P_{\xi}^{\perp}}\big(G(c)-G'(c)\big)\Big)\bigg)P_{\w}\quad\text{ by Lemma~\ref{lem:postivebottomrighttobottomright}}
\\
&=0.
\end{split}
\ee
This proves that composition of a.e.-equivalence classes of positive unital maps between finite-dimensional $C^*$-algebras is well-defined. 
\eprf

\br
By Lemma~\ref{lem:supportofstate} and Remark~\ref{eq:Wstaralgebrassupport}, Theorem~\ref{thm:compositionpositiveunital} holds if $\mA,\mB,$ and $\mC$ are $W^*$-algebras. 
\er

\begin{definition}
A \define{non-commutative probability space} is a pair $(\mA,\omega)$, with $\mA$ a $C^*$-algebra and $\omega$ a state on $\mA$. A \define{state-preserving map} $(\mB,\xi)\stoch(\mA,\omega)$ is a map (linear, positive, CP, $^*$-homomorphism, etc.) $\mB\xstoch{F}\mA$ such that $\xi=\omega\circ F$. 
\end{definition}

\begin{corollary}
\label{cor:classicalembedstoquantum}
The following facts hold. 
\begin{enumerate}[i.]
\itemsep0pt
\item
The collection of non-commutative probability spaces and state-preserving maps forms a category. 
\item
The collection of non-commutative probability spaces and a.e.\ equivalence classes of 2-positive unital maps forms a category.
\item
The collection of finite-dimensional non-commutative probability spaces (or non-commutative probability spaces on $W^*$-algebras) and a.e.\ equivalence classes of PU maps forms a category.
\item
\label{item:fpstoncfps}
The opposite of the category of finite probability spaces and probability-preserving stochastic maps embeds fully into the category of non-commutative probability spaces and state-preserving PU maps. It is an equivalence on the subcategory of finite-dimensional commutative $C^*$-algebras. 
\item
Two probability-preserving stochastic maps are a.e.\ equivalent if and only if their associated PU maps are a.e.\ equivalent. 
\end{enumerate}
\end{corollary}

The functor in item~\ref{item:fpstoncfps} is uniquely determined by sending a finite set $X$ to the $C^*$-algebra $\C^{X}$
and sending a stochastic map $f:X\stoch Y$ to the PU map 
$\C^{Y}\stoch \C^{X}$ uniquely determined by sending the basis vector $e_{y}$ 
to the function $\sum\limits_{x\in X}f_{yx}e_{x}\in\C^{X}$ (cf.\ Example~\ref{ex:aecommutativealgebras}). 

\bprf
This follows from Example~\ref{ex:aecommutativealgebras}, Proposition~\ref{prop:aecompositionalgebraswelldefined}, Theorem~\ref{thm:compositionpositiveunital}, 
and Section~2 of \cite{Pa17}. 
\eprf

\br
Note that when $f:X\to Y$ is a function, the 
functor from Corollary~\ref{cor:classicalembedstoquantum} item~\ref{item:fpstoncfps} 
produces the $^*$-homomorphism $\C^{Y}\to\C^{X}$ sending $\varphi\in\C^{Y}$ to $\varphi\circ f$, 
the pullback of $\vf$ along $f.$ 
\er

\br
\label{rmk:ChoJacobsNCae}
The diagrammatic definition of a.e.\ equivalence discussed in Remark~\ref{rmk:aeChoJacobs} cannot be transferred to our categories
of $C^*$-algebras and positive maps. 
To see this, first note that 
the cartesian product of sets also goes to the tensor product of $C^*$-algebras
(up to a natural isomorphism), i.e.\ 
the (contravariant) functor 
sending stochastic maps to PU maps extends to a monoidal functor (the product of stochastic maps is defined by the product of the associated probability measures).
In particular, 
the diagonal map $\D_{X}:X\to X\times X$ becomes
the multiplication map 
$\C^{X}\otimes\C^{X}\to\C^{X}.$ 
Therefore, a natural candidate for 
$\mathcal{A}\otimes\mathcal{A}\stoch\mA$ would be the linear map that takes the 
product of the elements. 
However, this map is not positive in general. This is closely related to the no-cloning/no-broadcasting theorem in quantum mechanics~\cites{Pa70,WoZu82,Di82,BCFJS,CoSp12,PaBayes}. 
As a result, it is not a morphism in any of our categories.
Our definition of a.e.\ equivalence in Definition~\ref{defn:aestates}, 
though not explicitly categorical, gives a direct
definition of a.e.\ equivalence in terms of null spaces 
and is suitable for our purposes of non-commutative probability.
Nevertheless, it has recently been proven that this result does agree with the categorical definition of a.e.\ equivalence when instantiated in quantum Markov categories~\cite[Theorem~5.12]{PaBayes}. 
\er

%%%%%%%%%%%%%%%%%%%%%%%%%%%%%%%%%%%%%
\section{Non-commutative disintegrations on matrix algebras} 
%\label{sec:ncdisint}
\label{sec:generalresults}
%%%%%%%%%%%%%%%%%%%%%%%%%%%%%%%%%%%%%

Here, we define optimal hypothesis, 
disintegration, and regular conditional probability
in the non-commutative setting. 
In Theorem~\ref{thm:diagonalimpliesseparable}, 
we provide a necessary and sufficient condition for a
disintegration to exist on matrix algebras. 
The state on
the initial algebra must be separable 
with the induced state as a factor.
This result holds for $^{*}$-homomorphisms of a special kind. 
In this same theorem, it is shown that a disintegration is \emph{unique} 
whenever one exists. 
In the proof, we construct an explicit formula for \emph{any} disintegration on matrix algebras. 
Theorem~\ref{thm:existenceofCPUdisintegration}
covers the more general case of arbitrary $^{*}$-homomorphisms between
matrix algebras. Briefly, the existence no longer requires the
initial state to be separable.
However, it is separable after a specific unitary operation 
that transforms the $^{*}$-homomorphism to one
of the kind discussed in Theorem~\ref{thm:diagonalimpliesseparable}.

\bd
\label{defn:non-commutativedisintegration}
Given a state-preserving $^*$-homomorphism
$(\mB,\xi)\xrightarrow{F}(\mA,\omega)$
on $C^*$-algebras, 
a \define{hypothesis} for $(\mB,\xi)\xrightarrow{F}(\mA,\omega)$
is a CPU map $R:\mA\stoch\mB$ such that%
%footnote
\footnote{This definition of hypothesis is a non-commutative generalization of the definition from~\cite{BaFr14}. In~\cite{BaFr14}, the definition also requires equality rather than a.e.\ equality, so our notion is also a weakening in this sense.
}
%end footnote
$R\circ F\underset{\raisebox{.6ex}[0pt][0pt]{\scriptsize$\xi$}}{=}\id_{\mB}.$
%\ee
A hypothesis for $(\mB,\xi)\xrightarrow{F}(\mA,\omega)$ is \define{optimal}
iff 
$\xi\circ R=\w.$
A CPU map $R:\mA\stoch\mB$ is a \define{disintegration of $\w$ over $\xi$} iff $\xi\circ R=\w$ holds.
A CPU map $R:\mA\stoch\mB$ is a \define{disintegration of $\w$ over $\xi$ consistent with $F$} iff $R$ is a 
disintegration of $\w$ over $\xi$ such that 
$R\circ F\underset{\raisebox{.6ex}[0pt][0pt]{\scriptsize$\xi$}}{=}\id_{\mB}$. 
More concisely, a \define{disintegration} refers to a disintegration of $\w$ over $\xi$ consistent with $F.$
\ed
 
The following example illustrates how Definition~\ref{defn:non-commutativedisintegration} 
extends the classical definition of a disintegration 
to the non-commutative setting.

\bx
\label{ex:reducingdisintegration}
Let $X$ and $Y$ be finite sets (with the discrete $\s$-algebras) with probability measures 
$p:\{\bullet\}\stoch X$ and $q:\{\bullet\}\stoch Y.$
Let $f:X\to Y$ be a function
and let $r:Y\stoch X$ be a stochastic map. 
Let $P:\C^{X}\stoch\C$, $Q:\C^{Y}\stoch\C$, 
$F:\C^{Y}\to\C^{X},$ and $R:\C^{X}\stoch\C^{Y}$ denote the corresponding PU maps (cf.\ Example~\ref{ex:aecommutativealgebras}).
Functoriality as discussed after Corollary~\ref{cor:classicalembedstoquantum} immediately implies the following. 
\begin{enumerate}[i.]
\itemsep0pt
\item
$F$ is state-preserving if and only if $f$ is measure-preserving.
\item
$R$ is a disintegration of $P$ over $Q$
consistent with $F$
in the sense of Definition~\ref{defn:non-commutativedisintegration}
if and only if $r$
is a disintegration of $p$ over $q$ consistent with $f$
in the sense of Appendix~\ref{app:optimalhyparedis} (see also Theorem~\ref{thm:classicalmeasprestocondition} later for a simpler description in terms of finite sets).
\end{enumerate}
\ex

Several natural questions arise when comparing our definition 
of disintegration
to the one from finite probability spaces, measure-preserving maps, 
and stochastic maps. First of all, given a state-preserving $^*$-homomorphism 
$(\mB,\xi)\xrightarrow{F}(\mA,\omega)$, does there \emph{exist} a
disintegration $R$ over $\w$ consistent with $\xi$? 
Second, if a disintegration exists, is it unique or at least unique up to a.e.\
equivalence? 
Third, is a disintegration of a $^*$-isomorphism (a.e.\ equivalent to) the inverse? 
All of these are true in the commutative case.
Before addressing the general case of finite-dimensional $C^*$-algebras, the present section focuses on the setting of matrix algebras. 

\bt
\label{thm:diagonalimpliesseparable}
Fix $n,p\in\N.$ 
Let $F$ be the $^{*}$-homomorphism given by
the block diagonal inclusion
\be
\label{eq:blockdiagonalinclusion}
\mathcal{M}_{n}(\C)\ni B\xmapsto{F}\begin{bmatrix}B&&0\\&\ddots&\\0&&B\end{bmatrix}\equiv\mathds{1}_{p}\otimes B\in\mathcal{M}_{np}(\C)\cong\mathcal{M}_{p}(\C)\otimes\mathcal{M}_{n}(\C)
\ee
and let 
$(\mathcal{M}_{n}(\C),\tr(\rho\;\cdot\;)\equiv\w)\xrightarrow{F}(\mathcal{M}_{np}(\C),\xi\equiv\tr(\s\;\cdot\;))$ be state-preserving. 
Then the following facts hold. 
\begin{enumerate}[i.]
\itemsep0pt
\item
\label{item:mainmatrixalgebrai}
A disintegration $R$ of $\w$ over $\xi$ consistent with $F$
exists if and only if there exists 
a density matrix $\t\in\mathcal{M}_{p}(\C)$
such that $\rho=\t\otimes\s.$ 
\item
\label{item:mainmatrixalgebraii}
When such a $\tau$ exists, the disintegration is unique and is given by the formula
\be
\label{eq:fromtautoR}
\mathcal{M}_{np}(\C)\ni A\equiv\begin{bmatrix}A_{11}&\cdots&A_{1p}\\\vdots&&\vdots\\A_{p1}&\cdots&A_{pp}\end{bmatrix}\mapsto R(A):=\sum_{j,k=1}^{p}\t_{kj}A_{jk}\equiv \tr_{\mM_{p}(\C)}\big((\t\otimes\mathds{1}_{n})A\big), 
\ee
where $A_{jk}$ is the $jk$-th $n\times n$ block of $A$
using the isomorphisms 
$\mathcal{M}_{np}(\C)\cong\mathcal{M}_{p}(\mathcal{M}_{n}(\C))\cong\mM_{p}(\C)\otimes\mM_{n}(\C)$
and $\tr_{\mM_{p}(\C)}:\mM_{p}(\C)\otimes\mM_{n}(\C)\stoch\mM_{n}(\C)$ is the \define{partial trace}, uniquely determined by sending $C\otimes B\in\mM_{p}(\C)\otimes\mM_{n}(\C)$ to $\tr(C)B$. 
Furthermore, $R\circ F=\id_{\mathcal{M}_{n}(\C)}.$
\item
When such a $\tau$ exists, a Kraus decomposition of $R$ is given by 
\be
R=\mathrm{Ad}_{\begin{bmatrix}\mathds{1}_{n}&0&\cdots&0\end{bmatrix}(\sqrt{\t}\otimes\mathds{1}_{n})}+\cdots+\mathrm{Ad}_{\begin{bmatrix}0&\cdots&0&\mathds{1}_{n}\end{bmatrix}(\sqrt{\t}\otimes\mathds{1}_{n})}.
\ee
\end{enumerate}
\et

As a consequence of uniqueness, 
we prove all disintegrations on matrix algebras
are strict left inverses of their associated $^{*}$-homomorphism. 
Note that uniqueness is meant in the 
literal sense, not in the a.e.\ sense. This is surprising due to 
Remark~\ref{rmk:aeequaldnimplyequal}, which says two a.e.\ equivalent 
CPU maps on matrix algebras need not be equal. The additional
conditions for a disintegration are strong enough to imply 
equality. The following proof also provides a construction of the density matrix $\tau$ from $R$.

\bprf
[Proof of Theorem~\ref{thm:diagonalimpliesseparable}.]
{\color{white}{you found me!}}
\begin{enumerate}[i.]
\itemsep0pt
\item
($\Rightarrow$)
Suppose a disintegration $R:\mathcal{M}_{np}(\C)\stoch\mathcal{M}_{n}(\C)$
exists. 
Let
\be
\label{eq:KrausdecompR}
R=\sum_{i=1}^{n^2p}\mathrm{Ad}_{\begin{bmatrix}V_{i1}&\cdots&V_{ip}\end{bmatrix}}
\ee
be a Kraus decomposition of $R$ with $V_{ij}\in\mathcal{M}_{n}(\C)$ 
for all $i\in\{1,\dots,n^2p\}$ and $j\in\{1,\dots,p\}$ 
($n^2p$ is the minimal number of Kraus operators needed in this case). 
For the moment, let $R_{i}:=\begin{bmatrix}V_{i1}&\cdots&V_{ip}\end{bmatrix}.$
Also note that $F$ has a Kraus decomposition $F=\sum\limits_{j=1}^{p}\mathrm{Ad}_{F_{j}},$
where the (adjoint of the) Kraus operators are given by 
\be
\label{eq:KrausdecompF}
F_{j}^{\dag}:=\begin{bmatrix}0&\cdots&\mathds{1}_{n}&\cdots&0\end{bmatrix}
\ee
with $\mathds{1}_{n}$ in the $j$-th $n\times n$ block.
By Corollary \ref{cor:ChoiKrausIdentity}, 
there exist
numbers $\{\a_{ij}\in\C\}_{\substack{i\in\{1,\dots,n^2p\},\\j\in\{1,\dots,p\}}}$ such that 
\be
\label{eq:Choiuniquenessalphas}
V_{ij}=R_{i}F_{j}=\a_{ij}\mathds{1}_{n}
\quad\forall\;i,j
%\aand
\qquad\text{and}\qquad
\sum_{i,j}|\a_{ij}|^2=1.
\ee
due to the form of our matrices in (\ref{eq:KrausdecompR}) 
and (\ref{eq:KrausdecompF}).
We now impose the condition $\xi\circ R=\w,$
which is equivalent to $R^{*}(\s)=\rho,$
where 
\be
\label{eq:Rkraus}
R^{*}=\sum_{i=1}^{n^2p}\mathrm{Ad}_{R_{i}^{\dag}}
\ee
is the dual or $R$ with respect to the Hilbert--Schmidt inner product 
(cf.\ Remark~\ref{rmk:notationQIT}). 
Therefore, 
\be
\label{eq:constructingtau}
\begin{split}
\rho&=R^{*}(\s)\overset{\text{(\ref{eq:Rkraus})}}{=\joinrel=\joinrel=\joinrel=}\sum_{i=1}^{n^2p}\mathrm{Ad}_{R_{i}^{\dag}}(\s)
\overset{\text{(\ref{eq:KrausdecompR})}}{=\joinrel=\joinrel=}\sum_{i=1}^{n^2p}\begin{bmatrix}V_{i1}^{\dag}\\\vdots\\V_{ip}^{\dag}\end{bmatrix}\s\begin{bmatrix}V_{i1}&\cdots&V_{ip}\end{bmatrix}
\\
&\overset{\text{(\ref{eq:Choiuniquenessalphas})}}{=\joinrel=\joinrel=}\sum_{i=1}^{n^2p}\begin{bmatrix}\overline{\a_{i1}}\mathds{1}_{n}\\\vdots\\\overline{\a_{ip}}\mathds{1}_{n}\end{bmatrix}
\s
\begin{bmatrix}\a_{i1}\mathds{1}_{n}&\cdots&\a_{ip}\mathds{1}_{n}\end{bmatrix}
=
\sum_{i=1}^{n^2p}
\begin{bmatrix}|\a_{i1}|^2\s&\cdots&\overline{\a_{i1}}\a_{ip}\s\\\vdots&&\vdots\\\overline{\a_{ip}}\a_{i1}\s&\cdots&|\a_{ip}|^2\s\end{bmatrix}
\\
&\overset{\text{(\ref{eq:kroneckerproduct})}}{=\joinrel=\joinrel=}
\underbrace{\left(\sum_{i=1}^{n^2p}\begin{bmatrix}|\a_{i1}|^2&\cdots&\overline{\a_{i1}}\a_{ip}\\\vdots&&\vdots\\\overline{\a_{ip}}\a_{i1}&\cdots&|\a_{ip}|^2\end{bmatrix}\right)}_{=:\t}\otimes\s
\end{split}
\ee
showing that $\rho$ is separable 
and has a tensor product factorization with $\s$ as a factor.
Now, $\t\in\mathcal{M}_{p}(\C)$ is a positive matrix because it is a positive sum of positive operators, namely
\be
\t=\sum_{i=1}^{n^2p}\begin{bmatrix}\overline{\a_{i1}}\\\vdots\\\overline{\a_{ip}}\end{bmatrix}\begin{bmatrix}\a_{i1}&\cdots&\a_{ip}\end{bmatrix}.
\ee
Furthermore, (\ref{eq:Choiuniquenessalphas}) implies
\be
\tr(\t)=\sum_{j=1}^{p}\sum_{i=1}^{n^2p}|\a_{ij}|^2=1,
\ee
which shows that $\t$ is a density matrix.  

%\vspace{3mm}
\noindent
($\Leftarrow$)
Conversely, suppose there exists a density matrix $\t$ such that 
$\rho=\t\otimes\s.$ 
Define $R:\mathcal{M}_{np}(\C)\stoch\mathcal{M}_{n}(\C)$ as in (\ref{eq:fromtautoR}). 
The map $R$ is linear by construction and unital since 
\be
R(\mathds{1}_{np})
=\sum_{j,k=1}^{p}\t_{kj}\de_{jk}\mathds{1}_{p}
=\sum_{j}^{p}\t_{jj}\mathds{1}_{p}
=\mathds{1}_{p}
\ee
because $\tr(\t)=1.$ 
A similar calculation shows
\be
R\big(F(B)\big)=\sum_{j,k=1}^{p}\t_{kj}\de_{jk}B=B
\ee
for all $B\in\mathcal{M}_{n}(\C).$ 
Hence, $R$ is actually a left inverse of $F.$
In order for $R$ to preserve the states,
it must be that $\w\circ R=\xi,$ i.e.\ $\tr(\rho A)=\tr(\s R(A))$ for all
$A\in\mathcal{M}_{np}(\C).$ This follows from 
\be
\begin{split}
\tr(\rho A)
&=\tr\big((\t\otimes\s)A\big)
%\\
%&
=\tr\left(\begin{bmatrix}\t_{11}\s&\cdots&\t_{1p}\s\\\vdots&&\vdots\\\t_{p1}\s&\cdots&\t_{pp}\s\end{bmatrix}\begin{bmatrix}A_{11}&\cdots&A_{1p}\\\vdots&&\vdots\\A_{p1}&\cdots&A_{pp}\end{bmatrix}\right)\\
&=\tr\left(\begin{bmatrix}\sum\limits_{j=1}^{p}\t_{1j}\s A_{j1}&\cdots&\sum\limits_{j=1}^{p}\t_{1j}\s A_{jp}\\\vdots&&\vdots\\\sum\limits_{j=1}^{p}\t_{pj}\s A_{j1}&\cdots&\sum\limits_{j=1}^{p}\t_{pj}\s A_{jp}\end{bmatrix}\right)
%\\
%&
=\sum_{j,k=1}^{p}\t_{kj}\tr(\s A_{jk})
=\tr(\s(R(A)).
\end{split}
\ee
The final step is to prove $R$ is CP.  
This follows from the fact that the partial trace satisfies a partially cyclic property, namely 
\be
\tr_{\mM_{p}(\C)}\big((\t\otimes\mathds{1}_{n})A\big)=\tr_{\mM_{p}(\C)}\big(A(\t\otimes\mathds{1}_{n})\big).
\ee
Thus, 
\be
\tr_{\mM_{p}(\C)}\big((\t\otimes\mathds{1}_{n})A\big)
=\tr_{\mM_{p}(\C)}\big((\sqrt{\t}\otimes\mathds{1}_{n})A(\sqrt{\t}\otimes\mathds{1}_{n})\big)
=\left(\tr_{\mM_{p}(\C)}\circ\mathrm{Ad}_{\sqrt{\t}\otimes\mathds{1}_{n}}\right)(A).
\ee
shows that $R$ is the composite of two CP maps, and is therefore CP. The claim also follows from showing that the Choi matrix associated to $R$ is $\Phi(R)=\t^{T}\otimes\Phi(\mathrm{id}_{\mathcal{M}_{n}(\C)})$, which is positive (the proof is omitted). 

\item
Suppose $R'$ is another disintegration of $\w$ over $\xi$
consistent with $F.$ 
Let $\{\a_{ij}\}$ and $\{\a'_{ij}\}$ be coefficients
obtained from Choi's theorem as in (\ref{eq:Choiuniquenessalphas}). 
Construct the density matrices $\t$ and $\t'$ as in the proof of 
part~\ref{item:mainmatrixalgebrai} of this theorem.
Then $\t'\otimes\s=\rho=\t\otimes\s,$
i.e.\ $(\t-\t')\otimes\s=0.$ Since $\s$ is non-zero, 
this means $\t-\t'=0,$ i.e.\ $\t'=\t.$ 
Hence, each of the entries of $\t$ and $\t'$ are equal, 
i.e.\ $\t_{jk}=\t'_{jk}$ for all $j,k,$ or in terms of the $\a$'s and $\a'$'s,
\be
\label{eq:uniquenessalphas}
\sum_{i=1}^{n^2p}\overline{\a_{ij}}\a_{ik}=\sum_{i=1}^{n^2p}\overline{\a'_{ij}}\a'_{ik}\qquad\forall\;j,k\in\{1,\dots,p\}.
\ee
Now, let $A$ be a matrix in $\mathcal{M}_{np}(\C)\cong\mathcal{M}_{p}(\mathcal{M}_{n}(\C))$ as in~(\ref{eq:fromtautoR})
so that each $A_{jk}\in\mathcal{M}_{n}(\C).$ 
Then, after some algebra
\be
\begin{split}
R(A)&=\sum_{i=1}^{n^2p}
\begin{bmatrix}\a_{i1}\mathds{1}_{n}&\cdots&\a_{ip}\mathds{1}_{n}\end{bmatrix}
\begin{bmatrix}A_{11}&\cdots&A_{1p}\\\vdots&&\vdots\\A_{p1}&\cdots&A_{pp}\end{bmatrix}
\begin{bmatrix}\overline{\a_{i1}}\mathds{1}_{n}\\\vdots\\\overline{\a_{ip}}\mathds{1}_{n}\end{bmatrix}
\\
&=\sum_{j,k=1}^{p}\left(\sum_{i=1}^{n^2p}\a_{ij}\overline{\a_{ik}}\right)A_{jk}
\overset{\text{(\ref{eq:uniquenessalphas})}}{=\joinrel=\joinrel=\joinrel=}\sum_{j,k=1}^{p}\left(\sum_{i=1}^{n^2p}\a'_{ij}\overline{\a'_{ik}}\right)A_{jk}
=R'(A),
\end{split}
\ee
which shows that $R=R'.$ 
Hence, disintegrations are unique when they exist. 
The fact that $R\circ F=\id_{\mathcal{M}_{n}(\C)}$ follows from
uniqueness of disintegrations and Corollary \ref{cor:ChoiKrausIdentity}. 

\item
The formula for the Kraus decomposition follows from the results just proven and a Kraus decomposition for the partial trace. \qedhere
\end{enumerate}
\eprf

\begin{example}
Let 
\be
\label{eq:EPRdensitymatrix}
\rho:=\frac{1}{2}\begin{bmatrix}0&0&0&0\\0&1&-1&0\\0&-1&1&0\\0&0&0&0\end{bmatrix}
\ee
be the density matrix on $\C^4$ corresponding to the projection 
operator onto the one-dimensional subspace of $\C^2\otimes\C^2$ 
spanned by the vector%
%footnote
\footnote{This is the spin EPR state discussed in Section~1.3.6 in Nielsen and Chuang~\cite{NiCh11}.}
%end footnote
\be
\vec{u}_{\mathrm{EPR}}:=\frac{1}{\sqrt{2}}\big(\vec{e}_{1}\otimes\vec{e}_{2}-\vec{e}_{2}\otimes\vec{e}_{1}\big).
\ee
Let $F:\mathcal{M}_{2}(\C)\to\mathcal{M}_{4}(\C)$ be the map defined by 
\be
\label{eq:Fdiagonal}
\mathcal{M}_{2}(\C)\ni B\mapsto F(B):=\begin{bmatrix}B&0\\0&B\end{bmatrix},
\ee
which corresponds to the assignment $\mathcal{M}_{2}(\C)\ni B\mapsto\mathds{1}_{2}\otimes B\in\mathcal{M}_{2}(\C)\otimes\mathcal{M}_{2}(\C)$
under the isomorphism from (\ref{eq:tensorproductiso}). 
Let $\s$ be the density matrix on $\C^{2}$ given by 
$
\s:=\frac{1}{2}\mathds{1}_{2}. 
$
Let $\w:=\tr(\rho\;\cdot\;)$ and $\xi:=\tr(\s\;\cdot\;)$ be the corresponding
states.
Then, $\mathcal{N}_{\xi}=\{0\}$ and 
$(\mathcal{M}_{2}(\C),\tr(\rho\;\cdot\;)\equiv\w)\xrightarrow{F}(\mathcal{M}_{4}(\C),\xi\equiv\tr(\s\;\cdot\;))$ is state-preserving, 
but there does not exist a 
disintegration of $\w$ over $\xi$ consistent with $F.$ 
\end{example}

\bx
Fix $p_{1},p_{2},p_{3},p_{4}\ge0$ with 
$p_{1}+p_{2}+p_{3}+p_{4}=1,$ 
$p_{1}+p_{3}>0,$ and $p_{2}+p_{4}>0.$
Let 
\be
\label{eq:rhoandsigma}
\rho=\begin{bmatrix}p_{1}&0&0&0\\0&p_{2}&0&0\\0&0&p_{3}&0\\0&0&0&p_{4}\end{bmatrix}
%\aand
\qquad\text{and}\qquad
\s=\begin{bmatrix}p_{1}+p_{3}&0\\0&p_{2}+p_{4}\end{bmatrix}
\ee
be density matrices with associated states given by 
$\w:=\tr(\rho\;\cdot\;)$ and
$\xi:=\tr(\s\;\cdot\;),$ respectively.
Let $F:\mathcal{M}_{2}(\C)\to\mathcal{M}_{4}(\C)$ be the 
diagonal inclusion from (\ref{eq:Fdiagonal}).
Then $\xi=\w\circ F$ and $\mathcal{N}_{\xi}=\{0\}.$
Furthermore, 
a CPU disintegration $R:\mathcal{M}_{4}(\C)\stoch\mathcal{M}_{2}(\C)$
of $\w$ over $\xi$ consistent with $F$ exists
if and only if 
$p_{1}p_{4}=p_{2}p_{3}.$
When this holds, the map 
\be
\label{eq:diagonaldensityRmap}
R=\mathrm{Ad}_{\sqrt{p_{1}+p_{2}}\begin{bmatrix}\mathds{1}_{2}&0\end{bmatrix}}
+\mathrm{Ad}_{\sqrt{p_{3}+p_{4}}\begin{bmatrix}0&\mathds{1}_{2}\end{bmatrix}},
\ee
is the unique disintegration of $\w$ over $\xi$ consistent with $F.$ 
Furthermore, the density matrix $\t\in\mathcal{M}_{2}(\C)$
given by 
\be
\t=\begin{bmatrix}[1.25]\frac{p_{1}}{p_{1}+p_{3}}&0\\0&\frac{p_{4}}{p_{2}+p_{4}}\end{bmatrix}
\ee
satisfies 
$
\t\otimes\s=\rho.
$
\ex

\br
Theorem~\ref{thm:diagonalimpliesseparable} reproduces a well-known result in quantum information theory in the special case when the density matrices $\rho$ and $\s$ are invertible (see Example~9.6 in Petz's text for example~\cite{Pe08}).%
%footnote
\footnote{The techniques we have used to prove our results do not use Takesaki's theorem nor the modular group (see~\cite[Theorem~9.2]{Pe08}). Instead, we worked directly with Kraus operators, a familiar tool in the quantum information theory community. A deeper analysis relating conditional expectations to disintegrations will be presented in forthcoming work.}
%end footnote 
The surprising result we have shown is the fact that this still holds regardless of the sizes of the null-spaces associated to the density matrices and, moreover, the disintegration is uniquely determined. 
\er

The following result is a generalization of 
Theorem~\ref{thm:diagonalimpliesseparable}
on the existence of disintegrations to allow for 
$^{*}$-homomorphisms $F$ that are not necessary of the block diagonal
form.

\bt
\label{thm:existenceofCPUdisintegration}
Fix $n,p\in\N.$ 
Let 
$(\mathcal{M}_{n}(\C),\tr(\rho\;\cdot\;)\equiv\w)\xrightarrow{F}(\mathcal{M}_{np}(\C),\xi\equiv\tr(\s\;\cdot\;))$ be a state-preserving $^*$-homomorphism. 
A disintegration of $\w$ over $\xi$ consistent with $F$
exists if and only if there exists a unitary 
$U\in \mathcal{M}_{np}(\C)$ and a density matrix $\t\in\mathcal{M}_{p}(\C)$ such that
$F=\mathrm{Ad}_{U}\circ i$ and
$U^{\dag}\rho U=\t\otimes\s.$ 
Here $i:\mathcal{M}_{n}(\C)\to\mathcal{M}_{np}(\C)$
is the block diagonal inclusion (\ref{eq:blockdiagonalinclusion}).
Furthermore, if a disintegration exists, it is unique. 
\et

\bprf
For any unital $^{*}$-homomorphism $F:\mathcal{M}_{n}(\C)\to\mathcal{M}_{np}(\C),$
there exists a unitary $U\in\mathcal{M}_{np}(\C)$ such that
$F=\mathrm{Ad}_{U}\circ i$ 
(cf.\ Section~1.1.2 of Fillmore~\cite{Fi96}).
Hence, 
the diagram 
\be
\xy0;/r.25pc/:
(0,-7.5)*+{\C}="C";
(-12.5,7.5)*+{\mathcal{M}_{np}(\C)}="H";
(12.5,7.5)*+{\mathcal{M}_{n}(\C)}="K";
{\ar@{~>}"H";"C"_{\w\circ\mathrm{Ad}_{U}=\tr(U^{\dag}\rho U\;\cdot\;)}};
{\ar@{~>}"K";"C"^{\xi\equiv\tr(\s\;\cdot\;)}};
{\ar"K";"H"_{i}};
\endxy
\ee
commutes. 
By Theorem~\ref{thm:diagonalimpliesseparable}, 
a disintegration $R:\mathcal{M}_{np}(\C)\stoch\mathcal{M}_{n}(\C)$
of $\w\circ\mathrm{Ad}_{U}$ over $\xi$ consistent with $i$
exists if and only if there exists a density matrix $\t$ such that
$U^{\dag}\rho U=\t\otimes \s.$ Explicitly, this means 
$\xi\circ R=\w\circ\mathrm{Ad}_{U}$ and 
$R\circ i\underset{\raisebox{.6ex}[0pt][0pt]{\scriptsize$\xi$}}{=}\id_{\mathcal{M}_{n}(\C)}.$
Setting
$R_{U}:=R\circ\mathrm{Ad}_{U^{\dag}}:\mathcal{M}_{np}(\C)\stoch\mathcal{M}_{n}(\C)$ 
and applying $\mathrm{Ad}_{U^{\dag}}$
to the right of $\xi\circ R=\w\circ\mathrm{Ad}_{U}$ gives
$\xi\circ R_{U}=\w.$ 
Similarly, $R\circ i\underset{\raisebox{.6ex}[0pt][0pt]{\scriptsize$\xi$}}{=}\id_{\mathcal{M}_{n}(\C)}$ holds if and only if
$R\circ\mathrm{Ad}_{U^{\dag}}\circ\mathrm{Ad}_{U}\circ i\underset{\raisebox{.6ex}[0pt][0pt]{\scriptsize$\xi$}}{=}\id_{\mathcal{M}_{n}(\C)}$ holds, i.e.\ 
$R_{U}\circ F\underset{\raisebox{.6ex}[0pt][0pt]{\scriptsize$\xi$}}{=}\id_{\mathcal{M}_{n}(\C)}.$
The map $R_{U}$ is CPU if and only if $R$ is CPU. Thus, $R_{U}$ defines
a disintegration of $\w$ over $\xi$ consistent with $F$ if and only if 
$U^{\dag}\rho U=\t\otimes \s.$
Finally, the uniqueness of $R_{U}$ follows from the uniqueness of $R$
by part~\ref{item:mainmatrixalgebraii} of Theorem~\ref{thm:diagonalimpliesseparable}.
\eprf

Note that an immediate consequence of this theorem is when $F$ is a $^*$-isomorphism, then $F^{-1}$ is the unique disintegration. Thus, a disintegration can be viewed as a generalization of time reversal. 

\br
\label{rmk:unitaryseparable}
Theorem~\ref{thm:existenceofCPUdisintegration},
says there exists a tensor factorization
$\t\otimes\s=U^{\dag}\rho U$ 
if and only if a disintegration exists. 
It is not necessary for $\rho$ to be separable
in this case (compare this to Theorem~\ref{thm:diagonalimpliesseparable},
where $\rho=\t\otimes\s$ was separable).
This is because $U^{\dag}\rho U$ is separable
does not imply $\rho$ is separable in general---the unitary evolution of a separable state can cause that state to become
entangled due to interactions between subsystems.
\er

\br
Theorem~\ref{thm:existenceofCPUdisintegration} bears a striking resemblance 
to Theorem~2.1 in the work of Nayak and Sen~\cite{NaSe07}. 
However, there are three main differences. 
First, they work with completely positive
trace-preserving (not necessarily unital) 
maps $F:\mathcal{M}_{m}(\C)\stoch\mathcal{M}_{n}(\C),$
where $m\le n,$ 
while we focus
on the class of unital $^{*}$-homomorphisms.
Second, they assume $R:\mathcal{M}_{n}(\C)\stoch\mathcal{M}_{m}(\C)$
is a strict left inverse of $F$ 
while we initially assume $R$ is a left inverse up to a.e.\ equivalence.
We showed this condition is actually equivalent for matrix algebras
in Corollary \ref{cor:ChoiKrausIdentity} but we will see that a.e.\ equivalence
is necessary for arbitrary finite-dimensional $C^*$-algebras. 
Third, and most importantly, 
Nayak and Sen do not require $R$ and $F$ to preserve any specified states
while we do. 
This forces an additional constraint that our map $R$ must satisfy
making it even less obvious whether such a CP map $R$ exists.
Therefore, it seems that neither of our results subsume each other
but are complementary and cover different situations.
\er

If a deterministic process (a $^*$-homomorphism) evolves a pure state into a mixed state between matrix algebras, is it possible for there to exist a disintegration that evolves the mixed state back into the pure state? The following corollary is a ``no-go theorem'' for such disintegrations. 

\bc
\label{cor:purestatesnodisint}
Given a state-preserving $^*$-homomorphism $(\mathcal{M}_{n}(\C),\tr(\s\;\cdot\;))\xrightarrow{F}(\mathcal{M}_{np}(\C),\tr(\rho\;\cdot\;))$,
with $\rho$ pure, if 
a disintegration exists, then $\s$ must necessarily be pure as well.
\ec

\bprf
By Theorem~\ref{thm:existenceofCPUdisintegration}, there exist a unitary $U\in\mathcal{M}_{np}(\C)$ and a density matrix $\tau\in\mathcal{M}_{p}(\C)$ such that $F(A)=U\mathrm{diag}(A,\dots,A)U^{\dag}$ and $\rho=U^{\dag}(\tau\otimes\sigma)U.$ Since $\rho$ is pure, it is a rank 1 projection operator. Its rank also equals $\mathrm{rank}(\rho)=\mathrm{rank}(\tau)\mathrm{rank}(\sigma),$ which equals $1$ if and only if both $\mathrm{rank}(\tau)$ and $\mathrm{rank}(\sigma)$ are equal to 1. Hence $\tau$ and $\sigma$ are pure.
\eprf

\br
One might object to the conclusion of Corollary \ref{cor:purestatesnodisint} and ask a more elementary question without referring to disintegrations. Namely, does there exist a mixed state $\xi:\mathcal{M}_{n}(\C)\stoch\C$ and a CPU map $\vf:\mathcal{M}_{m}(\C)\stoch\mathcal{M}_{n}(\C)$ such that $\xi\circ\vf$ is a pure state? 
The reason to ask such a question is that if its answer is no, then one does not even need a disintegration for it to be impossible to evolve a mixed state into a pure state.
The following example addresses this. Let $\w:\mathcal{M}_{m}(\C)\stoch\C$ be \emph{any} pure state and let $\xi:\mathcal{M}_{n}(\C)\stoch\C$ be \emph{any} mixed state. Set $\vf(A):=\w(A)\mathds{1}_{n}$, which is a CPU map satisfying $\w=\xi\circ\vf.$ 
Note that this situation is described by the diagram 
\be
\xy0;/r.25pc/:
(-20,0)*+{\mathcal{M}_{m}(\C)}="m";
(0,0)*+{\C}="1";
(20,0)*+{\mathcal{M}_{n}(\C)}="n";
(0,-10)*+{\C}="C";
{\ar@{~>}"m";"1"^(0.65){\w}};
{\ar@{~>}"m";"C"_{\w}};
{\ar"1";"n"^(0.35){!}};
{\ar"1";"C"^{!}};
{\ar@{~>}"n";"C"^{\xi}};
{\ar@{~>}@/^2.0pc/"m";"n"^{\vf}};
\endxy
,
\ee
i.e.\ $\vf$ factors through $\C.$ In this diagram, $!$ is the unique unital map from $\C$ into any (unital) $C^*$-algebra. 
\er

%%%%%%%%%%%%%%%%%%%%%%%%%%%%%%%%%%%%%
\section{Disintegrations on finite-dimensional $C^*$-algebras} 
\label{sec:fdcalg}
%%%%%%%%%%%%%%%%%%%%%%%%%%%%%%%%%%%%%

In the present section, we will extend Theorems~\ref{thm:diagonalimpliesseparable} and \ref{thm:existenceofCPUdisintegration} to the case of $^*$-homomorphisms between arbitrary finite-dimensional $C^*$-algebras, which are all isomorphic to finite direct sums of matrix algebras. 
We begin by analyzing CP maps between such direct sums in Lemma~\ref{lem:CPondirectsums}, their adjoints with respect to a generalized Hilbert--Schmidt inner product in Lemma~\ref{lem:vfadjointdirectsum}, and the general form of states on direct sums in Lemma~\ref{lem:xionsum}. After these preliminary results are established, we study the structure of Kraus decompositions of hypotheses in Lemma~\ref{cor:ChoiKrausIdentitydirectsum} and Lemma~\ref{thm:formofdisintoncalg}. Proposition~\ref{prop:Fpreservesstatesds} provides a generalization of the ``tracing out'' operation for direct sums, i.e.\ the induced state via pull-back from a $^*$-homomorphism and a state on the target. After all this preparation, our main result, Theorem~\ref{thm:theoremaegregium}, is provided. 
Theorem~\ref{thm:theoremaegregiumarbitraryF} generalizes this disintegration theorem to arbitrary (unital) $^*$-homomorphisms. 
But first, we recall the classical disintegration theorem. 

\bt
\label{thm:classicalmeasprestocondition}
Let 
$(X,p)\xrightarrow{f}(Y,q)$ be a probability-preserving function. 
Then the following facts hold.
\begin{enumerate}[i.]
\itemsep0pt
\item
The assignment
\be
\label{eq:fromfandptor}
X\times Y\ni(x,y)\mapsto r_{xy}:=
\begin{cases}
p_{x}\de_{yf(x)}/q_{y}&\mbox{ if } q_{y}>0\\
1/|X|&\mbox{ otherwise }
\end{cases}
\ee
defines a disintegration $r:Y\stoch X$ of $p$ over $q$ consistent with $f.$ 
\item
The stochastic map $r$ is the unique one up to a set of measure 
zero with respect to $q$ satisfying $f\circ r \underset{\raisebox{.6ex}[0pt][0pt]{\scriptsize$q$}}{=}\id_{Y}$ and
$r\circ q=p$, i.e.\ 
$r\underset{\raisebox{.6ex}[0pt][0pt]{\scriptsize$q$}}{=}r'$ for any other 
disintegration $r':Y\stoch X.$  
\item
Suppose $f'$ is another measure-preserving function satisfying 
$f\underset{\raisebox{.6ex}[0pt][0pt]{\scriptsize$p$}}{=}f'.$ 
Let $r$ be a disintegration of $f$ and let $r'$ be a disintegration of $f'.$ 
Then $r\underset{\raisebox{.6ex}[0pt][0pt]{\scriptsize$q$}}{=}r'.$
\end{enumerate}
\et

\begin{wrapfigure}{r}{0.27\textwidth}
  \centering
    \begin{tikzpicture}[decoration=snake]
\node at (-1.75,3.25) {$X$};
\node at (-1.75,0) {$Y$};
\draw[blue,thick,fill=blue,fill opacity=0.4] (-1,3.75) circle (1.0ex);
\draw[blue,thick,fill=blue,fill opacity=0.4] (-1,3.25) circle (0.5ex);
\draw[blue,thick,fill=blue,fill opacity=0.4] (-1,2.75) circle (1.25ex);
\draw[blue,thick,fill=blue,fill opacity=0.4] (0,4.5) circle (0.25ex);
\draw[blue,thick,fill=blue,fill opacity=0.4] (0,4.0) circle (0.75ex);
\draw[blue,thick,fill=blue,fill opacity=0.4] (0,3.5) circle (1.0ex);
\draw[blue,thick,fill=blue,fill opacity=0.4] (0,3) circle (0.5ex);
\draw[blue,thick,fill=blue,fill opacity=0.4] (1,4) circle (1.75ex); 
\draw[blue,thick,fill=blue,fill opacity=0.4] (1,3.5) circle (0.75ex); 
\draw[blue,thick,fill=blue,fill opacity=0.4] (1,3) circle (1.0ex); 
\draw[blue,thick,fill=blue,fill opacity=0.4] (1,2.5) circle (1.25ex); 
\draw[blue,thick,fill=blue,fill opacity=0.4] (2,3.75) circle (0.75ex); 
\draw[blue,thick,fill=blue,fill opacity=0.4] (2,3.25) circle (0.25ex); 
\draw[blue,thick,fill=blue,fill opacity=0.4] (2,2.75) circle (0.5ex); 
\draw[blue,thick,fill=blue,fill opacity=0.4] (2,2.25) circle (0.75ex); 
\draw[-{>[scale=2.5,
          length=2,
          width=3]},thick] (-0.5,2.0) -- node[left]{$f$} (-0.5,0.5);
\draw[-{>[scale=2.5,
          length=2,
          width=3]},decorate,thick] (1.5,0.5) -- node[right,xshift=0.1cm]{$r$} (1.5,2.0);
\draw[blue,thick,fill=blue,fill opacity=0.4] (-1,0) circle (1.677ex);%sqrt(1^2+0.5^2+1.25^2)
\draw[blue,thick,fill=blue,fill opacity=0.4] (0,0) circle (1.3693ex);
\draw[blue,thick,fill=blue,fill opacity=0.4] (1,0) circle (2.48746ex);
\draw[blue,thick,fill=blue,fill opacity=0.4] (2,0) circle (1.19895ex);
\end{tikzpicture}
%  \end{center}
%  \caption{Birds}
\end{wrapfigure}
We will omit the details of this proof, which are neither difficult nor new. 
However, some of the lemmas used in proving it provide insight into the proof of our main theorem
on non-commutative disintegrations in Theorem~\ref{thm:theoremaegregium}. 
These lemmas motivate the formula (\ref{eq:fromfandptor})
and also assist in proving a.e.\ uniqueness. 
They show that a measure-preserving function $(X,p)\xrightarrow{f}(Y,q)$ 
is surjective onto a set of full measure,
and they illustrate that a hypothesis $r:Y\stoch X$ forces $r_{y}$ to be supported on $f^{-1}(\{y\})$ 
for almost all $y\in Y.$ 
This allows us to think of a disintegration more visually as follows.%
%footnote
\footnote{We learned this point of view from Gromov~\cite{Gr14}.} 
%end footnote 
First, a probability space can be viewed as a finite number of water droplets, each of which has some volume (probability); the total volume is normalized to one. 
One can visualize a morphism $(X,p)\xrightarrow{f}(Y,q)$ 
as combining some of the water droplets, summing their volumes in the process.
A hypothesis is a choice of physically splitting the water droplets back to the original set, but possibly with different volumes.
A perfect splitting of the water droplets in which the volumes are reproduced exactly is an optimal hypothesis. 
From a topologist's point of view, a hypothesis is a stochastic section of $f$, which
assigns a probability measure on the fiber (as opposed to a specific element) over each point that has non-zero $q$ measure. 

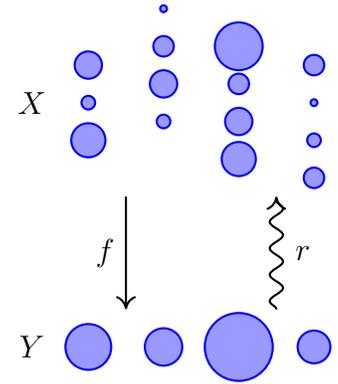
\blem
\label{lem:supportonfiber}
Let $r:Y\stoch X$ be a hypothesis for $(X,p)\xrightarrow{f}(Y,q).$ 
Then the probability measure $r_{y}$ is supported on $f^{-1}(\{y\})$ for all $y\in Y\setminus N_{q}.$  
\elem

\blem
\label{lem:aesurjective}
Let $(X,p)\xrightarrow{f}(Y,q)$ be a morphism in $\FinProb.$ 
Then, for each $y\in Y\setminus N_{q},$ there exists an $x\in X$ 
such that $f(x)=y,$ i.e.\ $f$ is surjective onto a set of full $q$-measure.
\elem

\blem
\label{lem:optimality}
Let $r:Y\stoch X$ be a disintegration of $(X,p)\xrightarrow{f}(Y,q).$ 
Then 
\be
r_{xf(x)}q_{f(x)}=p_{x}\qquad\forall\;x\in X.
\ee
\elem

\begin{notation}
\label{not:dsumsection}
Throughout the rest of this section, let 
\be
\label{eq:ABsum}
\mathcal{A}:=\mathcal{M}_{m_{1}}(\C)\oplus\cdots\oplus\mathcal{M}_{m_{s}}(\C)
\quad\text{ and }\quad
\mathcal{B}:=\mathcal{M}_{n_{1}}(\C)\oplus\cdots\oplus\mathcal{M}_{n_{t}}(\C)
\ee
denote direct sums of matrix algebras. 
An element $\vec{A}\in\mathcal{A}$ will be denoted as a column vector 
\be
\vec{A}\equiv
\begin{pmatrix}
A_{1}\\
\vdots\\
A_{s}
\end{pmatrix}
\ee
and similarly for elements of $\mathcal{B}.$ The vector notation is often used for emphasis. 
An arbitrary linear map $\vf:\mathcal{A}\stoch\mathcal{B}$ will be written 
in matrix form as 
\be
\label{eq:vfsum}
\vf\equiv
\begin{pmatrix}
\vf_{11}&\cdots&\vf_{1s}\\
\vdots&&\vdots\\
\vf_{t1}&\cdots&\vf_{ts}
\end{pmatrix}, 
\ee
where $\vf_{ji}:\mathcal{M}_{m_{i}}(\C)\stoch\mathcal{M}_{n_{j}}(\C)$
is a linear map for all $i,j.$ The notation indicates the action of $\vf$ on $\vec{A}$ as
\be
\vf(\vec{A})
=
\begin{pmatrix}
\vf_{11}&\cdots&\vf_{1s}\\
\vdots&&\vdots\\
\vf_{t1}&\cdots&\vf_{ts}
\end{pmatrix}
\begin{pmatrix}
A_{1}\\
\vdots\\
A_{s}
\end{pmatrix}
=
\begin{pmatrix}
\ds\sum_{i=1}^{s}\vf_{1i}(A_{i})\\
\vdots\\
\ds\sum_{i=1}^{s}\vf_{ti}(A_{i})\\
\end{pmatrix}
.
\ee
Let 
$(\mB,\xi)\xrightarrow{F}(\mA,\omega)$ be a state-preserving $^*$-homomorphism defined by%
%footnote
\footnote{We will work with more general $^{*}$-homomorphisms later, 
but we will see that all (unital) $^{*}$-homomorphisms are unitarily
equivalent to ones of this form. Hence, we do not lose much
generality by focusing on these.}
%end footnote
\be
\label{eq:FBratteli}
\mathcal{B}\ni \vec{B}\mapsto 
F\left(\begin{pmatrix}B_{1}\\\vdots\\B_{t}\end{pmatrix}\right)
:=\begin{pmatrix}\mathrm{diag}(\overbrace{B_{1},\dots,B_{1}}^{c_{11}\text{ times}},\dots,\overbrace{B_{t},\dots,B_{t}}^{c_{1t}\text{ times}})\\
\vdots\\
\mathrm{diag}(\underbrace{B_{1},\dots,B_{1}}_{c_{s1}\text{ times}},\dots,\underbrace{B_{t},\dots,B_{t}}_{c_{st}\text{ times}})
\end{pmatrix}
,
\ee
where the non-negative integer $c_{ij}$ is called the
\define{multiplicity} of $F$ of the factor $\mathcal{M}_{n_{j}}(\C)$
inside $\mathcal{M}_{m_{i}}(\C)$ (cf.\ Section~1.1.2 and 1.1.3 in Fillmore~\cite{Fi96}). 
In particular, the dimensions are related by the formula
\be
\label{eq:multiplicity}
m_{i}=\sum_{j=1}^{t}c_{ij}n_{j}\qquad\forall\;i\in\{1,\dots,s\}.
\ee
Since $F$ is linear,
it also has a matrix representation
\be
F\equiv
\begin{pmatrix}
F_{11}&\cdots&F_{1t}\\
\vdots&&\vdots\\
F_{s1}&\cdots&F_{st}
\end{pmatrix}
\ee
with $F_{ij}:\mathcal{M}_{n_{j}}(\C)\to\mathcal{M}_{m_{i}}(\C)$
a (not necessarily unital) $^{*}$-homomorphism for all $i,j.$ 
\end{notation}

\blem
\label{lem:CPondirectsums}
Let $\mathcal{A},\mathcal{B},$ and $\vf$ be as in 
(\ref{eq:ABsum}) and (\ref{eq:vfsum}). Then $\vf$ is CP if and only if $\vf_{ji}$ is CP for all $i\in\{1,\dots,s\}$ and $j\in\{1,\dots,t\}.$
Furthermore, $\vf$ is unital if and only if 
\be
\label{eq:unitalitydirectsum}
\mathds{1}_{n_{j}}=\sum_{i=1}^{s}\vf_{ji}(\mathds{1}_{m_{i}})\qquad\forall\;j\in\{1,\dots,t\}.
\ee
\elem

A Kraus decomposition of $\vf_{ji}$ in this case will be expressed as
\be
\label{eq:krausdecompvfji}
\vf_{ji}=\sum_{l_{ji}=1}^{m_{i}n_{j}}\mathrm{Ad}_{V_{ji;l_{ji}}}
\ee
where the $V_{ji;l_{ji}}:\C^{m_{i}}\to\C^{n_{j}}$ are linear maps.
This allows the unitality condition (\ref{eq:unitalitydirectsum})
to be expressed as
\be
\label{eq:unitalitydirectsumKraus}
\mathds{1}_{n_{j}}=\sum_{i=1}^{s}\sum_{l_{ji}=1}^{m_{i}n_{j}}V_{ji;l_{ji}}V_{ji;l_{ji}}^{\dag}\qquad\forall\;j\in\{1,\dots,t\}.
\ee 

The following facts are easy to check and are analogous to what happens
in the usual matrix algebra case. We include them here for completeness. 

\blem
\label{lem:HilbertSchmidt}
Let $\mathcal{A}:=\mathcal{M}_{m_{1}}(\C)\oplus\cdots\oplus\mathcal{M}_{m_{s}}(\C).$ Then the assignment
\be
\mathcal{A}\times\mathcal{A}\ni\left(\vec{A},\vec{A}'\right)\mapsto\left\<\vec{A},\vec{A}'\right\>
:=\sum_{i=1}^{s}\tr(A_{i}^{\dag}A_{i}')
\ee
defines an inner product on $\mathcal{A}.$ This is called the
\define{Hilbert--Schmidt} (a.k.a. \define{Frobenius}) inner
product on $\mathcal{A}.$ 
\elem

\blem
\label{lem:vfadjointdirectsum}
Let $\mathcal{A},\mathcal{B},$ and $\vf$ be as in 
(\ref{eq:ABsum}) and (\ref{eq:vfsum}). Then there exists a unique
linear map $\vf^*:\mathcal{B}\stoch\mathcal{A}$ satisfying
\be
\left\<\vec{B},\vf\big(\vec{A}\big)\right\>=\left\<\vf^*\big(\vec{B}\big),\vec{A}\right\>
\qquad\forall\;\vec{A}\in\mathcal{A},\;\vec{B}\in\mathcal{B}.
\ee
The linear map $\vf^*$ is called the \define{adjoint} of $\vf.$ 
Furthermore, 
\be
\vf^*=\begin{pmatrix}
\vf_{11}^*&\cdots&\vf_{t1}^*\\
\vdots&&\vdots\\
\vf_{1s}^*&\cdots&\vf_{ts}^*
\end{pmatrix}
,
\ee
where $\vf_{ji}^*:\mathcal{M}_{n_{j}}(\C)\stoch\mathcal{M}_{m_{i}}(\C)$
is the usual (Hilbert--Schmidt) adjoint of $\vf_{ji}.$%
%\footnote
\footnote{$\vf_{ji}^{*}$ will be the notation used for the dual of $\vf_{ji}$ as opposed to the more precise $(\vf_{ji})^{*}.$ It is the dual of the $ij$-th entry of $\vf^{*},$ which itself could be denoted by $(\vf^{*})_{ij}.$ Hence, $(\vf^{*})_{ij}=\vf_{ji}^{*}=(\vf_{ji})^{*}.$}
%end footnote
 In particular, if $\vf$ is a CP
map where $\vf_{ji}$ has Kraus decomposition as in 
(\ref{eq:krausdecompvfji}), then 
\be
\vf_{ji}^*=\sum_{l_{ji}=1}^{m_{i}n_{j}}\mathrm{Ad}_{V_{ji;l_{ji}}^{\dag}}.
\ee
Finally, $\vf$ is CPU
if and only if $\vf^*$ is CP and trace-preserving 
in the sense that 
\be
\<\vec{B},1_{\mathcal{B}}\>=\<\vf^*\big(\vec{B}\big),1_{\mA}\>
\qquad\forall\;\vec{B}\in\mathcal{B}, 
\ee
i.e.\
\be
\sum_{j=1}^{t}\tr(B_{j})=\sum_{i=1}^{s}\sum_{j=1}^{t}\tr\big(\vf_{ji}^*(B_{j})\big)
\ee
in terms of the components of $\vf^*$ and $\vec{B}.$ 
\elem

\blem
\label{lem:xionsum}
Let 
$\xi:\mB\stoch\C$ be a state with $\mB$ as in (\ref{eq:ABsum}). 
Then there exists unique non-negative real numbers 
$q_{1},\dots,q_{t}$ and (not necessarily unique) density matrices 
$\s_{1}~\in~\mathcal{M}_{n_{1}}(\C),\dots,$ $\s_{t}~\in~\mathcal{M}_{n_{t}}(\C)$
such that 
\be
\label{eq:sumofqs}
\sum_{j=1}^{t}q_{j}=1
\qquad\text{ and }\qquad
\xi(\vec{B})=\sum_{j=1}^{t}q_{j}\tr(\s_{j}B_{j})
\quad\text{ for all }\vec{B}\in\mathcal{B}. 
\ee
Furthermore, for every $j$ such that $q_{j}>0,$
the density matrix $\s_{j}$ is the unique one
satisfying these conditions. 
\elem

\bprf
Since $\xi$ is a state, it is CPU. The adjoint 
$\xi^*:\C\stoch\mathcal{M}_{n_{1}}(\C)\oplus\cdots\oplus\mathcal{M}_{n_{t}}(\C)$ of 
$\xi$ is CP and trace-preserving by Lemma~\ref{lem:vfadjointdirectsum}. 
Let 
%\be
$\vec{\varsigma}:=\xi^*(1)$,
%\ee
let $\varsigma_{j}\in\mathcal{M}_{n_{j}}(\C)$ 
denote the $j$-th component of $\vec{\varsigma},$ 
and set $q_{j}:=\tr(\varsigma_{j})$. 
By the trace-preserving condition of $\xi^*,$ 
the first equation in (\ref{eq:sumofqs}) holds. 
By the positivity of $\xi^*,$ each 
$\varsigma_{j}$ is a non-negative matrix. 
If $q_{j}>0$ set
\be
\sigma_{j}:=\frac{\varsigma_{j}}{q_{j}}. 
\ee
Otherwise, if $q_{j}=0$, then $\varsigma_{j}$ is the zero matrix. In this case,  let $\sigma_{j}$ be \emph{any} density matrix. 
The conclusions of this lemma follow from these assignments. 
\eprf

\begin{notation}
\label{eq:nullspaceforindices}
Due to Lemma~\ref{lem:xionsum}, a state $\xi$ as above might occasionally 
be denoted by $\xi\equiv\sum\limits_{j=1}^{t}q_{j}\tr(\s_{j}\;\cdot\;)$, 
where $\xi(\vec{B})$ is understood to be given as in (\ref{eq:sumofqs}). 
Furthermore, the subset
$N_{q}:=\big\{j\in\{1,\dots,t\}\;:\;q_{j}=0\big\}$
will occasionally be used. 
\end{notation}

\blem
\label{lem:xionsumsupport}
Using the same notation from Lemma~\ref{lem:xionsum}, 
the support $P_{\xi}$ of $\xi$ is given by the vector of matrices
whose $j$-th component is given by 
\be
(P_{\xi})_{j}
=
\begin{cases}
P_{\xi_{j}}&\mbox{ if }q_{j}>0\\
0&\mbox{ if }q_{j}=0
\end{cases}
,
\ee
where $P_{\xi_{j}}$ is the support of $\s_{j}$ on 
$\mathcal{M}_{n_{j}}(\C).$ 
\elem

\begin{notation}
Let $\mathcal{B}$ be as in (\ref{eq:ABsum}). 
For each $j,k\in\{1,\dots,t\},$ let 
\be
\label{eq:inclproj}
\i_{j}:\mathcal{M}_{n_{j}}(\C)\hookrightarrow\mathcal{B}
\quad\text{ and }\quad
\pi_{k}:\mathcal{B}\twoheadrightarrow\mathcal{M}_{n_{k}}(\C)
\ee
be the inclusion of the $j$-th factor and projection of the $k$-th factor, 
respectively. 
\end{notation}

\blem
\label{lem:RPkpikRFj}
Given $\mathcal{A},\mathcal{B},F,\xi,P_{\xi},(P_{\xi})_{k},\i_{j},$ and $\pi_{k}$
as in (\ref{eq:ABsum}), (\ref{eq:FBratteli}), (\ref{eq:inclproj}), 
and Lemma~\ref{lem:xionsumsupport}, 
a CPU map $R:\mathcal{A}\stoch\mathcal{B}$ 
satisfies $R\circ F\underset{\raisebox{.6ex}[0pt][0pt]{\scriptsize$\xi$}}{=}\id_{\mathcal{B}}$ if and only if 
\be
\label{eq:breakupRFkj}
\mathcal{R}_{(P_{\xi})_{k}}\circ\pi_{k}\circ R\circ F\circ\i_{j}
=
\begin{cases}
\mathcal{R}_{(P_{\xi})_{k}}&\mbox{ if $k=j$}\\
0&\mbox{ if $k\ne j$}\\
\end{cases}
\ee
for all $j,k\in\{1,\dots,t\}.$ Here, 
$\mathcal{R}_{(P_{\xi})_{k}}$ is the 
right-multiplication map defined by 
$\mathcal{R}_{(P_{\xi})_{k}}(B_{k}):=B_{k}(P_{\xi})_{k}$ for all
$B_{k}\in\mathcal{M}_{n_{k}}(\C).$ 
\elem

\bprf
The condition $R\circ F\underset{\raisebox{.6ex}[0pt][0pt]{\scriptsize$\xi$}}{=}\id_{\mathcal{B}}$ holds if and only if (cf.\ Lemma~\ref{lem:aeidentitysupport})
$
\mathcal{R}_{P_{\xi}}\circ R\circ F=\mathcal{R}_{P_{\xi}},
$
which holds if and only if 
$\mathcal{R}_{P_{\xi}}\circ R\circ F\circ\i_{j}=\mathcal{R}_{P_{\xi}}\circ\i_{j}$ 
for all $j\in\{1,\dots,t\}.$
Finally, this is equivalent to 
\be
\xy0;/r.22pc/:
(-24,8)*+{\pi_{k}\circ\mathcal{R}_{P_{\xi}}\circ R\circ F\circ\i_{j}}="1";
(24,8)*+{\pi_{k}\circ\mathcal{R}_{P_{\xi}}\circ\i_{j}}="2";
(-24,-8)*+{\mathcal{R}_{(P_{\xi})_{k}}\circ\pi_{k}\circ R\circ F\circ\i_{j}}="3";
(24,-8)*+{\mathcal{R}_{(P_{\xi})_{k}}\circ\pi_{k}\circ\i_{j}}="4";
{\ar@{=}@/^1.5pc/"1";"2"};
{\ar@{=}@/^0.6pc/"2";"4"};
{\ar@{=}@/^0.6pc/"3";"1"};
\endxy
\qquad\forall\;j,k\in\{1,\dots,t\},
\ee
which is equivalent to the claim (\ref{eq:breakupRFkj}) because 
\be
\pi_{k}\circ\i_{j}=
\begin{cases}
\id_{\mathcal{M}_{n_{k}}(\C)}&\mbox{ if $k=j$}\\
0&\mbox{ if $k\ne j$}\\
\end{cases}
.
\qedhere
\ee
\eprf

It may be helpful to visualize the map $\pi_{k}\circ R\circ F\circ\i_{j}\equiv(R\circ F)_{kj}$
as the following composite of CP (not necessarily unital) maps
\be
\xy0;/r.25pc/:
(-50,0)*+{\mathcal{M}_{m_{1}}(\C)\oplus\cdots\oplus\mathcal{M}_{m_{s}}(\C)}="A";
(0,10)*+{\mathcal{M}_{n_{1}}(\C)\oplus\cdots\oplus\mathcal{M}_{n_{t}}(\C)}="Btop";
(0,-10)*+{\mathcal{M}_{n_{1}}(\C)\oplus\cdots\oplus\mathcal{M}_{n_{t}}(\C)}="Bbot";
(50,10)*+{\mathcal{M}_{n_{j}}(\C)}="j";
(50,-10)*+{\mathcal{M}_{n_{k}}(\C)}="k";
(-47,10)*{}="top1";
(-50,7)*{}="top2";
(-47,-10)*{}="bot1";
(-50,-7)*{}="bot2";
{\ar@{_{(}->}"j";"Btop"_(0.35){\i_{j}}};
{\ar@{-}"Btop";"top1"};
{\ar@{-}@/_0.25pc/"top1";"top2"_{F}};
{\ar"top2";"A"};
{\ar@{~>}"bot1";"Bbot"};
{\ar@{~}@/_0.25pc/"bot2";"bot1"_{R}};
{\ar@{~}"A";"bot2"};
{\ar@{->>}"Bbot";"k"_(0.65){\pi_{k}}};
\endxy
\ee

The following is an analogue of Lemma~\ref{lem:CPUtoidentity}
to direct sums of matrix algebras. 

\blem
\label{cor:ChoiKrausIdentitydirectsum}
Using the same notation 
as in Lemma~\ref{lem:RPkpikRFj} and assuming $R\circ F\underset{\raisebox{.6ex}[0pt][0pt]{\scriptsize$\xi$}}{=}\id_{\mB},$ 
write
\be
R=\begin{pmatrix}
R_{11}&\cdots&R_{1s}\\
\vdots&&\vdots\\
R_{t1}&\cdots&R_{ts}
\end{pmatrix}
\quad\text{ and }\quad
F=
\begin{pmatrix}
F_{11}&\cdots&F_{1t}\\
\vdots&&\vdots\\
F_{s1}&\cdots&F_{st}
\end{pmatrix}
\ee
where the completely positive (not necessarily unital) maps 
$R_{ki}:\mathcal{M}_{m_{i}}(\C)\stoch\mathcal{M}_{n_{k}}(\C)$ 
and $F_{ij}:\mathcal{M}_{n_{j}}(\C)\to\mathcal{M}_{m_{i}}(\C)$
have Kraus decompositions,%
%footnote
\footnote{We will see in the text surrounding (\ref{eq:Fijgammaij})
that there exists a Kraus decomposition of 
$F_{ij}$ such that the index $\g_{ij}$ runs from $1$ to $c_{ij}$ instead of 
$m_{i}n_{j}=\sum\limits_{k=1}^{t}c_{ik}n_{k}n_{j}.$ 
This just means that $F_{ij;\g_{ij}}$ is zero 
when $\g_{ij}$ exceeds $c_{ij}.$}
%end footnote
\be
\label{eq:RFkrausdecomp}
R_{ki}=\sum_{\b_{ki}=1}^{n_{k}m_{i}}\mathrm{Ad}_{R_{ki;\b_{ki}}}
\quad\text{ and }\quad
F_{ij}=\sum_{\g_{ij}=1}^{m_{i}n_{j}}\mathrm{Ad}_{F_{ij;\g_{ij}}}
\ee
with $R_{ki;\b_{ki}}:\C^{m_{i}}\to\C^{n_{k}}$ and 
$F_{ij;\g_{ij}}:\C^{n_{j}}\to\C^{m_{i}}$ linear maps. 
\begin{enumerate}[i.]
\itemsep0pt
\item
For each $k\in\{1,\dots,t\}\setminus N_{q},$ 
there exist 
a collection of complex numbers
$\{\a_{k;i,\b_{ki},\g_{ik}}\},$ indexed by $\g_{ik}\in\{1,\dots,c_{ik}\},\b_{ki}\in\{1,\dots,n_{k}m_{i}\},i\in\{1,\dots,s\},$
such that 
\be
\label{eq:PRFdirectsum1}
P_{\xi_{k}}R_{ki;\b_{ki}}F_{ik;\g_{ik}}=\a_{k;i,\b_{ki},\g_{ik}}P_{\xi_{k}}
\ee
for all $\b_{ki}\in\{1,\dots,n_{k}m_{i}\},
\;\g_{ik}\in\{1,\dots,n_{k}m_{i}\},
\;i\in\{1,\dots,s\}$
and
\be
\label{eq:PRFdirectsum2}
\sum_{i=1}^{s}\sum_{\b_{ki}=1}^{n_{k}m_{i}}
\sum_{\g_{ik}=1}^{m_{i}n_{k}}|\a_{k;i,\b_{ki},\g_{ik}}|^2=1.
\ee
\item
For every pair $j\in\{1,\dots,t\}$ and $k\in\{1,\dots,t\}\setminus N_{q}$ 
with $j\ne k$,  
\be
P_{\xi_{k}}R_{ki,\b_{ki}}F_{ij,\g_{ij}}=0
\ee
for all $\b_{ki}\in\{1,\dots,n_{k}m_{i}\},
\;\g_{ij}\in\{1,\dots,n_{j}m_{i}\},
\;i\in\{1,\dots,s\}.$
\end{enumerate}
\elem

\bprf
Computing $\pi_{k}\circ R\circ F\circ\i_{j}$ for $j,k\in\{1,\dots,t\}$ gives
\be
\label{eq:KrausdecompRFkj}
\pi_{k}\circ R\circ F\circ\i_{j}
=\sum_{i=1}^{s}R_{ki}\circ F_{ij}
=\sum_{i=1}^{s}
\sum_{\b_{ki}=1}^{n_{k}m_{i}}
\sum_{\g_{ij}=1}^{m_{i}n_{j}}
\mathrm{Ad}_{R_{ki;\b_{ki}}F_{ij;\g_{ij}}}.
\ee
Suppose $q_{k}>0.$ Then, (\ref{eq:breakupRFkj}) becomes
\be
\label{eq:breakupRFkjcomponent}
\mathcal{R}_{P_{\xi_{k}}}\circ\pi_{k}\circ R\circ F\circ\i_{j}
=
\begin{cases}
\mathcal{R}_{P_{\xi_{k}}}&\mbox{ if }k=j\\
0&\mbox{ if }k\ne j
\end{cases}
.
\ee
\begin{enumerate}[i.]
\itemsep0pt
\item
In the case $k=j,$ (\ref{eq:breakupRFkjcomponent}) entails
\be
\mathrm{Ad}_{P_{\xi_{k}}}\circ\pi_{k}\circ R\circ F\circ\i_{k}=\mathrm{Ad}_{P_{\xi_{k}}}
\ee
upon multiplying by $P_{\xi_{k}}$ on the left. 
Combining this with (\ref{eq:KrausdecompRFkj}) and 
Lemma~\ref{lem:ChoiKraus} (by following a similar proof to that of Lemma~\ref{lem:CPUtoidentity}), 
there exist complex numbers
$\a_{k;i,\b_{ki},\g_{ik}}$ 
satisfying (\ref{eq:PRFdirectsum1}) and (\ref{eq:PRFdirectsum2}). 

\item
In the case $k\ne j,$ (\ref{eq:breakupRFkjcomponent}) becomes
\be
\mathrm{Ad}_{P_{\xi_{k}}}\circ\pi_{k}\circ R\circ F\circ\i_{j}=0
\ee
upon multiplying by $P_{\xi_{k}}$ on the left. 
This implies
\be
P_{\xi_{k}}R_{ki;\b_{ki}}F_{ij;\g_{ij}}=0
\ee
by (\ref{eq:KrausdecompRFkj}) and 
Lemma~\ref{lem:ChoiKraus}. 
\qedhere
\end{enumerate}
\eprf

At this point, it is helpful to make the conclusions of 
Lemma~\ref{cor:ChoiKrausIdentitydirectsum} even more explicit
by further explicating $P_{\xi_{k}}R_{ki;\b_{ki}}F_{ij;\g_{ij}}.$
The Kraus operator $R_{ki;\b_{ki}}:\C^{m_{i}}\to\C^{n_{k}}$ can be partitioned into
block sums of matrices based on the multiplicity of $F$ in the following way
\be
\label{eq:KrausdecompRkibetaki}
R_{ki;\b_{ki}}=
\big[
\overbrace{\underbrace{V_{ki;\b_{ki};11}}_{n_{k}\times n_{1}\text{ matrix}}\;\;\;\cdots\;\;\;
\underbrace{V_{ki;\b_{ki};1c_{i1}}}_{n_{k}\times n_{1}\text{ matrix}}}^{n_{k}\times (c_{i1}n_{1})\text{ matrix}}\;\;\;\cdots\;\;\;
\overbrace{\underbrace{V_{ki;\b_{ki};t1}}_{n_{k}\times n_{t}\text{ matrix}}\;\;\;\cdots\;\;\;\underbrace{V_{ki;\b_{ki};tc_{it}}}_{n_{k}\times n_{t}\text{ matrix}}}^{n_{k}\times (c_{it}n_{t})\text{ matrix}}
\big]
\ee
due to (\ref{eq:multiplicity}). 
Based on this partitioning, the unitality condition on $R$ reads
\be
\label{eq:unitalitydirectsumKrauspartitioned}
\mathds{1}_{n_{k}}=\sum_{i=1}^{s}\sum_{\b_{ki}=1}^{m_{i}n_{k}}
\sum_{j=1}^{t}\sum_{\g_{ij}=1}^{c_{ij}}
V_{ki;\b_{ki};j\g_{ij}}V_{ki;\b_{ki};j\g_{ij}}^{\dag}\qquad\forall\;k\in\{1,\dots,t\}.
\ee
due to (\ref{eq:unitalitydirectsumKraus}).
Furthermore, the definition of $F$ from (\ref{eq:FBratteli})
says
\be
F_{ij}(B_{j})=\mathrm{diag}(0,\dots,0,\dots,B_{j},\dots,B_{j},\dots,0,\dots,0)\qquad\forall\;B_{j}\in\mathcal{M}_{n_{j}}(\C).
\ee
This implies that the (adjoint of the) Kraus operators 
$F_{ij;\g_{ij}}:\C^{n_{j}}\to\C^{m_{i}}$ of 
$F_{ij}:\mathcal{M}_{n_{j}}(\C)\to\mathcal{M}_{m_{i}}(\C)$
have the following partitioned form
\be
\label{eq:Fijgammaij}
F_{ij;\g_{ij}}^{\dag}=
\big[
\overbrace{\underbrace{0}_{n_{j}\times n_{1}}\;\;\;\cdots\;\;\;
\underbrace{0}_{n_{j}\times n_{1}}}^{n_{j}\times (c_{i1}n_{1})}\;\;\;\cdots\;\;\;
\overbrace{\underbrace{0}_{n_{j}\times n_{j}}\;\;\;\cdots\;\;\;\mathds{1}_{n_{j}}\;\;\;\cdots\;\;\;\underbrace{0}_{n_{j}\times n_{j}}}^{n_{j}\times (c_{ij}n_{j})}
\;\;\;\cdots\;\;\;
\overbrace{\underbrace{0}_{n_{j}\times n_{t}}\;\;\;\cdots\;\;\;\underbrace{0}_{n_{j}\times n_{t}}}^{n_{j}\times (c_{it}n_{t})}
\big],
\ee
where the identity matrix $\mathds{1}_{n_{j}}$ is in the $\g_{ij}$-th $n_{j}\times n_{j}$ subblock
inside the $n_{j}\times(c_{ij}n_{j})$ block indicated (and all other entries are $0$). 
In particular, the index $\g_{ij}$ runs from $1$ to $c_{ij}$ (as opposed to $m_{i}n_{j}$). 
Therefore, the product $R_{ki;\b_{ki}}F_{ij;\g_{ij}}$ is
\be
\label{eq:RFisV}
R_{ki;\b_{ki}}F_{ij;\g_{ij}}=V_{ki;\b_{ki};j\g_{ij}},
\ee
which is an $n_{k}\times n_{j}$ matrix. The following result is a generalization of Equation~(\ref{eq:Choiuniquenessalphas}) to the direct sum case. 

\blem
\label{thm:formofdisintoncalg}
Under the same assumptions as in
Lemma~\ref{cor:ChoiKrausIdentitydirectsum}, 
for every ${k\in\{1,\dots,t\}\setminus N_{q}}$, there exist 
a collection of complex numbers $\{\a_{k;i,\b_{ki},\g_{ik}}\},$ indexed by $\g_{ik}\in\{1,\dots,c_{ik}\},$ $\b_{ki}\in\{1,\dots,n_{k}m_{i}\},$ and $i\in\{1,\dots,s\},$ 
such that 
\be
\label{eq:PRds}
R_{ki;\b_{ki}}=
\big[
\overbrace{\underbrace{0}_{n_{k}\times n_{1}}\;\;\;\cdots\;\;\;
\underbrace{0}_{n_{k}\times n_{1}}}^{n_{k}\times (c_{i1}n_{1})}\;\;\;\cdots\;\;\;
\overbrace{\underbrace{\a_{k;i,\b_{ki},1}\mathds{1}_{n_{k}}}_{n_{k}\times n_{k}}\;\;\;\cdots\;\;\;\underbrace{\a_{k;i,\b_{ki},c_{ik}}\mathds{1}_{n_{k}}}_{n_{k}\times n_{k}}}^{n_{k}\times (c_{ik}n_{k})}
\;\;\;\cdots\;\;\;
\overbrace{\underbrace{0}_{n_{k}\times n_{t}}\;\;\;\cdots\;\;\;\underbrace{0}_{n_{k}\times n_{t}}}^{n_{k}\times (c_{it}n_{t})}
\big],
\ee
for all $\b_{ki}\in\{1,\dots,n_{k}m_{i}\},i\in\{1,\dots,s\}$ 
and %such that 
\be
\label{eq:sumalphaisidentity}
\sum_{i=1}^{s}\sum_{\b_{ki}=1}^{n_{k}m_{i}}
\sum_{\g_{ik}=1}^{c_{ik}}|\a_{k;i,\b_{ki},\g_{ik}}|^2=1.
\ee
\elem

\bprf
In analogy to the proof of Theorem~\ref{thm:CPUtoidentity}, 
for every $k\in\{1,\dots,t\}\setminus N_{q}$, 
let $\mathcal{E}_{k}$ 
be the pre-Hilbert $\mathcal{M}_{n_{k}}(\C)$-module 
consisting of vectors of $n_{k}\times n_{k}$ matrices whose vector 
components are labelled by 
the triple of indices $\g_{ik},\b_{ki},i.$
Let $\vec{V}_{k}$ be the vector whose vector components are the $n_{k}\times n_{k}$ matrices $V_{ki;\b_{ki};k\g_{ik}}.$
The first case of 
Lemma~\ref{cor:ChoiKrausIdentitydirectsum} implies 
there exists a vector $\vec{\a}_{k}\in\mathcal{E}_{k}$ whose vector
components are 
constant multiples of the identity matrix satisfying
\be
\vec{V}_{k}=P_{\xi_{k}}\vec{\a}_{k}+\vec{V}_{k}^{\mathrm{bl}}+\vec{V}_{k}^{\mathrm{br}},
\ee
where
\be
\vec{V}_{k}^{\mathrm{bl}}:=P_{\xi_{k}}^{\perp}\vec{V}_{k}P_{\xi_{k}},
\qquad
%\quad\text{and}\quad
\vec{V}_{k}^{\mathrm{br}}:=P_{\xi_{k}}^{\perp}\vec{V}_{k}P_{\xi_{k}}^{\perp},
\qquad\text{and}\qquad
%\ee
%and
%\be
\<\!\<\vec{\a}_{k},\vec{\a}_{k}\>\!\>=\mathds{1}_{n_{k}}.
\ee
Similarly, for every pair $(k,j)\in\{1,\dots,t\}\times\{1,\dots,t\}$ such that $j\ne k$
and such that $q_{k}>0,$ let $\mathcal{E}_{kj}$
be the pre-Hilbert $\mathcal{M}_{n_{k}}(\C)$-module
consisting of vectors of $n_{k}\times n_{j}$ matrices 
whose vector components are labelled by 
the triple of indices $\g_{ij},\b_{ki},i.$ 
Let $\vec{V}_{kj}$ be the vector of the $n_{k}\times n_{j}$ matrices
whose components are given by $V_{ki;\b_{ki};j\g_{ij}}.$
The second case of 
Lemma~\ref{cor:ChoiKrausIdentitydirectsum} implies
$
\vec{V}_{kj}=\vec{V}_{kj}^{\mathrm{bl}}+\vec{V}_{kj}^{\mathrm{br}},
$
where 
\be
\vec{V}_{kj}^{\mathrm{bl}}:=P_{\xi_{k}}^{\perp}\vec{V}_{kj}P_{\xi_{j}}
\qquad\text{and}\qquad
\vec{V}_{kj}^{\mathrm{br}}:=P_{\xi_{k}}^{\perp}\vec{V}_{kj}P_{\xi_{k}}^{\perp}.
\ee
The equalities 
\be
\label{eq:Vbasicidentitiesdirectsum}
\<\!\<P_{\xi_{k}}\vec{\a}_{k},\vec{V}_{k}^{\mathrm{br}}\>\!\>=
\<\!\<\vec{V}_{k}^{\mathrm{bl}},\vec{V}_{k}^{\mathrm{br}}\>\!\>=
\<\!\<\vec{V}_{kj}^{\mathrm{bl}},\vec{V}_{kj}^{\mathrm{br}}\>\!\>=0
\ee
all follow immediately from the definitions. 
Unitality of $R$ takes on the form 
\be
\label{eq:unitalitydscase}
\mathds{1}_{n_{k}}=\<\!\<\vec{V}_{k},\vec{V}_{k}\>\!\>+%\sum_{\substack{j=1\\j\ne k}}^{t}\<\!\<\vec{V}_{kj},\vec{V}_{kj}\>\!\>
\sum_{j=1, j\ne k}^{t}\<\!\<\vec{V}_{kj},\vec{V}_{kj}\>\!\>
\ee
by (\ref{eq:unitalitydirectsumKrauspartitioned}). 
By expanding out (\ref{eq:unitalitydscase}) and multiplying
on the right by $P_{\xi_{k}},$ 
completely similar arguments to those in the proof of Theorem~\ref{thm:CPUtoidentity}, specifically the discussion surrounding
Equations~(\ref{eq:iunitalityfactorcase}) through  (\ref{eq:Viblalphaiszero}),
prove 
$
\<\!\<\vec{V}_{k}^{\mathrm{bl}},\vec{\a}_{k}\>\!\>={0}.
$
Hence, the unitality condition (\ref{eq:unitalitydscase}) simplifies to 
\be
\label{eq:unitalitydscaseblbr}
P_{\xi_{k}}^{\perp}=\<\!\<\vec{V}_{k}^{\mathrm{bl}},\vec{V}_{k}^{\mathrm{bl}}\>\!\>+\<\!\<\vec{V}_{k}^{\mathrm{br}},\vec{V}_{k}^{\mathrm{br}}\>\!\>+
%\sum_{\substack{j=1\\j\ne k}}^{t}
\sum_{j=1,j\ne k}^{t}
\left(\<\!\<\vec{V}_{kj}^{\mathrm{bl}},\vec{V}_{kj}^{\mathrm{bl}}\>\!\>+\<\!\<\vec{V}_{kj}^{\mathrm{br}},\vec{V}_{kj}^{\mathrm{br}}\>\!\>\right)
\ee
analogously to (\ref{eq:iunitalityfactorcase}). 
Now, computing $\pi_{k}\circ R\circ F\circ\i_{j}$ in terms of the
pre-Hilbert module inner product gives 
\be
(\pi_{k}\circ R\circ F\circ\i_{j})(A_{j})=\<\!\<\vec{V}_{kj}A_{j},\vec{V}_{kj}\>\!\>
\qquad\forall\;A_{j}\in\mathcal{M}_{n_{j}}(\C)
\ee
for all $k,j$ (when $j=k,$ remove one of the indices from $\vec{V}_{kk}$)
by (\ref{eq:RFisV}). When $j=k,$ multiplying this equation on the right by $P_{\xi_{k}}$ (which equals $(P_{\xi})_{k}$ since $q_{k}>0$) and combining this with 
Lemma~\ref{lem:RPkpikRFj} gives 
$
\<\!\<\vec{V}_{k}^{\mathrm{br}},\vec{\a}_{k}\>\!\>=P_{\xi_{k}}^{\perp}
$
by following an argument exactly analogous to (\ref{eq:idisintexplicit}) and
the text surrounding this equation. 
Similarly, combining this
result with the Paschke--Cauchy--Schwarz
inequality gives
$
P_{\xi_{k}}^{\perp}\le\<\!\<\vec{V}_{k}^{\mathrm{br}},\vec{V}_{k}^{\mathrm{br}}\>\!\>.
$
On the other hand, (\ref{eq:unitalitydscaseblbr}) says 
$
P_{\xi_{k}}^{\perp}\ge\<\!\<\vec{V}_{k}^{\mathrm{br}},\vec{V}_{k}^{\mathrm{br}}\>\!\>.
$
Therefore, following analogous lines of thought to those 
from (\ref{eq:iVbrVbrVblVbl})
to (\ref{eq:VbrisPperpalpha})
gives
\be
\vec{V}_{k}^{\mathrm{br}}=P_{\xi_{k}}^{\perp}\vec{\a}_{k},
\qquad
\vec{V}_{k}^{\mathrm{bl}}=\vec{0},
\qquad
\vec{V}_{kj}^{\mathrm{bl}}=\vec{0},
\quad\text{and}\quad
\vec{V}_{kj}^{\mathrm{br}}=\vec{0}.
\ee
Therefore, 
$\vec{V}_{k}=\vec{\a}_{k}$ and $\vec{V}_{kj}=\vec{0}$.
Expanding out the vector entries coming from the definitions of 
$\mathcal{E}_{k}$ and $\mathcal{E}_{kj}$ completes the proof.
\eprf

Given a state-preserving $^{*}$-homomorphism $(\mB,\xi)\xrightarrow{F}(\mA,\omega)$, it may be useful to know
how the density matrices associated to $\xi$ and $\w$ 
are related. The following fact describes this relationship.
It is a generalization of the ``tracing out degrees of freedom'' method
in quantum theory. 

\bn
\label{prop:Fpreservesstatesds}
Let $\mathcal{A},\mathcal{B},F,\w,$ and $\xi$ be as in 
Notation~\ref{not:dsumsection}, and let 
\be
\label{eq:omegaxidecomp}
\w\equiv\sum_{i=1}^{s}p_{i}\tr(\rho_{i}\;\cdot\;)
\quad\text{ and }\quad
\xi\equiv\sum_{j=1}^{t}q_{j}\tr(\s_{j}\;\cdot\;)
\ee
be decompositions of the states $\w$ and $\xi$ as described in 
Lemma~\ref{lem:xionsum}. Then the following facts hold. 
\begin{enumerate}[i.]
\itemsep0pt
\item
\label{item:Fpreservesstatesdsi}
For each $i\in\{1,\dots,s\},$ there exists a $j\in\{1,\dots,t\}$ 
such that $c_{ij}>0.$ 
\item
\label{propitem:injectiveF}
If there exists a $j\in\{1,\dots,s\}$ such that $c_{ij}=0$ for all $i\in\{1,\dots,s\},$
then $q_{j}=0.$ 
\item
Finally, 
\be
\label{eq:expressingsigmajintermsofrho}
q_{j}\sigma_{j}=\sum_{i=1}^{s}\sum_{\g_{ij}=1}^{c_{ij}}p_{i}\rho_{i;jj;\g_{ij}\g_{ij}}
\qquad\forall\;j\in\{1,\dots,t\},
\ee
where $\rho_{i;jj;\g_{ij}\g_{ij}}$ is the $n_{j}\times n_{j}$ matrix obtained from $\rho_{i}$ in the following way. 
Since $m_{i}=\sum\limits_{k=1}^{t}c_{ik}n_{k},$ each 
$m_{i}\times m_{i}$ matrix $\rho_{i}$ has
a block matrix decomposition 
\be
\label{eq:blockdecomposerhoi}
\rho_{i}=\begin{bmatrix}\rho_{i;11}&\cdots&\rho_{i;1t}\\
\vdots&&\vdots\\
\rho_{i;t1}&\cdots&\rho_{i;tt}\end{bmatrix}
,
\ee
where $\rho_{i;jk}$ is a $(c_{ij}n_{j})\times(c_{ik}n_{k})$ matrix.
This matrix further breaks up into subblocks 
\be
\rho_{i;jk}=
\begin{bmatrix}\rho_{i;jk;11}&\cdots&\rho_{i;jk;1c_{ik}}\\
\vdots&&\vdots\\
\rho_{i;jk;c_{ij}1}&\cdots&\rho_{i;jk;c_{ij}c_{ik}}\end{bmatrix}
,
\ee
where $\rho_{i;jk;\g_{ij}\g_{ik}}$ is an $n_{j}\times n_{k}$ matrix. 
\end{enumerate}
\en

\br
\label{rmk:injectivityae}
The contrapositive of part \ref{propitem:injectiveF} of Proposition~\ref{prop:Fpreservesstatesds}
will be used occasionally in certain technical points later. It states
that if $q_{j}>0,$ there exists at least one $i\in\{1,\dots,s\}$ such that
$c_{ij}>0.$ 
In other words, $F$ is injective almost everywhere. This should be compared to Lemma~\ref{lem:aesurjective}. 
Furthermore, using partial traces, 
Equation~(\ref{eq:expressingsigmajintermsofrho}) becomes
\be
q_{j}\sigma_{j}=\sum_{i=1}^{s}p_{i}\tr_{\mathcal{M}_{c_{ij}}(\C)}(\rho_{i;jj}),
\ee
where $\rho_{i}$ is decomposed as in (\ref{eq:blockdecomposerhoi}). 
\er

\bprf[Proof of Proposition~\ref{prop:Fpreservesstatesds}]
{\color{white}{you found me!}}
\begin{enumerate}[i.]
\itemsep0pt
\item
Since $m_{i}>0$ and $m_{i}=\sum\limits_{j=1}^{t}c_{ij}n_{j},$ 
there must exist a non-zero $c_{ij}$ for some $j\in\{1,\dots,t\}.$ 
\item
Suppose there exists a $j\in\{1,\dots,s\}$ such that $c_{ij}=0$ for all $i\in\{1,\dots,s\}.$
Then $F_{ij}(\mathds{1}_{n_{j}})=0$ for all $i\in\{1,\dots,s\}.$
Since $\w\circ F=\xi,$ this shows $\xi(\mathds{1}_{n_{j}})=0.$
But $\xi(\mathds{1}_{n_{j}})=q_{j}\tr(\s_{j})=q_{j}$ so that $q_{j}=0.$ 
\item
This follows from taking the adjoint of the equation $\w\circ F=\xi,$
which gives
\be
F^{*}\circ\w^{*}=\xi^*
\implies
F^{*}\big(\w^{*}(1)\big)=\xi^*(1).
\ee
Expanding out these expressions and extracting the $j$-th term gives
\be
\label{eq:qjsigmaj}
q_{j}\s_{j}=\sum_{i=1}^{s}p_{i}F_{ij}^{*}(\rho_{i})\qquad\forall\;j\in\{1,\dots,t\}.
\ee
Applying (\ref{eq:RFkrausdecomp}) and (\ref{eq:Fijgammaij}) gives the desired result. \qedhere
\end{enumerate}
\eprf

A consequence of Lemma~\ref{thm:formofdisintoncalg} is the following
fact regarding the existence and 
uniqueness of disintegrations on finite-dimensional
$C^*$-algebras.
It is a generalization of Theorem~\ref{thm:diagonalimpliesseparable} 
to direct sums of matrix algebras and is the main theorem of the present work. 

\bt
\label{thm:theoremaegregium}
Let $\mathcal{A},\mathcal{B},F,\w,$ and $\xi$ be as in 
Notation~\ref{not:dsumsection} and Proposition~\ref{prop:Fpreservesstatesds}. 
\begin{enumerate}[i.]
\itemsep0pt
\item
A disintegration $R$ of $\w$ over $\xi$ consistent with $F$ 
exists if and only if 
for each $i\in\{1,\dots,s\}$ and $j\in\{1,\dots,t\}$ there exist
non-negative matrices $\t_{ji}\in\mathcal{M}_{c_{ij}}(\C)$ such that%
%footnote
\footnote{${N}_{q}$ was introduced in 
Notation~\ref{eq:nullspaceforindices}.}
\be
\label{eq:taujimatrices}
\tr\left(\sum_{i=1}^{s}\t_{ji}\right)=1\qquad\forall\;j\in\{1,\dots,t\}\setminus{N}_{q}
\ee
and
\be
\label{eq:separabilitydisrectsumcase}
p_{i}\rho_{i}=
\mathrm{diag}(q_{1}\t_{1i}\otimes\s_{1},\dots,q_{t}\t_{ti}\otimes\s_{t})
\qquad\forall\;i\in\{1,\dots,s\}.
\ee
\item
Furthermore, if $R'$ is another disintegration 
of $\w$ over $\xi$ consistent with $F,$ then 
$R'\underset{\raisebox{.6ex}[0pt][0pt]{\scriptsize$\xi$}}{=}R$
and 
\be
\label{eq:uniquenessdisintds}
R'_{ji}=R_{ji}\qquad\forall\;i\in\{1,\dots,s\}
\qquad\forall\;j\in\{1,\dots,t\}\setminus{N}_{q}.
\ee
\item
Finally, if such a disintegration $R$ exists, a formula for the $ji$-th component of the disintegration is given by 
\be
\label{eq:disintformuladsumcase}
R_{ji}(A_{i})=\tr_{\mathcal{M}_{c_{ij}}(\C)}\big((\t_{ji}\otimes\mathds{1}_{n_{j}})A_{i;jj}\big)
\ee
for all $j\in\{1,\dots,t\}\setminus N_{q}$ and for all $i\in\{1,\dots,s\}.$ Here, $A_{i;jj}$ is uniquely defined by the decomposition 
as a 
$t\times t$ matrix 
\be
\label{eq:breakupAi}
A_{i}\equiv
\begin{bmatrix}
A_{i;11}&\cdots&A_{i;1t}\\
\vdots&&\vdots\\
A_{i;t1}&\cdots&A_{i;tt}\\
\end{bmatrix}
\ee
where the $kl$-th subblock, $A_{i;kl},$ is a $(c_{ik}n_{k})\times(c_{il}n_{l})$ matrix.
\end{enumerate}
\et

\bprf
Proving the first item will provide proofs of the subsequent claims. 

\noindent
($\Rightarrow$)
Suppose a disintegration $R$ exists. The condition $\xi\circ R=\w$ is equivalent to $R^*\big(\xi^*(1)\big)=\w^*(1)$
by Lemma~\ref{lem:vfadjointdirectsum}. Hence, using the notation 
from (\ref{eq:omegaxidecomp}), this equation gives
\be
R^*\big(\xi^*(1)\big)\equiv
\begin{pmatrix}
R_{11}^*&\cdots&R_{t1}^*\\
\vdots&&\vdots\\
R_{1s}^*&\cdots&R_{ts}^*
\end{pmatrix}
\begin{pmatrix}
q_{1}\s_{1}\\
\vdots\\
q_{t}\s_{t}\\
\end{pmatrix}
=
\begin{pmatrix}
p_{1}\rho_{1}\\
\vdots\\
p_{s}\rho_{s}\\
\end{pmatrix}
\equiv
\w^*(1)
,
\ee
which is equivalent to 
\be
\label{eq:rhoi}
p_{i}\rho_{i}=\sum_{j=1}^{t}q_{j}R_{ji}^*(\s_{j})\qquad\forall\;i\in\{1,\dots,s\}.
\ee
To compute $R_{ji}^*(\s_{j}),$ 
we can follow an analogous computation
to that from (\ref{eq:constructingtau}). First, when $q_{j}>0,$ we obtain
\be
\begin{split}
R_{ji}^*(\s_{j})&=
\sum_{\b_{ji}=1}^{n_{j}m_{i}}\mathrm{Ad}_{R_{ji;\b_{ji}}^{\dag}}(\s_{j})\\
&\overset{\text{(\ref{eq:PRds})}}{=\joinrel=\joinrel=\joinrel=}
\sum_{\b_{ji}=1}^{n_{j}m_{i}}
\begin{bmatrix}0\\
\overline{\a_{j;i,\b_{ji},1}}\mathds{1}_{n_{j}}\\
\vdots\\
\overline{\a_{j;i,\b_{ji},c_{ij}}}\mathds{1}_{n_{j}}\\
0\end{bmatrix}
\s_{j}
\begin{bmatrix}0&
{\a_{j;i,\b_{ji},1}}\mathds{1}_{n_{j}}&
\cdots&
{\a_{j;i,\b_{ji},c_{ij}}}\mathds{1}_{n_{j}}&
0\end{bmatrix},
\end{split}
\ee
where the top $0$ block in the left matrix is a
$\left(\sum\limits_{k=1}^{j-1}c_{ik}n_{k}\right)\times n_{j}$ matrix
and the bottom $0$ block in the left matrix is a
$\bigg(\sum\limits_{k=j+1}^{t}c_{ik}n_{k}\bigg)\times n_{j}$ matrix. 
Keeping track of these sizes, we obtain 
\be
\label{eq:Rjipullsbackstate}
R_{ji}^*(\s_{j})=\sum_{\b_{ji}=1}^{n_{j}m_{i}}
\begin{bmatrix}
0&0&\cdots&0&0\\
0&|\a_{j;i,\b_{ji},1}|^2\s_{j}&\cdots&\overline{\a_{j;i,\b_{ji},1}}\a_{j;i,\b_{ji},c_{ij}}\s_{j}&0\\
\vdots&\vdots&&\vdots&\vdots\\
0&\overline{\a_{j;i,\b_{ji},c_{ij}}}\a_{j;i,\b_{ji},1}\s_{j}&\cdots&|\a_{j;i,\b_{ji},c_{ij}}|^2\s_{j}&0\\
0&0&\cdots&0&0\\
\end{bmatrix}
,
\ee
where the top-left $0$ matrix is a $\left(\sum\limits_{k=1}^{j-1}c_{ik}n_{k}\right)\times\left(\sum\limits_{k=1}^{j-1}c_{ik}n_{k}\right)$ matrix and 
the bottom-right $0$ matrix is a $\bigg(\sum\limits_{k=j+1}^{t}c_{ik}n_{k}\bigg)\times\bigg(\sum\limits_{k=j+1}^{t}c_{ik}n_{k}\bigg)$ matrix.
Define the $c_{ij}\times c_{ij}$ matrix $\t_{ji}$ to be 
\be
\t_{ji}:=\sum_{\b_{ji}=1}^{n_{j}m_{i}}
\begin{bmatrix}
|\a_{j;i,\b_{ji},1}|^2&\cdots&\overline{\a_{j;i,\b_{ji},1}}\a_{j;i,\b_{ji},c_{ij}}\\
\vdots&&\vdots\\
\overline{\a_{j;i,\b_{ji},c_{ij}}}\a_{j;i,\b_{ji},1}&\cdots&|\a_{j;i,\b_{ji},c_{ij}}|^2
\end{bmatrix}
\ee
so that the $\g_{ij}\h_{ij}$-th entry of $\t_{ji}$ is given by 
\be
\label{eq:entriesoftau}
\t_{ji;\g_{ij}\h_{ij}}=\sum_{\b_{ji}=1}^{n_{j}m_{i}}\overline{\a_{j;i;\b_{ji},\g_{ij}}}\a_{j;i;\b_{ji},\h_{ij}}.
\ee
Notice that $\t_{ji}$ is defined only when $c_{ij}>0$ and when $q_{j}>0.$
Furthermore, when it is defined, $\t_{ji}$ is a non-negative matrix and 
\be
\label{eq:tjisumidensity}
\sum_{i=1}^{s}\tr(\t_{ji})
=\sum_{i=1}^{s}\sum_{\b_{ji}=1}^{n_{j}m_{i}}\sum_{\g_{ij}=1}^{c_{ij}}|\a_{j;i,\b_{ji},\g_{ij}}|^2
\overset{\text{(\ref{eq:sumalphaisidentity})}}{=\joinrel=\joinrel=\joinrel=}1,
\ee
which shows $\sum\limits_{i=1}^{s}\t_{ji}$ is a density matrix 
(again, when $q_{j}>0$). 
The sum in (\ref{eq:tjisumidensity}) is guaranteed to have at least one term 
due to 
Remark~\ref{rmk:injectivityae}. 
Second, when $q_{j}=0,$ then $q_{j}R_{ji}^*(\s_{j})=0$ so that this term does not
contribute to the sum in (\ref{eq:rhoi}). 
Therefore, in this case,
$\t_{ji}$ can be chosen to be an arbitrary non-negative matrix
provided that $c_{ij}>0.$ If $c_{ij}=0,$ then $\t_{ji}$ does not exist
and any expression involving such a $\t_{ji}$ should be excluded. 
Then, 
\be
p_{i}\rho_{i}=\sum_{j=1}^{t}q_{j}R_{ji}^*(\s_{j})
\overset{\text{(\ref{eq:Rjipullsbackstate})}}{=\joinrel=\joinrel=\joinrel=}\begin{bmatrix}
q_{1}\t_{1i}\otimes\s_{1}&&0\\
&\ddots&\\
0&&q_{t}\t_{ti}\otimes\s_{t}
\end{bmatrix}
\qquad\forall\;i\in\{1,\dots,s\}.
\ee
Note that this sum after the first equality is not empty by part~\ref{item:Fpreservesstatesdsi} of Proposition~\ref{prop:Fpreservesstatesds}.

\noindent
($\Leftarrow$)
For the converse, suppose the non-negative matrices
$\t_{ji}\in\mathcal{M}_{c_{ij}}(\C)$ 
satisfying (\ref{eq:taujimatrices}) and 
(\ref{eq:separabilitydisrectsumcase}) exist. 
Denote the $\g_{ij}\h_{ij}$-th entry of $\t_{ji}$ by $\t_{ji;\g_{ij}\h_{ij}}.$ 
For each pair of indices $i\in\{1,\dots,s\}$ and $j\in\{1,\dots,t\},$ 
define $R'_{ji}:\mathcal{M}_{m_{i}}(\C)\stoch\mathcal{M}_{n_{j}}(\C)$ in the 
following way. Write an element $A_{i}\in\mathcal{M}_{m_{i}}(\C)$ as in~(\ref{eq:breakupAi}). 
Then, write $A_{i;kl}$ as a $c_{ik}\times c_{il}$ matrix consisting of 
$n_{k}\times n_{l}$ matrices
indexed as in $A_{i;kl;\g_{ik}\h_{il}}.$ Set%
%footnote
\footnote{Note that the swapping of the $\g_{ij}$ and $\h_{ij}$ indices
in Equation~(\ref{eq:canonicaldisintegerationds}) is not a typo. In addition, note that if $q_{j}>0$ and $c_{ij}=0,$ then the sum in the top case is empty and gives, by definition of an empty sum, $0.$}
%end footnote
\be
\label{eq:canonicaldisintegerationds}
R_{ji}'(A_{i}):=
\begin{cases}
\sum\limits_{\g_{ij},\h_{ij}=1}^{c_{ij}}\t_{ji;\h_{ij}\g_{ij}}A_{i;jj;\g_{ij}\h_{ij}}&
\mbox{ if $q_{j}>0$}\\
\frac{1}{sm_{i}}\tr(A_{i})\mathds{1}_{n_{j}}&
\mbox{ if $q_{j}=0$ }\\
\end{cases}
.
\ee
A direct calculation shows that this formula equals~(\ref{eq:disintformuladsumcase}) when $j\in \{1,\dots,t\}\setminus N_{q}$. 
Set $R':\mathcal{A}\stoch\mathcal{B}$ to be the $t\times s$ matrix of linear maps whose
$ji$-th entry is $R_{ji}'$ from (\ref{eq:canonicaldisintegerationds}). Then $R'$ is a disintegration of $\w$ over $\xi$ consistent with $F.$ 
The proof of this is similar to the proof of Theorem~\ref{thm:diagonalimpliesseparable}
though one must keep track of indices more carefully. 
Unitality of $R'$ follows from
\be
\sum_{i=1}^{s}R'_{ji}(\mathds{1}_{m_{i}})
=\sum_{i=1}^{s}\sum_{\g_{ij},\h_{ij}=1}^{c_{ij}}\t_{ji;\h_{ji}\g_{ij}}\mathds{1}_{m_{i};jj;\g_{ij}\h_{ij}}
=\sum_{i=1}^{s}\sum_{\g_{ij}=1}^{c_{ij}}\t_{ji;\g_{ji}\g_{ij}}\mathds{1}_{n_{j}}
=\sum_{i=1}^{s}\tr(\t_{ji})\mathds{1}_{n_{j}}
=\mathds{1}_{n_{j}}
\ee
whenever $q_{j}>0$ 
because $\sum\limits_{i=1}^{s}\tr(\t_{ji})=1.$ 
When $q_{j}=0,$ one obtains 
\be
\sum_{i=1}^{s}R'_{ji}(\mathds{1}_{m_{i}})
=\sum_{i=1}^{s}\frac{1}{sm_{i}}\tr(\mathds{1}_{m_{i}})\mathds{1}_{n_{j}}
=\mathds{1}_{n_{j}}.
\ee
We will now show 
$\pi_{j}\circ R'\circ F=\pi_{j}$ for all $j$ satisfying $q_{j}>0.$
First, note that
\be
(R'\circ F)(\vec{B})
=
\begin{pmatrix}
R'_{11}&\cdots&R'_{1s}\\
\vdots&&\vdots\\
R'_{t1}&\cdots&R'_{ts}
\end{pmatrix}
\begin{pmatrix}\mathrm{diag}(\overbrace{B_{1},\dots,B_{1}}^{c_{11}\text{ times}},\dots,\overbrace{B_{t},\dots,B_{t}}^{c_{1t}\text{ times}})\\
\vdots\\
\mathrm{diag}(\underbrace{B_{1},\dots,B_{1}}_{c_{s1}\text{ times}},\dots,\underbrace{B_{t},\dots,B_{t}}_{c_{st}\text{ times}})
\end{pmatrix}
.
\ee
Focusing on the $j$-th term when $q_{j}>0$, one obtains
\be
\sum_{i=1}^{s}R'_{ji}\Big(\mathrm{diag}(\underbrace{B_{1},\dots,B_{1}}_{c_{i1}\text{ times}},\dots,\underbrace{B_{t},\dots,B_{t}}_{c_{it}\text{ times}})\Big)
=
\sum_{i=1}^{s}\sum_{\g_{ij}=1}^{c_{ij}}\t_{ji;\g_{ji}\g_{ij}}B_{j}
=\sum_{i=1}^{s}\tr(\t_{ji})B_{j}
=B_{j}.
\ee
Although the equality $\pi_{j}\circ R'\circ F=\pi_{j}$ fails when $q_{j}=0,$ the equation $R'\circ F\underset{\raisebox{.6ex}[0pt][0pt]{\scriptsize$\xi$}}{=}\id_{\mB}$
still holds.
Furthermore, $R'$ is state-preserving because
\be
\begin{split}
\w\big(\vec{A}\big)&=\sum_{i=1}^{s}\tr(p_{i}\rho_{i}A_{i})
\overset{\text{(\ref{eq:separabilitydisrectsumcase})}}{=\joinrel=\joinrel=\joinrel=}
\sum_{i=1}^{s}\sum_{j=1}^{t}q_{j}\tr\big((\t_{ji}\otimes\s_{j})A_{i;jj}\big)\\
&=\sum_{i=1}^{s}\sum_{\substack{j=1\\q_{j}>0}}^{t}q_{j}\tr\left(\sum_{\g_{ij}=1,\h_{ij}=1}^{c_{ij}}\t_{ji;\h_{ij}\g_{ij}}\s_{j}A_{i;jj;\g_{ij}\h_{ij}}\right)\\
&\overset{\text{(\ref{eq:canonicaldisintegerationds})}}{=\joinrel=\joinrel=\joinrel=}\sum_{i=1}^{s}\sum_{\substack{j=1\\q_{j}>0}}^{t}q_{j}\tr\big(\s_{j}R'_{ji}(A_{i})\big)=(\xi\circ R')\big(\vec{A}\big)
\end{split}
\ee
for all $\vec{A}\in\mA.$ To show $R'$ is CP, 
it suffices to show each $R'_{ji}$ is CP by Lemma~\ref{lem:CPondirectsums}.  
This follows from the equality between the formulas~(\ref{eq:disintformuladsumcase}) and~(\ref{eq:canonicaldisintegerationds}) when $j\in \{1,\dots,t\}\setminus N_{q}$. The case when $j\in N_{q}$ gives a CP map as well since the trace in~(\ref{eq:canonicaldisintegerationds}) is a CP map. 

Finally, we prove the uniqueness condition (\ref{eq:uniquenessdisintds}) for disintegrations. 
The condition $R'\underset{\raisebox{.6ex}[0pt][0pt]{\scriptsize$\xi$}}{=}R$
is equivalent to  
\be
\sum_{i=1}^{s}R'_{ji}(A_{i})(P_{\xi})_{j}=\sum_{i=1}^{s}R_{ji}(A_{i})(P_{\xi})_{j}
\qquad\forall\;\vec{A}\in%\mathcal{M}_{m_{i}}(\C),\;
\mA,\;
%\text{ and }\forall\;
j\in\{1,\dots,t\}.%
\ee
When $q_{j}=0,$ this equality holds trivially because 
$(P_{\xi})_{j}=0.$ 
When $q_{j}\ne0,$ Lemma~\ref{thm:formofdisintoncalg} 
guarantees the existence of complex
numbers $\{\a_{j;i,\b_{ji},\g_{ij}}\}$ and Kraus operators
$\{R_{ji;\b_{ji}}\}$ for $R_{ji}$ 
satisfying the conditions in the statement of that theorem. 
Therefore, by carefully working out the matrix operations,
one obtains
\be
\begin{split}
R_{ji}(A_{i})&\overset{\text{Lemma~\ref{thm:formofdisintoncalg}}}{=\joinrel=\joinrel=\joinrel=\joinrel=\joinrel=\joinrel=\joinrel=}\sum_{\b_{ji}=1}^{n_{j}m_{i}}\sum_{\g_{ij}=1}^{c_{ij}}\sum_{\h_{ij}=1}^{c_{ij}}\a_{j;i;\b_{ji},\g_{ij}}\overline{\a_{j;i;\b_{ji},\h_{ij}}}A_{i;jj;\g_{ij}\h_{ij}}
\\
&\overset{\text{(\ref{eq:entriesoftau})}}{=\joinrel=\joinrel=\joinrel=}
\sum_{\g_{ij}=1}^{c_{ij}}\sum_{\h_{ij}=1}^{c_{ij}}
\t_{ji;\h_{ij}\g_{ij}}A_{i;jj;\g_{ij}\h_{ij}}
\overset{\text{(\ref{eq:canonicaldisintegerationds})}}{=\joinrel=\joinrel=\joinrel=}
R_{ji}'(A_{i})
\qquad\forall\;A_{i}\in\mathcal{M}_{m_{i}}(\C).
\end{split}
\ee
This concludes the proof of the theorem.
\eprf

\br
By applying the trace to both sides of (\ref{eq:separabilitydisrectsumcase}), one obtains
$p_{i}=\sum\limits_{j=1}^{t}q_{j}\tr(\t_{ji})$ for all $i\in\{1,\dots,s\}.$
\er

An immediate corollary of Theorem~\ref{thm:theoremaegregium}
is the standard existence and uniqueness theorem of 
regular conditional probabilities from classical finite probability. 
We work this out in full detail as an example. 

\bx
\label{ex:classicaldisintegrations}
Using the notation from Theorem~\ref{thm:theoremaegregium}, 
suppose $m_{i}=1$ and $n_{j}=1$ for all $i,j.$ 
Then, $\rho_{i}=1=\sigma_{j}$ for all $i,j.$ 
Furthermore, since each $m_{i}=1,$ the multiplicity 
is drastically restricted since $m_{i}=\sum\limits_{j=1}^{t}c_{ij}n_{j}.$ 
By this equality, for each $i,$ there exists a unique $j$ such that
$c_{ij}=1$ and all other $c_{ik}=0.$ In other words, 
there exists a unique function $f:\{1,\dots,s\}\to\{1,\dots,t\}$ 
such that 
\be
c_{ij}=\de_{f(i)j}
\equiv
\begin{cases}
1&\mbox{ if }f(i)=j\\
0&\mbox{ otherwise}
\end{cases}
.
\ee
This implies 
\be
\label{eq:Fcommutativecase}
F=
\begin{bmatrix}
\de_{f(1)1}&\cdots&\de_{f(1)t}\\
\vdots&&\vdots\\
\de_{f(s)1}&\cdots&\de_{f(s)t}\\
\end{bmatrix}
\quad\text{ and }\quad
F^{*}=
\begin{bmatrix}
\de_{f(1)1}&\cdots&\de_{f(s)1}\\
\vdots&&\vdots\\
\de_{f(1)t}&\cdots&\de_{f(s)t}\\
\end{bmatrix}
.
\ee
Hence, 
\be
\label{eq:commutativestatepreserving}
q_{j}\overset{\text{(\ref{eq:qjsigmaj})}}{=\joinrel=\joinrel=\joinrel=}
\sum_{i=1}^{s}p_{i}F^{*}_{ij}(1)
\overset{\text{(\ref{eq:Fcommutativecase})}}{=\joinrel=\joinrel=\joinrel=}
\sum_{i=1}^{s}p_{i}\de_{f(i)j}
=\sum_{i\in f^{-1}(j)}p_{i},
\ee
which reproduces the probability-preserving condition $(X,p)\xrightarrow{f}(Y,q)$, assuming 
$X=\{1,\dots,s\}$ and $Y=\{1,\dots,t\}.$ 
In what follows, we will construct, without any additional
assumptions, non-negative matrices
$\t_{ji}\in\mathcal{M}_{c_{ij}}(\C)$ satisfying 
(\ref{eq:taujimatrices}) and (\ref{eq:separabilitydisrectsumcase})
as well as a disintegration
\be
R=\begin{bmatrix}
r_{11}&\cdots&r_{1s}\\
\vdots&&\vdots\\
r_{t1}&\cdots&r_{ts}\\
\end{bmatrix}
.
\ee
This will prove that a disintegration automatically exists in this case. 
First note that if $j\ne f(i),$ the set $\mathcal{M}_{c_{ij}}(\C)$ is just a singleton
so that we only have a chance of constructing $\t_{ji}$ when $j=f(i).$ 
In this case, $c_{if(i)}=1$ and such a matrix will be a $1\times1$ 
matrix, i.e.\ a non-negative number. 
We set 
\be
\label{eq:taujicommutativecase}
\t_{ji}:=
\begin{cases}
p_{i}/q_{j}&\mbox{ if $q_{j}>0$ and $j=f(i)$}\\
\#&\mbox{ if $q_{j}=0$ and $j=f(i)$}\\
\text{DNE}&
\mbox{ if $c_{ij}=0$}\\
\end{cases}
,
%.
\ee
where $\#$ can be chosen to be any non-negative number. 
Note that if there exists a $j\in\{1,\dots,t\}$ for which 
$c_{ij}=0$ for all $i\in\{1,\dots,s\},$ then $q_{j}=0$
by part~\ref{propitem:injectiveF} of Proposition~\ref{prop:Fpreservesstatesds}. 
For such $j,$ $\t_{ij}$ cannot be defined for any $i\in\{1,\dots,s\}.$ 
Nevertheless, 
\be
\tr\left(\sum_{i=1}^{s}\t_{ji}\right)
=\sum_{i=1}^{s}\t_{ji}
=\sum_{i\in f^{-1}(j)}\frac{p_{i}}{q_{j}}
\overset{\text{(\ref{eq:commutativestatepreserving})}}{=\joinrel=\joinrel=\joinrel=}
1
\qquad\forall\;j\in\{1,\dots,s\}\setminus{N}_{q}
\ee
proves (\ref{eq:taujimatrices}). 
Secondly, because there exists a unique $j$ for each $i$ such that $c_{ij}=1,$ 
\be
\begin{split}
\mathrm{diag}(q_{1}\t_{1i}\otimes\s_{1},\dots,q_{t}\t_{ti}\otimes\s_{t})
&=q_{f(i)}\t_{f(i)i}
=q_{f(i)}
\begin{cases}
p_{i}/q_{f(i)}&\mbox{ if $q_{f(i)}>0$}\\
\#&\mbox{ if $q_{f(i)}=0$}\\
\end{cases}
\\
&=
\begin{cases}
p_{i}&\mbox{ if $q_{f(i)}>0$}\\
0&\mbox{ if $q_{f(i)}=0$}\\
\end{cases}
\qquad\forall\;i\in\{1,\dots,s\}.
\end{split}
\ee
Note that if $q_{f(i)}=0,$ then $p_{i}=0$ by (\ref{eq:commutativestatepreserving}). Hence, this proves
(\ref{eq:separabilitydisrectsumcase}). 
Although this already proves a disintegration exists via Theorem~\ref{thm:theoremaegregium}, it is
fruitful to construct it based on the proof of 
Theorem~\ref{thm:theoremaegregium} and compare it to the classical 
disintegration from Theorem~\ref{thm:classicalmeasprestocondition}. 
Using the construction of a disintegration from 
(\ref{eq:canonicaldisintegerationds}), we get
\be
\begin{split}
r_{ji}
&\overset{\text{(\ref{eq:canonicaldisintegerationds})}}{=\joinrel=\joinrel=\joinrel=}
\begin{cases}
\t_{ji}&\mbox{ if $q_{j}>0$ and $c_{ij}=1$}\\
0&\mbox{ if $q_{j}>0$ and $c_{ij}=0$}\\
1/s&\mbox{ if $q_{j}=0$}\\
\end{cases}
\\
&\overset{\text{(\ref{eq:taujicommutativecase})}}{=\joinrel=\joinrel=\joinrel=}
\begin{cases}
p_{i}/q_{j}&\mbox{ if $q_{j}>0$ and $j=f(i)$}\\
0&\mbox{ if $q_{j}>0$ and $j\ne f(i)$}\\
1/s&\mbox{ if $q_{j}=0$}\\
\end{cases}
\\
&=
\begin{cases}
p_{i}\de_{f(i)j}/q_{j}&\mbox{ if $q_{i}>0$}\\
1/s&\mbox{ if $q_{j}=0$}
\end{cases}
.
\end{split}
\ee
This reproduces formula (\ref{eq:fromfandptor}) for an ordinary disintegration. 
\ex

Finally, we end this section with a generalization of Theorems~\ref{thm:theoremaegregium} and \ref{thm:existenceofCPUdisintegration}
by allowing for arbitrary (unital) $^*$-homomorphisms $F:\mB\to\mA.$ 

\bt
\label{thm:theoremaegregiumarbitraryF}
Let $\mathcal{A},\mathcal{B},F,\w,$ and $\xi$ be as in 
Notation~\ref{not:dsumsection}  
and Proposition~\ref{prop:Fpreservesstatesds} except that $F$ is now
an arbitrary (unital) $^{*}$-homomorphism, but not necessarily of the form~(\ref{eq:FBratteli}). Then, a disintegration $R$ exists 
if and only if 
there exist unitary matrices $U_{i}\in\mathcal{M}_{m_{i}}(\C)$ and 
non-negative matrices $\t_{ji}\in\mathcal{M}_{c_{ij}}(\C)$ such that
$\mathrm{Ad}_{\vec{U}}\circ F$ is of the form (\ref{eq:FBratteli}), 
\be
\label{eq:taujimatricesarbitraryF}
\tr\left(\sum_{i=1}^{s}\t_{ji}\right)=1\qquad\forall\;j\in\{1,\dots,t\}\setminus{N}_{q}
\ee
and
\be
\label{eq:separabilitydisrectsumcasearbitraryF}
p_{i}U_{i}^{\dag}\rho_{i}U_{i}=
\mathrm{diag}(q_{1}\t_{1i}\otimes\s_{1},\dots,q_{t}\t_{ti}\otimes\s_{t})
\qquad\forall\;i\in\{1,\dots,s\}.
\ee
Furthermore, any two such disintegrations are unique $\xi$-a.e.
\et

\bprf
This follows from an argument analogous to the proof of Theorem~\ref{thm:existenceofCPUdisintegration}. 
\eprf

%%%%%%%%%%%%%%%%%%%%%%%%%%%%%%%%%%%%%
\section{Example: measurement in quantum mechanics} 
\label{sec:measurement}
%%%%%%%%%%%%%%%%%%%%%%%%%%%%%%%%%%%%%

It is instructive to work out the following example due to its connection with measurement in quantum mechanics (it may be helpful at this point to review Example~\ref{ex:aecommutativealgebras} for notation). 
We will also avoid using the results of Theorem~\ref{thm:theoremaegregium} and will instead provide a self-contained analysis since this is simple enough in this special case. 
Fix $m\in\N$, let $A\in\mathcal{M}_{m}(\C)$ be a self-adjoint matrix with spectrum $\s(A)\subseteq\R$, and let $\w=\tr(\rho\;\cdot\;):\mathcal{M}_{m}(\C)\stoch\C$ be a state. The matrix $A$ induces the $^*$-homomorphism uniquely determined by 
\be
\label{eq:measurementtospectrum}
\begin{split}
\C^{\s(A)}&\xrightarrow{F}\mathcal{M}_{m}(\C)\\
e_{\l}&\mapsto P_{\l}, 
\end{split}
\ee
where $P_{\l}$ is the orthogonal projection onto the $\l$ eigenspace associated to $A.$ 
This pulls back the state $\w$ to a probability measure $q$ on $\s(A)$ whose evaluation on $\l\in\s(A)$ will be denoted by $q_{\l}.$ The pullback state will be denoted by $\<q,\;\cdot\;\>$, where $\<\;\cdot\;,\;\cdot\;\>$ is the natural inner product on $\C^{\s(A)}$ induced by the basis $\{e_{\l}\}_{\l\in\s(A)}.$ Physically, the number $q_{\l}$ is interpreted as the probability that the state $\w$ takes the value $\l$ when the observable $A$ is measured. 
If a disintegration $R:\mathcal{M}_{m}(\C)\stoch\C^{\s(A)}$ exists, it is uniquely determined by the collection of PU maps $R_{\l}$ defined by
\be
\begin{split}
\mathcal{M}_{m}(\C)&\xstoch{R}\C^{\s(A)}\xrightarrow{\mathrm{ev}_{\l}}\C\\
A&\xmapsto{\hspace{11.7mm}R_{\lambda}\hspace{11.7mm}}\<e_{\l},R(A)\>
\end{split}
\ee
and indexed by $\l\in\s(A).$  Because these are states on $\mathcal{M}_{m}(\C)$, they uniquely determine a density matrix $\rho_{\l}\in\mathcal{M}_{m}(\C)$, i.e.\ 
\be
\label{eq:Rpropertiesmeasurement}
R_{\l}=\tr(\rho_{\l}\;\cdot\;),\quad
\rho_{\l}\ge0,\quad
\tr(\rho_{\l})=1,\quad\text{ and }\quad
R=\sum_{\l\in\s(A)}\mathrm{ev}_{\l}^*\circ R_{\l}.
\ee 
Because $R$ must be state-preserving to be a disintegration, this entails
\be
\label{eq:Rlambdastatepreserving}
\tr(\rho\;\cdot\;)
=\<{q},R(\;\cdot\;)\>
\overset{\text{(\ref{eq:Rpropertiesmeasurement})}}{=\joinrel=\joinrel=}
\sum_{\l\in\s(A)}q_{\l}\tr(\rho_{\l}\;\cdot\;)
\;\;\implies\;\;
\rho=\sum_{\l\in\s(A)}q_{\l}\rho_{\l}.
\ee
The other condition for $R$ to be a disintegration is 
$R\circ F\underset{\raisebox{.6ex}[0pt][0pt]{\scriptsize$\<q,\;\cdot\;\>$}}{=\joinrel=\joinrel=}\id_{\C^{\s(A)}}$, 
which says
\be
\sum_{\l\in\s(A)}b_{\l}e_{\l}-R\left(F\left(\sum_{\l\in\s(A)}b_{\l}e_{\l}\right)\right)\in\mathcal{N}_{\<q,\;\cdot\;\>}\qquad\forall\;\sum_{\l\in\s(A)}b_{\l}e_{\l}\in\C^{\s(A)}.
\ee
Expanding this out, relabeling indices, and using part \ref{partioflem:aeidentitysupport} of Lemma~\ref{lem:aeidentitysupport} gives 
\be
\sum_{\l\in\s(A)\setminus N_{q}}b_{\l}e_{\l}
=\sum_{\l\in\s(A)\setminus N_{q}}\left(\sum_{\mu\in\s(A)}b_{\mu}\tr(\rho_{\l}P_{\mu})\right)e_{\l}.
\ee
Linear independence of the  $e_{\l}$ then gives the constraints
\be
b_{\l}=\sum_{\mu\in\s(A)}b_{\mu}\tr(\rho_{\l}P_{\mu})
\qquad\forall\;\l\in\s(A)\setminus N_{q}.
\ee
Since the $b$'s can be chosen arbitrarily and independently (indeed, set $b_{\mu}:=\de_{\mu\nu}$ for various $\nu$), we conclude 
\be
\tr(\rho_{\l}P_{\mu})=\de_{\l\mu}
\qquad\forall\;\mu,\l\in\s(A)\setminus N_{q}. 
\ee 
Since $\rho_{\l}$ is a positive matrix, 
$\tr(\rho_{\l}P_{\mu})=\tr(P_{\mu}\rho_{\l}P_{\mu})=0$ if and only if $P_{\mu}\rho_{\l}P_{\mu}=0$ whenever $\mu\ne\l.$ 
In what follows, we will prove $\rho_{\l}=P_{\l}\rho_{\l}P_{\l}.$ To see this, first let $\vec{u}\in\mathrm{Im}(P_{\mu})$ and $\vec{v}\in\mathrm{Im}(P_{\nu})$, where $\mu,\nu\in\s(A)\setminus\{\l\}$. Then $P_{\mu},P_{\nu}\le P_{\l}^{\perp}$ and
\be
\begin{split}
0&\le\<\vec{u}+\vec{v},P_{\l}^{\perp}\rho_{\l}P_{\l}^{\perp}(\vec{u}+\vec{v})\>\\
&=\<\vec{u},P_{\l}^{\perp}\rho_{\l}P_{\l}^{\perp}\vec{u}\>+\<\vec{u},P_{\l}^{\perp}\rho_{\l}P_{\l}^{\perp}\vec{v}\>+\<\vec{v},P_{\l}^{\perp}\rho_{\l}P_{\l}^{\perp}\vec{u}\>+\<\vec{v},P_{\l}^{\perp}\rho_{\l}P_{\l}^{\perp}\vec{v}\>\\
&=\<\vec{u},P_{\mu}\rho_{\l}P_{\mu}\vec{u}\>+\<\vec{u},P_{\mu}\rho_{\l}P_{\nu}\vec{v}\>+\<\vec{v},P_{\nu}\rho_{\l}P_{\mu}\vec{u}\>+\<\vec{v},P_{\nu}\rho_{\l}P_{\nu}\vec{v}\>\\
&=2\Re\<\vec{u},P_{\mu}\rho_{\l}P_{\nu}\vec{v}\>,
\end{split}
\ee 
where we have freely used the facts $P_{\l}^{\perp}\vec{u}=P_{\mu}\vec{u}=\vec{u}$ and $P_{\l}^{\perp}\vec{v}=P_{\nu}\vec{v}=\vec{v}$ together with the self-adjointness and orthogonality of these projections. Since $\vec{u}$ and $\vec{v}$ can be arbitrary, positivity of $P_{\l}^{\perp}\rho_{\l}P_{\l}^{\perp}$ guarantees that $P_{\mu}\rho_{\l}P_{\nu}=0$ for all $\mu,\nu\in\s(A)\setminus\{\l\}.$ 
So far, we have shown $\rho_{\l}=P_{\l}\rho_{\l}P_{\l}+P_{\l}^{\perp}\rho_{\l}P_{\l}+P_{\l}\rho_{\l}P_{\l}^{\perp}.$ 
What is left to show is that $P_{\mu}\rho_{\l}P_{\l}=0$ for all $\mu\in\s(A)\setminus\{\l\}$ (which would imply $P_{\l}\rho_{\l}P_{\mu}=0$ by taking the adjoint). 
Now, let $\vec{u}\in\mathrm{Im}(P_{\mu})$ and $\vec{v}\in\mathrm{Im}(P_{\l})$, where $\mu\in\s(A)\setminus\{\l\}$. Positivity of $\rho_{\l}$ gives
\be
0\le\<\vec{u}+\vec{v},\rho_{\l}(\vec{u}+\vec{v})\>
=2\Re\<\vec{u},P_{\mu}\rho_{\l}P_{\l}\vec{v}\>+\<\vec{v},\rho_{\l}\vec{v}\>
\ee
by a similar calculation and using the previous result. Since $\vec{u}$ can be chosen freely, it can be chosen so that the left term becomes arbitrarily negative unless $P_{\mu}\rho_{\l}P_{\l}=0.$ This concludes the argument that $\rho_{\l}=P_{\l}\rho_{\l}P_{\l}.$
Thus, $\rho_{\l}$ and $\rho_{\l'}$ have mutually orthogonal supports for $\l\ne\l'$ provided that $\l,\l'\in\s(A)\setminus N_{q}.$ 
Hence, although we have no restrictions on $\rho_{\l}$ when $\l\in N_{q},$ we still obtain 
\be
\label{eq:rhobreaksupintorholambda}
\rho
\overset{\text{(\ref{eq:Rlambdastatepreserving})}}{=\joinrel=\joinrel=}
\sum_{\l\in\s(A)}q_{\l}\rho_{\l}
=\sum_{\l\in\s(A)\setminus N_{q}}q_{\l}\rho_{\l}
=\sum_{\l\in\s(A)\setminus N_{q}}q_{\l}P_{\l}\rho_{\l}P_{\l},
\ee
which agrees with the result (\ref{eq:separabilitydisrectsumcase}) with respect to a spectral basis, or more accurately (\ref{eq:separabilitydisrectsumcasearbitraryF}), in this special case since $s=1$ so that there is only one $i$ index and $\s_{j}=1$ for all $j$ because $\s_{j}$ is a $1\times1$ matrix. Thus, the $\tau_{ji}$ matrices reduce to the $\rho_{\l}$ matrices. To make a more explicit connection to quantum information theory, we recall the definition of a L\"uders projection, which is a model for the ensemble of the induced states of a system after a measurement has taken place \cite{Lu06}. 

\bd
Let $\rho\in\mathcal{M}_{m}(\C)$ be a density matrix and let $A\in\mathcal{M}_{m}(\C)$ be self-adjoint with spectrum $\s(A).$ The \define{L\"uders projection of $\rho$ with respect to the measurement of $A$} is the density matrix 
\be
\rho':=\sum_{\l\in\s(A)}P_{\l}\rho P_{\l}. 
\ee
\ed

In summary, we have obtained the following theorem based on our above analysis. 

\bt
Let $A\in\mathcal{M}_{m}(\C)$ be a self-adjoint matrix with spectrum $\s(A)$, let $F:\C^{\s(A)}\to\mathcal{M}_{m}(\C)$ be as in (\ref{eq:measurementtospectrum}), and let $\w=\tr(\rho\;\cdot\;):\mathcal{M}_{m}(\C)\stoch\C$ be a state with $\<q,\;\cdot\;\>:=\omega\circ F$ the induced state on $\C^{\s(A)}$. 
Then $F$ has a disintegration of $\w$ over $\<q,\;\cdot\;\>$ consistent with $F$ if and only if $\rho$ equals its L\"uders projection with respect to the measurement of $A$. 
\et

\bprf
We will use the same notation as earlier in this section. 

\noindent
($\Rightarrow$)
Assume a disintegration exists. By (\ref{eq:rhobreaksupintorholambda}), 
$P_{\l}\rho P_{\l}=q_{\l}\rho_{\l}$ for all $\l\in\s(A)\setminus N_{q}.$ Hence, 
\be
\label{eq:rholambdaisLuders}
\rho_{\l}=\frac{P_{\l}\rho P_{\l}}{q_{\l}}=\frac{P_{\l}\rho P_{\l}}{\tr(\rho P_{\l})}\qquad\forall\;\l\in\s(A)\setminus N_{q}
\ee
because $\rho_{\l}$ is a density matrix. 
Furthermore, since a disintegration exists,
\be
\rho
\overset{\text{(\ref{eq:Rlambdastatepreserving})}}{=\joinrel=\joinrel=}
\sum_{\l\in\s(A)\setminus N_{q}}q_{\l}\rho_{\l}
\overset{\text{(\ref{eq:rholambdaisLuders})}}{=\joinrel=\joinrel=}
\sum_{\l\in\s(A)\setminus N_{q}}q_{\l}\frac{P_{\l}\rho P_{\l}}{q_{\l}}
=\sum_{\l\in\s(A)\setminus N_{q}}P_{\l}\rho P_{\l}.
\ee

\noindent
($\Leftarrow$)
Suppose $\rho$ equals its L\"uders projection, i.e.\ suppose 
\be
\rho=\sum_{\l\in\s(A)}P_{\l}\rho P_{\l}. 
\ee
For each $\l\in\s(A)$, set $R_{\l}:\mathcal{M}_{m}(\C)\stoch\C$ to be the linear map defined by 
\be
\label{eq:disintegrationforLuders}
\mathcal{M}_{m}(\C)\ni B\mapsto
R_{\l}(B):=\begin{cases}
\tr\left(\frac{P_{\l}\rho P_{\l}}{q_{\l}}B\right)&\mbox{ if }q_{\l}>0\\
\frac{1}{m}\tr(B)&\mbox{ if }q_{\l}=0\\
\end{cases}
.
\ee
Then the linear map $R:\mathcal{M}_{m}(\C)\stoch\C^{\s(A)}$ defined by 
$R:=\sum\limits_{\l\in\s(A)}\mathrm{ev}_{\l}^*\circ R_{\l}$ is a disintegration of $F.$ To see this, first notice that $R$ is positive, which implies it is CP since $\C^{\s(A)}$ is commutative (cf.\ Theorem~3 in Stinespring~\cite{St55}). Second, $R$ is unital because
\be
R(\mathds{1}_{m})
=\sum_{\l\in\s(A)\setminus N_{q}}R_{\l}(\mathds{1}_{m})e_{\l}+\sum_{\l\in N_{q}}R_{\l}(\mathds{1}_{m})e_{\l}
\overset{\text{(\ref{eq:disintegrationforLuders})}}{=\joinrel=\joinrel=}
\sum_{\l\in\s(A)\setminus N_{q}}e_{\l}+\sum_{\l\in N_{q}}e_{\l}=\sum_{\l\in\s(A)}e_{\l}
\ee
since $\tr(P_{\l}\rho P_{\l})=q_{\l}$ for all $\l\in\s(A)$. 
To show $R$ satisfies $R\circ F\underset{\raisebox{.6ex}[0pt][0pt]{\scriptsize$\<q,\;\cdot\;\>$}}{=\joinrel=\joinrel=}\id_{\C^{\s(A)}},$ we will show $R(F(e_{\mu}))-e_{\mu}\in\mathcal{N}_{\<q,\;\cdot\;\>}$ for all $\mu\in\s(A)$. Setting $d_{\mu}:=\tr(P_{\mu}),$ the degeneracy/mutliplicity of $\mu\in\s(A),$ we obtain 
\be
\begin{split}
R\big(F(e_{\mu})\big)
&
\overset{\text{(\ref{eq:measurementtospectrum})}}{=\joinrel=\joinrel=}
R(P_{\mu})
=\sum_{\l\in\s(A)\setminus N_{q}}R_{\l}(P_{\mu})e_{\l}+\sum_{\l\in N_{q}}R_{\l}(P_{\mu})e_{\l}\\
&
\overset{\text{(\ref{eq:disintegrationforLuders})}}{=\joinrel=\joinrel=}
\sum_{\l\in\s(A)\setminus N_{q}}\de_{\mu\l}e_{\l}+\sum_{\l\in N_{q}}\frac{d_{\mu}}{m}e_{\l}
=\begin{cases}
e_{\mu}+\frac{d_{\mu}}{m}\sum\limits_{\scriptscriptstyle\l\in N_{q}}e_{\l}&\mbox{ if }q_{\mu}>0\\
\frac{d_{\mu}}{m}\sum\limits_{\scriptscriptstyle\l\in N_{q}}e_{\l}&\mbox{ if }q_{\mu}=0\\
\end{cases}
\end{split}
\ee
Hence, $R\big(F(e_{\mu})\big)-e_{\mu}\in\C^{N_{q}},$ which is the null space of $\<q,\;\cdot\;\>$. Thus, $R$ is a disintegration. 
\eprf

\appendix

%%%%%%%%%%%%%%%%%%%%%%%%%%%%%%%%%%%%%
\section{Equivalent definitions of disintegration} 
\label{app:optimalhyparedis}
%%%%%%%%%%%%%%%%%%%%%%%%%%%%%%%%%%%%%

In this appendix, we review the definition of a disintegration 
from measure theory (cf.\ Definition~452E in Fremlin~\cite{FrV4}). 
Tables~\ref{table:disintegration} and \ref{table:consistentdisintegration}
provide two, a-priori different, definitions of a disintegration with varying input data and consistency conditions. 
This appendix serves to explain how these definitions are related to each other.
More precisely, 
Theorems~\ref{thm:disintegrationequivalence}
and \ref{thm:consistentdisintegrationequivalence}
state that the definitions in the respective tables are equivalent. 
Theorem~\ref{prop:conditionalprobability} says that this diagrammatic
definition of a disintegration is equivalent to the definition 
of a regular conditional probability (cf.\ Definition~2.1 in Panagaden~\cite{Pa99}). 

In all that follows, $(X,\S,\mu)$ and $(Y,\W,\nu)$ are measure spaces with no additional assumptions other than $\mu$ and $\nu$ are non-negative measures. Furthermore, $f:X\to Y$ is taken to be measure-preserving so that the pushforward $f_{*}\mu$ of $\mu$ along $f$ is $\nu,$ i.e.\ $\nu(F)=\mu(f^{-1}(F))$ for all $F\in\W.$ 

\bd
\label{defn:transitionkernel}
Let $(X,\S)$ and $(Y,\W)$ be measurable spaces. 
A \define{transition kernel}
$r$ from $(Y,\W)$ to $(X,\S)$, written $r:Y\stoch X,$ is a function 
$r:Y\times\S\to[0,\infty]$ such that 
\begin{enumerate}[i.]
\itemsep0pt
\item
$r(y,\;\cdot\;):\S\to[0,\infty]$ is a measure for all $y\in Y$ and
\item
$r(\;\cdot\;,E):Y\to[0,\infty]$ is measurable for all $E\in\S.$
\end{enumerate}
The notation $r_{y}(E):=r(y,E)$ will be implemented. A transition kernel as above is called a \define{stochastic map} (also \define{Markov kernel}) when $r_{y}$ is a probability measure for all $y\in Y$.  
\ed

Transition kernels are generalizations of measurable functions in that they assign to each point in the source/domain a measure
on the target/codomain (cf.\ Example~\ref{ex:fromfunctiontotransitionkernel}). If a function is to be thought of as a deterministic
process, a transition kernel whose associated measures are probability measures can be interpreted as a non-deterministic (i.e.\ stochastic)
process, where one only knows the probabilities associated with the possible outcomes of that process. 

\bx
Let $(X,\S)$ be a measurable space and 
let $\{\bullet\}$ denote a one element set with the unique $\s$-algebra.
There is a bijection between the set of 
measures on $(X,\S)$ and the set of transition kernels 
$\{\bullet\}\stoch X$ from $\{\bullet\}$ to $X.$
This allows measures to be viewed as morphisms.%
%%footnote
%\footnote{This is an idea that goes back to Lawvere~\cite{La62}.}
%%end footnote
\ex

\bd
\label{defn:compositionoftransitionkernels}
Let $(X,\S),$ $(Y,\W),$ and $(Z,\Xi)$ be measurable spaces. 
Let $\mu:X\stoch Y$ and $\nu:Y\stoch Z$ be two transition kernels. 
The \define{composite} of $\mu$ followed by $\nu,$
written as $\nu\circ\mu:X\stoch Z,$ is defined by
\be
X\times\Xi\ni(x,E)\mapsto(\nu\circ\mu)_{x}(E)
:=\int_{Y}\nu_{y}(E)\;d\mu_{x}(y).
\ee
This equation is known as the \define{Chapman--Kolmogorov} equation.
\ed

The fact that the composite of transition kernels defines a 
transition kernel follows from the monotone convergence theorem~\cite[Theorem~1.26]{Ru87}. Rather than proving this here, 
we will recall techniques from analysis that can be used to prove this when we prove Theorem~\ref{thm:disintegrationequivalence} below. 

\bx
\label{ex:fromfunctiontotransitionkernel}
Let $(X,\S)$ and $(Y,\W)$ be measurable spaces and let $\mu:\{\bullet\}\stoch X$ be a 
measure on $X$ and let $f:X\to Y$ be a measurable function. Then
$f$ can be viewed as the transition kernel $f:X\stoch Y$ given by 
\be
X\times\W\ni(x,E)\mapsto f_{x}(E):=\chi_{E}\big(f(x)\big):=\begin{cases}1&\mbox{if }f(x)\in E\\0&\mbox{otherwise}\end{cases}.
\ee
Furthermore, $f\circ\mu$ is the pushforward $f_{*}\mu$ of the measure $\mu$ along the map $f$ because
\be
\W\ni E\mapsto(f\circ\mu)(E)
=\int_{X}f_{x}(E)\;d\mu(x)
=\int_{X}(\chi_{E}\circ f)\;d\mu
=\int_{Y}\chi_{E}\;d(f_{*}\mu)
=\mu\big(f^{-1}(E)\big).
\ee
A special case of this occurs for the diagonal map
$\D_{Y}:Y\to Y\times Y$. This pushes forward a
probability measure $q:\{\bullet\}\stoch Y$ to the diagonal subset
\be
\D_{Y}(Y):=\big\{(y,y)\in Y\times Y\;:\;y\in Y\big\}
\ee
of $Y\times Y$ so that $(\D_{Y}\circ q)(A\times B)=q(A\cap B)$ for all $A,B\in\Omega$. 
This map is used to instantiate a categorical formulation of a.e.\ equivalence (cf.\ Remark~\ref{rmk:aeChoJacobs}). 
\ex

\bx
Let $X,Y,$ and $Z$ be finite sets equipped with the discrete $\s$-algebra
and suppose that all transition kernels are stochastic maps.
Then Definitions~\ref{defn:transitionkernel}
and \ref{defn:compositionoftransitionkernels}
reproduce the notion of \define{stochastic matrices} including their compositions. Indeed, since the $\s$-algebra on $Y$ is discrete, 
\be
f_{x}(E)=\sum_{y\in E}f_{x}(\{y\})
\ee
so that the probability measure $f_{x}$ is determined by its values
on points of $Y.$ We therefore write $f_{yx}:=f_{x}(\{y\})$ to denote the
$yx$ entry of $f$ in matrix form. Second, the composite $X\xstoch{f}Y\xstoch{g}Z$ sends $x\in X$ to the probability
measure on $Z$ determined by
\be
\label{eq:compositionstochasticmaps}
\begin{split}
Z \ni z \xmapsto{(g\circ f)(x)} (g\circ f)_{zx}
:= \sum_{y\in Y}g_{zy}f_{yx}.
\end{split}
\ee
Following Example~\ref{ex:fromfunctiontotransitionkernel}, when $Y$ is a finite set equipped with the discrete $\s$-algebra, the pushforward of $q:\{\bullet\}\stoch Y$ along $\Delta_{Y}:Y\to Y\times Y$ simplifies to $(\D_{Y}\circ q)_{(y,y')}=\de_{yy'}q_{y}.$
\ex

With these definitions in place, we can compare several definitions of disintegrations. Table~\ref{table:disintegration} below describes three equivalent
definitions of a disintegration of one measure over another together with a list of references that use said definition.

\begin{center}
\begin{tabular}{|c|c|c|c|}
\hline
 & Functional & Measure-theoretic & Diagrammatic\\
\hline
Data
&
%$r:Y\times\S\to[0,\infty]$
\begin{tabular}{c}
transition kernel\\
$r:Y\stoch X$\\
\end{tabular}
&
%$r:Y\times\S\to[0,\infty]$
\begin{tabular}{c}
transition kernel\\
$r:Y\stoch X$\\
\end{tabular}
&
\begin{tabular}{c}
transition kernel\\
$r:Y\stoch X$\\
\end{tabular}
\\
\hline
Conditions
&
\begin{tabular}{c}
$\int_{X}h\;d\mu=\int_{Y}\left(\int_{X}h\;dr_{y}\right)d\nu(y)$\\
$\forall$ measurable $h:X\to[0,\infty]$
\end{tabular}
&
\begin{tabular}{c}
$\mu(E)=\int_{Y}r_{y}(E)\;d\nu(y)$ \\
$\forall\;E\in\S$
\end{tabular}
&
\begin{tabular}{c}
$\xy0;/r.25pc/:
(0,7.5)*+{\{\bullet\}}="o";
(-10,-7.5)*+{X}="X";
(10,-7.5)*+{Y}="Y";
{\ar@{~>}"o";"X"_{\mu}};
{\ar@{~>}"o";"Y"^{\nu}};
{\ar@{~>}"Y";"X"^{r}};
{\ar@{=}(-3,0);(5,-4.5)};
\endxy$
\\
i.e.\ $r\circ\nu=\mu$
\end{tabular}
\\
\hline
References
&
\cites{Ro52,GaPa18}
&
\cites{We94,FrV4,Mi11,Ch73}
&
\cite{CDDG17}
\\
\hline
\end{tabular}
\captionof{table}{Three definitions of a disintegration of 
$(X,\S,\mu)$ over $(Y,\W,\nu)$.}
\label{table:disintegration}
\end{center}
%%%%%%%%%%%%%%%%

\bt
\label{thm:disintegrationequivalence}
Given a transition kernel $Y\xstoch{r}X$ from a measure space $(Y,\W,\nu)$ to a measure space $(X,\S,\mu),$ 
the three conditions in Table~\ref{table:disintegration} are equivalent.
\et

\bprf
The equivalence between the measure-theoretic definition and the diagrammatic definition is immediate from the definition of the composition of transition kernels. Therefore, it suffices to prove the equivalence between the measure-theoretic and functional definitions. By setting $h:=\chi_{E}$ with $E\in\S,$ the measure-theoretic condition follows from the functional definition. The only slightly non-trivial part of the proof of this equivalence is showing that the measure-theoretic definition implies the functional one. First, a straightforward computation, using $r\circ\nu=\mu$, shows  
\be
\label{eq:simplefunctions}
\int_{X}s\;d\mu=\int_{Y}\left(\int_{X}s\;dr_{y}\right)d\nu(y)
\ee
for all simple functions $s:X\to[0,\infty).$ The general case for arbitrary measurable $h:X\to[0,\infty]$ follows from the monotone convergence theorem, though one needs to be careful about how to choose a monotone sequence of simple functions $(s_{n})$ converging pointwise to $h.$ Such a sequence can be obtained as in the proof of Theorem~2.10 in Folland~\cite{Fo99} (cf.\ Lemma~4.10 in \cite{Pa17}). From such a choice, it follows that 
\be
\N\ni n\mapsto\left(Y\ni y\mapsto\int_{X}s_{n}\;dr_{y}\right)
\ee
is a monotone increasing sequence of measurable functions on $Y$ (see Equation~(4.94) in the proof of part~iii of Proposition~4.79 of \cite{Pa17} for details). Using all of these facts gives
\be
\begin{split}
\int_{X}h\;d\mu&=\lim_{n\to\infty}\int_{X}s_{n}\;d\mu\quad\text{ by definition of $\int$ w.r.t. $\mu$}\\
&=\lim_{n\to\infty}\int_{Y}\left(\int_{X}s_{n}\;dr_{y}\right)d\nu(y)\quad\text{ by (\ref{eq:simplefunctions})  for simple $s_{n}$}\\
&=\int_{Y}\lim_{n\to\infty}\left(\int_{X}s_{n}\;dr_{y}\right)d\nu(y)\quad\text{ by the monotone convergence theorem}\\
&=\int_{Y}\left(\int_{X}h\;dr_{y}\right)d\nu(y)\quad\text{ by definition of $\int$ w.r.t. $r_{y}$}
\end{split}
\ee
for arbitrary measurable $h:X\to[0,\infty].$ 
\eprf

When one is equipped with the additional datum of a measure-preserving map $f:X\to Y,$ 
there is another 
coherence condition that can be enforced on disintegrations.
This assumption is to demand that a disintegration $r:Y\stoch X$ be \emph{consistent} with 
the map $f.$ From our diagrammatic perspective, 
this means $r$ is a (stochastic) section of 
$f$ a.e. This is described in Table~\ref{table:consistentdisintegration}.

\bd
\label{defn:aeequivalencetransitionkernels}
Let $(X,\S,\mu)$ and $(Y,\W,\nu)$ be two measure spaces. 
Two transition kernels $f,g:X\stoch Y$ are said to be 
\define{$\mu$-a.e.\ equivalent}, written as 
$f\underset{\raisebox{.6ex}[0pt][0pt]{\scriptsize$\mu$}}{=}g,$
iff for each $F\in\W,$ there exists a measurable set $N_{F}\in\S$ such that
\be
f_{x}(F)=g_{x}(F)\quad\forall\;x\in X\setminus N_{F}
\qquad\text{and}\qquad
\mu(N_{F})=0.
\ee
\ed

\bx
The definition of a.e.\ equivalence takes a particularly simple form for finite sets. 
Let $X$ and $Y$ be finite sets and let $\mu$ be a measure on $X.$
Let 
\be
N_{\mu}:=\{x\in X\;:\;\mu_{x}=0\}
\ee
denote the \define{null-set} of $(X,\mu)$. 
Two transition kernels $f,g:X\stoch Y$ are 
\define{$\mu$-a.e.\ equivalent} iff 
\be
\big\{x\in X\;:\;f_{x}\ne g_{x}\big\}\subseteq N_{\mu}.
\ee
The notation $f\underset{\raisebox{.6ex}[0pt][0pt]{\scriptsize$\mu$}}{=}g$ is
used whenever $f$ and $g$ are $\mu$-a.e.\ equivalent.
Here, $f_{x}\ne g_{x}$ means 
$f_{x}$ and $g_{x}$ are different measures on $Y,$ i.e.\
there exists a $y\in Y$ such that $f_{yx}\ne g_{yx}.$
\ex

\br
\label{rmk:aeChoJacobs}
The definition of a.e.\ equivalence in Definition~\ref{defn:aeequivalencetransitionkernels} is a bit subtle in the general measure-theoretic case. Another reasonable option would be to say $f$ and $g$ are $\mu$-a.e.\ equivalent iff there exists an $N\in\S$ such that $f_{x}=g_{x}$ (equality of measures) for all $x\in X\setminus N$ and $\mu(N)=0$. However, this definition is too strong for the conditions in Table~\ref{table:consistentdisintegration} to be equivalent for arbitrary measure spaces. The definition we have chosen agrees with the diagrammatic definition of Cho and Jacobs~\cite[Section~5]{ChJa18}, which says that the diagram 
\be
\xy;/r.25pc/:
(-30,0)*+{\{\bullet\}}="0";
(-15,7.5)*+{X}="Xt";
(-15,-7.5)*+{X}="Xb";
(7.5,7.5)*+{X\times X}="XXt";
(7.5,-7.5)*+{X\times X}="XXb";
(30,0)*+{X\times Y}="XY";
{\ar@{~>}"0";"Xt"^{\mu}};
{\ar@{~>}"0";"Xb"_{\mu}};
{\ar"Xt";"XXt"^(0.45){\D_{X}}};
{\ar"Xb";"XXb"_(0.45){\D_{X}}};
{\ar@{~>}"XXt";"XY"^{\id_{X}\times f}};
{\ar@{~>}"XXb";"XY"_{\id_{X}\times g}};
\endxy
\ee
commutes (the product of stochastic maps can be defined using joint probability measures as is done in Section~2 of \cite{ChJa18}).
\er

\begin{center}
\begin{tabular}{|c|c|c|}
\hline
 & Measure-theoretic & Diagrammatic\\
\hline
Data
&
\begin{tabular}{c}
transition kernel\\
$r:Y\stoch X$\\
\end{tabular}
&
\begin{tabular}{c}
transition kernel\\
$r:Y\stoch X$\\
\end{tabular}
\\
\hline
\begin{tabular}{c}Conditions\\besides $r$ is a \\disintegration of\\$\mu$ over $\nu$\end{tabular}
&
\begin{tabular}{c}
for each $F\in \W$\\
$\exists$ $\nu$-null set $N_{F}\in\W$\\
s.t. $r_{y}(f^{-1}(F))=1$\\
$\forall\;y\in(Y\setminus N_{F})\cap F$\\
\end{tabular}
&
\begin{tabular}{c}
$\xy0;/r.25pc/:
(0,7.5)*+{X}="X";
(10,-7.5)*+{Y}="Y1";
(-10,-7.5)*+{Y}="Y2";
{\ar@{~>}"Y1";"X"_{r}};
{\ar"X";"Y2"_{f}};
{\ar"Y1";"Y2"^{\id_{Y}}};
{\ar@{=}(-3,0);(5,-4.5)_{\nu}};
\endxy$
\\
i.e.\ $f\circ r\underset{\raisebox{.6ex}[0pt][0pt]{\scriptsize$\nu$}}{=}\id_{Y}$
\end{tabular}
\\
\hline
References
&
\cites{FrV4,Mi11}
&
\cite{CDDG17}
\\
\hline
\end{tabular}
\captionof{table}{Two definitions of a disintegration of $(X,\S,\mu)$ over $(Y,\W,\nu)$ consistent with a measure-preserving measurable map $f:X\to Y$.}
\label{table:consistentdisintegration}
\end{center}
%%%%%%%%%%%%%%%% 
More explicitly, the condition $f\circ r\underset{\raisebox{.6ex}[0pt][0pt]{\scriptsize$\nu$}}{=}\id_{Y}$ says that for each $F\in\W,$
there exists a $\nu$-null set $M_{F}\in\W$
such that $(f\circ r)_{y}(F)=\chi_{F}(y)$ for all
$y\in Y\setminus M_{F}.$ Expanding out $(f\circ r)_{y}(F)$
using Example~\ref{ex:fromfunctiontotransitionkernel} and
the definition of transition kernels,
this is equivalent
to $r_{y}\big(f^{-1}(F)\big)=\chi_{F}(y)$ for all
$y\in Y\setminus M_{F}.$ Therefore, it is immediate that
the diagrammatic definition implies the measure-theoretic one.

\bt
\label{thm:consistentdisintegrationequivalence}
Let $(Y,\W,\nu)$ and $(X,\S,\mu)$ be measure spaces. Given a measure-preserving measurable map $X\xrightarrow{f}Y,$  together with a disintegration $Y\xstoch{r}X$ of $\mu$ over $\nu,$ 
the conditions in Table~\ref{table:consistentdisintegration} are equivalent.
\et

\bprf
By the comment preceding the statement of this theorem, 
the equivalence will follow from proving the measure-theoretic 
definition implies the diagrammatic one, i.e.\ for each $F\in\W,$ there exists a 
$\nu$-null set $M_{F}\in\W$ such that $r_{y}\big(f^{-1}(F)\big)=\chi_{F}(y)$ 
for all $y\in Y\setminus M_{F}$ (cf.\ \cite[Proposition~452G]{FrV4}). 
In more detail, 
by assumption, there exist $\nu$-null sets 
$N_{F},N_{Y\setminus F},N_{Y}\in\W$
such that 
\be
\label{eq:consistentimpliesprobabilitymeasure}
r_{y}(X)=r_{y}\big(f^{-1}(Y)\big)=1\qquad\forall\;y\in(Y\setminus N_{Y})\cap Y\equiv Y\setminus N_{Y},
\ee
\be
r_{y}\big(f^{-1}(F)\big)=1\qquad\forall\; y\in (Y\setminus N_{F})\cap F,
\ee
and
\be
r_{y}\big(f^{-1}(Y\setminus F)\big)=1\qquad\forall\;y\in (Y\setminus N_{Y\setminus F})\cap (Y\setminus F)
\equiv Y\setminus (N_{Y\setminus F}\cup F).
\ee
Therefore, 
\be
\begin{split}
1-r_{y}\big(f^{-1}(F)\big)=r_{y}(X)-r_{y}\big(f^{-1}(F)\big)=r_{y}\big(f^{-1}(Y\setminus F)\big)=1\\
\forall\;y\in (Y\setminus N_{Y})\cap(Y\setminus(N_{Y\setminus F}\cup F)),\qquad\qquad\quad
\end{split}
\ee
i.e.\ 
\be
\label{eq:Fzero}
r_{y}\big(f^{-1}(F)\big)=0
\qquad\forall\;y\in (Y\setminus N_{Y})\cap(Y\setminus(N_{Y\setminus F}\cup F))
\equiv Y\setminus(N_{Y}\cup N_{Y\setminus F}\cup F).
\ee
Set 
\be
M_{F}:=N_{F}\cup N_{Y\setminus F}\cup N_{Y},
\ee
which, being the finite union of $\nu$-null sets, is $\nu$-null. 
If $y\in(Y\setminus M_{F})\cap F,$ then, in particular, $y\in (Y\setminus N_{F})\cap F$
so that $r_{y}\big(f^{-1}(F)\big)=1.$ 
If $y\in(Y\setminus M_{F})\cap(Y\setminus F)
\equiv Y\setminus(N_{Y}\cup N_{Y\setminus F}\cup N_{Y}\cup F),$ then, in particular, 
$y\in Y\setminus(N_{Y}\cup N_{Y\setminus F}\cup F)$ so that $r_{y}\big(f^{-1}(F)\big)=0.$ 
Putting these together, 
\be
r_{y}\big(f^{-1}(F)\big)=\chi_{F}(y)\qquad\forall\;y\in Y\setminus M_{F}.
\ee
Therefore, the measure-theoretic definition implies the diagrammatic one.
\eprf

\br
\label{rmk:consistentimpliesprobabilitymeasure}
The equality 
(\ref{eq:consistentimpliesprobabilitymeasure}) says
$r_{y}$ is a probability measure for all $y\in Y\setminus N_{Y}.$ 
\er

A consistent disintegration is also related to the notion of a 
regular conditional probability. 

\bd
Let $(X,\S,\mu)$ and $(Y,\W,\nu)$ be measure spaces and let 
$f:X\to Y$ be a measure-preserving map. 
A \define{regular conditional probability}
is a transition kernel $r:Y\stoch X$ for which 
there exists a $\nu$-null set $N\in\W$ such that 
$r_{y}$ is a probability measure for all $y\in Y\setminus N$ and 
\be
\label{eq:regcondprob}
\mu\big(E\cap f^{-1}(F)\big)=\int_{F}r_{y}(E)\;d\nu(y)
\qquad\forall\;E\in\S\text{ and }\forall\;F\in\W.
\ee
\ed

\bt
\label{prop:conditionalprobability}
Let $(X,\S,\mu)$ and $(Y,\W,\nu)$ be measure spaces and let 
$f:X\to Y$ be a measure-preserving map. $r:Y\stoch X$ is a regular conditional probability
if and only if it is a disintegration of $\mu$ over $\nu$ 
consistent with $f.$
\et

\bprf
{\color{white}{you found me!}}

\noindent
($\Rightarrow$)
Suppose $r$ is a regular conditional probability. Then 
\be
\mu(E)
=\mu(E\cap X)
=\mu\big(E\cap f^{-1}(Y)\big)
\overset{\text{(\ref{eq:regcondprob})}}{=\joinrel=\joinrel=}\int_{Y}r_{y}(E)\;d\nu(y)
=(r\circ\nu)(E)
\ee
for all $E\in\S.$ Now, fix $F\in\W$ 
and let $N\in\W$ be a $\nu$-null set
such that $r_{y}$ is a probability measure for all
$y\in Y\setminus N.$ Then
\be
\label{eq:regcondprobimpliesdisint}
\xy0;/r.22pc/:
(11.7557,-14.1803)*+{\ds\int_{F}d\nu}="1";
(-11.7557,-14.1803)*+{\nu(F)}="5";
(19.0211,5.1803)*+{\quad\ds\int_{F}r_{y}\big(f^{-1}(F)\big)\;d\nu(y)}="2";
(-19.0211,5.1803)*+{\mu\big(f^{-1}(F)\big)}="4";
(0,20)*+{\mu\big(f^{-1}(F)\cap f^{-1}(F)\big)}="3";
{\ar@{=}@/_0.6pc/"5";"1"}
{\ar@{=}@/_0.8pc/"2";"3"_(0.3){\text{(\ref{eq:regcondprob}) with $E=f^{-1}(F)$}}}
{\ar@{=}@/_0.8pc/"3";"4"}
{\ar@{=}@/_0.7pc/"4";"5"_{\text{since $\nu=f\circ\mu$}}}
\endxy
.
\ee
Since $r_{y}$ is a probability measure $\nu$-a.e.,
$r_{y}\big(f^{-1}(F)\big)\le1$ for all $y\in (Y\setminus N)\cap F$
so that the quantity in (\ref{eq:regcondprobimpliesdisint}) is finite. This allows us to meaningfully take the difference of these terms. 
Therefore, (\ref{eq:regcondprobimpliesdisint}) implies
$\int_{F}\big(1-r_{y}\big(f^{-1}(F)\big)\big)\;d\nu(y)=0.$
Furthermore, since the integrand is non-negative, 
there exists a $\nu$-null set $M_{F}\in\W$ such that 
\be
r_{y}\big(f^{-1}(F)\big)=1\qquad\forall\;y\in (Y\setminus (N\cup M_{F}))\cap F.
\ee
Hence, $f\circ r\underset{\raisebox{.6ex}[0pt][0pt]{\scriptsize$\nu$}}{=}\id_{Y}$ so that $r$ is a consistent disintegration. 

\noindent
($\Leftarrow$)
Conversely, suppose $r$ is a consistent disintegration. 
By Remark~\ref{rmk:consistentimpliesprobabilitymeasure}, 
$r_{y}$ is a probability measure $\nu$-a.e. 
Hence, 
\be
\xy0;/r.35pc/:
(-20,16)*+{\mu(E\cap f^{-1}(F))}="1";
(-30,0)*+{\int_{Y}r_{y}(E\cap f^{-1}(F))\;d\nu(y)}="2";
(-26,-16)*+{\int_{F}r_{y}(E\cap f^{-1}(F))\;d\nu(y)}="3";
(26,-16)*+{\begin{tabular}{c}$\int_{F}r_{y}(E\cap f^{-1}(F))\;d\nu(y)$\\$+\int_{F}r_{y}(E\cap f^{-1}(Y\setminus F))\;d\nu(y)$\end{tabular}}="4";
(30,0)*+{\int_{F}r_{y}\big((E\cap f^{-1}(F))\cup(E\cap f^{-1}(Y\setminus F))\big)\;d\nu(y)}="5";
(20,16)*+{\int_{F}r_{y}(E)\;d\nu(y)}="6";
{\ar@{=}@/_0.85pc/"1";"2"_{\mu=r\circ\nu}};
{\ar@{=}@/_0.55pc/"2";"3"_(0.45){\text{(\ref{eq:Fzero}) for $F$}}};
{\ar@{=}@/_0.85pc/"3";"4"_(0.45){\text{(\ref{eq:Fzero}) for $Y\setminus F$}}};
{\ar@{=}@/_0.55pc/"4";"5"_(0.65){\text{$r_{y}$ is countably additive}}};
{\ar@{=}@/_0.85pc/"5";"6"_{\text{set theory}}};
\endxy
\ee
for arbitrary $E\in\S$ and $F\in\W$. This proves $r$ is a regular conditional probability. 
\eprf

From this perspective, the results in this
paper can be viewed as an approach to non-commutative
regular conditional probabilities.

\section*{Acknowledgements}

The majority of this work was completed when AJP was an Assistant Research Professor at the University of Connecticut. 
We thank
Iddo Ben-Ari, 
David Fremlin,  
and Ambar Sengupta
for discussions on disintegrations and regular conditional probabilities. %. 
We also thank
Juha Javanainen for discussions on the physical consequences 
of our main theorem and for suggesting the possible relationship to measurement in quantum mechanics.
Finally, we thank John Baez, Tobias Fritz, and Tom Leinster for the inspiration for this project, which came from a thorough investigation of the calculations in Section~3 of~\cite{BFL}. 

%%%%%%%%BIBLIOGRAPHY%%%%%%%%%%%%
\addcontentsline{toc}{section}{\numberline{}Bibliography}
\bibliographystyle{hplainParzygnatv1}
\bibliography{ncprob}

\Addresses
%%based on egreg's code (see after title and authors in preamble)

\end{document}